\documentclass[a4paper,fleqn,usenatbib,useAMS]{mnras}

\usepackage{newtxtext,newtxmath}
\usepackage[T1]{fontenc}
\usepackage{ae,aecompl}

\usepackage{graphicx}	% Including figure files
\usepackage{amssymb}	% Extra maths symbols
\usepackage{multicol}        % Multi-column entries in tables
\usepackage{bm}		% Bold maths symbols, including upright Greek
\usepackage{enumitem}
\usepackage[usenames, dvipsnames]{color}

\title[On the drivers of wind structures in $\zeta$~Pup]{\emph{BRITE-Constellation} high-precision time-dependent photometry of the early-O-type supergiant $\zeta$~Puppis unveils the photospheric drivers of its small- and large-scale wind structures}

\author[Ramiaramanantsoa et al.]{Tahina Ramiaramanantsoa,$^{1,2}$\thanks{E-mail: tahina@astro.umontreal.ca}  Anthony F. J. Moffat,$^{1,2}$ Robert Harmon,$^{3}$\newauthor Richard Ignace,$^{4}$ Nicole St-Louis,$^{1,2}$ Dany Vanbeveren,$^{5}$ Tomer Shenar,$^{6}$\newauthor Herbert Pablo,$^{1,2}$ Noel D. Richardson,$^{7}$ Ian D. Howarth,$^{8}$ Ian R. Stevens,$^{9}$\newauthor Caroline Piaulet,$^{1}$  Lucas St-Jean,$^{1}$ Thomas Eversberg,$^{10}$ Andrzej Pigulski,$^{11}$ \newauthor Adam Popowicz,$^{12}$ Rainer Kuschnig,$^{13}$ El\.{z}bieta Zoc\l{}o\'nska,$^{14}$ Bram Buysschaert,$^{15,16}$\newauthor Gerald Handler,$^{14}$ Werner W. Weiss,$^{13}$ Gregg A. Wade,$^{17}$ Slavek M. Rucinski,$^{18}$\newauthor Konstanze Zwintz,$^{19}$ Paul Luckas,$^{20}$ Bernard Heathcote,$^{21}$ Paulo Cacella,$^{22}$\newauthor Jonathan Powles,$^{23}$ Malcolm Locke,$^{24}$ Terry Bohlsen,$^{25}$ Andr\'e-Nicolas Chen\'e,$^{26}$\newauthor Brent Miszalski,$^{27,28}$ Wayne L. Waldron,$^{29}$ Marissa M. Kotze,$^{27,28}$ Enrico J. Kotze$^{27}$\newauthor and Torsten B\"{o}hm$^{30,31}$
\\\\
Affiliations are listed at the end of the paper
}

\date{Accepted 2017 October 11. Received 2017 September 22; in original form 2017 May 23}

\pubyear{2017}

\begin{document}
\label{firstpage}
\pagerange{\pageref{firstpage}--\pageref{lastpage}}
\maketitle

\begin{abstract}
From $5.5$ months of dual-band optical photometric monitoring at the $1$~mmag level, \emph{BRITE-Constellation} has revealed two simultaneous types of variability in the O4I(n)fp star $\zeta$~Puppis: one single periodic non-sinusoidal component superimposed on a stochastic component. The monoperiodic component is the $1.78$~d signal previously detected by \emph{Coriolis}/SMEI, but this time along with a prominent first harmonic. The shape of this signal changes over time, a behaviour that is incompatible with stellar oscillations but consistent with rotational modulation arising from evolving bright surface inhomogeneities. By means of a constrained non-linear light curve inversion algorithm we mapped the locations of the bright surface spots and traced their evolution. Our simultaneous ground-based multi-site spectroscopic monitoring of the star unveiled cyclical modulation of its He~{\sc ii}~$\lambda4686$ wind emission line with the $1.78$-day rotation period, showing signatures of Corotating Interaction Regions (CIRs) that turn out to be driven by the bright photospheric spots observed by \emph{BRITE}. Traces of wind clumps are also observed in the He~{\sc ii}~$\lambda4686$ line and are correlated with the amplitudes of the stochastic component of the light variations probed by \emph{BRITE} at the photosphere, suggesting that the \emph{BRITE} observations additionally unveiled the photospheric drivers of wind clumps in $\zeta$~Pup and that the clumping phenomenon starts at the very base of the wind. The origins of both the bright surface inhomogeneities and the stochastic light variations remain unknown, but a subsurface convective zone might play an important role in the generation of these two types of photospheric variability.
\end{abstract}


%%%%%%%%%%%%%%%%%%%%%%%%%%%%%%%%%%%%%%%%%%%%%%%%%%

%%%%%%%%%%%%%%%%% BODY OF PAPER %%%%%%%%%%%%%%%%%%

%%%%%%%%%%%%%%%%%%%%%%%%%%%%%%%%%%%%%%%%%%%%%%%%%%%%%%%%%
\section{Introduction}
\label{sec:Naos_Intro}

Spaceborne high-precision photometric monitoring of stars as a means to probe their intrinsic light variations and trace back the underlying physical mechanisms has only emerged over the past two decades \citep{2000mons.proc....9B,2000ApJ...532L.133B}. Particularly, this practice is even less frequent on hot luminous massive O-type stars (with $M_{\rm init}\gtrsim 20 M_{\sun}$), partly because of their scarcity which is reflected in the decreasing Initial stellar Mass Function (IMF) following a power law of index $-2.35$ for masses between $1.25$ and $150M_{\sun}$ \citep{1955ApJ...121..161S,2010ARA&A..48..339B}. Nevertheless, with their strong radiation-driven stellar winds and their termination as supernovae, O stars and their descendant Wolf-Rayet stars are important drivers of the chemical enrichment of galaxies and the Universe. Understanding the physical origin of their intrinsic variations may provide  constraints not only on their internal properties (e.g. convective core overshoot, possible existence of a sub-surface convection zone, magnetic fields) but also their global wind properties, and the complex physics of their photosphere-wind interface.

\subsection{Space photometry of O stars}
\label{subsec:Naos_Intro_spacephot_Ostars}

So far a total of eighteen O stars have been monitored through space photometry (Table~3 of \citeauthor{2015MNRAS.453...89B}~\citeyear{2015MNRAS.453...89B}, adding the recently reported photometric observations of the $\delta$~Orionis system by \citeauthor{2015ApJ...809..134P}~\citeyear{2015ApJ...809..134P}, the $\iota$~Orionis binary system by \citeauthor{2017MNRAS.467.2494P}~\citeyear{2017MNRAS.467.2494P}, the magnetic hot supergiant $\zeta$~Orionis Aa by \citeauthor{2017A&A...602A..91B}~\citeyear{2017A&A...602A..91B}, and the late-O-type supergiant HD 188209 by \citeauthor{2017A&A...602A..32A}~\citeyear{2017A&A...602A..32A}). The \emph{Wide-field InfraRed Explorer} \citep[\emph{WIRE}: ][]{2000mons.proc....9B} was the pioneer of this practice by including the late-O-type subgiant $\zeta$~Oph in its sample of 90 targets \citep{2007CoAst.150..326B}. Then followed contributions from the \emph{Microvariability and Oscillations of STars} microsatellite \citep[\emph{MOST}: ][]{2003PASP..115.1023W}, the \emph{COnvection ROtation and planetary Transits} satellite \citep[\emph{CoRoT};][]{2006ESASP1306...33B,2009A&A...506..411A}, the \emph{Solar Mass Ejection Imager} \citep[\emph{SMEI}: ][]{2003SoPh..217..319E} instrument onboard the \emph{Coriolis} satellite, and the \emph{Kepler} spacecraft \citep{2010Sci...327..977B,2014PASP..126..398H}. Various types of intrinsic variations were reported from these space-based photometric observations, namely $\beta$ Cep-like pulsations for the late-O-type stars ($\zeta$~Oph: \citeauthor{2005ApJ...623L.145W}~\citeyear{2005ApJ...623L.145W},\citeauthor{2014MNRAS.440.1674H}~\citeyear{2014MNRAS.440.1674H}; HD 46202: \citeauthor{2011A&A...527A.112B}~\citeyear{2011A&A...527A.112B}; HD 256035: \citeauthor{2015MNRAS.453...89B}~\citeyear{2015MNRAS.453...89B}), one case of the presence of p-modes with finite lifetimes analoguous to solar-like oscillations \citep[HD 46149: ][]{2010A&A...519A..38D}, two cases of tidally influenced pulsations ($\delta$ Ori Aa: \citeauthor{2015ApJ...809..134P}~\citeyear{2015ApJ...809..134P}; $\iota$~Ori: \citeauthor{2017MNRAS.467.2494P}~\citeyear{2017MNRAS.467.2494P}), one case of a possible oscillatory convection g-mode \citep[$\zeta$ Pup: ][]{2014MNRAS.445.2878H}, six cases of possible internal gravity waves (HD46223, HD46150, HD 46966: \citeauthor{2011A&A...533A...4B}~\citeyear{2011A&A...533A...4B}, \citeauthor{2015ApJ...806L..33A}~\citeyear{2015ApJ...806L..33A}; HD44597, HD 255055: \citeauthor{2015MNRAS.453...89B}~\citeyear{2015MNRAS.453...89B}; HD 188209: \citeauthor{2017A&A...602A..32A}~\citeyear{2017A&A...602A..32A}) and five cases of possible rotational modulation (HD 46149: \citeauthor{2010A&A...519A..38D}~\citeyear{2010A&A...519A..38D}; HD 46223: \citeauthor{2011A&A...533A...4B}~\citeyear{2011A&A...533A...4B}; $\xi$~Per: \citeauthor{2014MNRAS.441..910R}~\citeyear{2014MNRAS.441..910R}; EPIC 202060097, EPIC 202060098: \citeauthor{2015MNRAS.453...89B}~\citeyear{2015MNRAS.453...89B}). Notably, some stars show simultaneously different types of intrinsic variations. With its sample of $36$ O stars brighter than $V=6$, the recently launched, commissioned and now operational nanosatellites forming \emph{BRIght Target Explorer} (\emph{BRITE-Constellation}: \citeauthor{2014PASP..126..573W}~\citeyear{2014PASP..126..573W}; \citeauthor{2016PASP..128l5001P}~\citeyear{2016PASP..128l5001P}) will almost triple the current numbers and improve the view on this broad range of O star intrinsic variability. 

\subsection{$\zeta$~Puppis}
\label{subsec:Naos_Intro_variability}

\begin{table}
\caption{Stellar parameters for $\zeta$~Puppis (HD 66811). The parameter $\beta$ denotes the exponent of the usual $\beta$-law velocity generally adopted in hot stellar winds, while $f_\infty$ denotes the asymptotic value of the clumping filling factor at the outer boundary of the stellar wind.}
{\normalsize
\begin{center}
\begin{tabular}{l r l c}
\hline
\hline
 Parameter & & Value & Ref.  \\
\hline
%RA (J2000)		& [h:m:s] 				& 	~~$08:03:35.047$ 			& (1)	\\
%Dec (J2000) 		& [d:m:s] 				& 	$-40:00:11.332$ 			& (1)	\\
Spectral Type		&					&	O4I(n)fp 					& (1)	\\
$V$				&					&	$2.256\pm0.019$			& (2)	\\
$B-V$			&					&	$-0.276\pm0.012$			& (2)	\\
$U-B$			&					&	$-1.108\pm0.013$			& (2)	\\
$\varv_{\rm e}\sin i$ 	&[km~s$^{-1}$]			&	$219 \pm 18$				& (3)	\\
$\varv_{\rm rad}$	&[km~s$^{-1}$]			&	$-23.9 \pm2.9$				& (4)	\\
$\varv_{\rm tan}$	&[km~s$^{-1}$]			&	$55.0 \pm16.6$			& (5)	\\
Distance 			&[kpc]				&	$0.46 \pm 0.04$			& (6)	\\
Age				& [Myr]				&	$3.2^{+0.4}_{-0.2}$			& (6)	\\
$\log(L/L_{\sun})$ 	&[cgs dex]				&	$5.91 \pm 0.1$				& (6)	\\
$T_{\rm eff}$ 		&[kK]					&	$40.0 \pm 1.0$				& (6)	\\
$\log g$ 			&[cm s$^{-2}$]			&	$3.64 \pm 0.1$				& (6)	\\
$R$				&[$R_{\sun}$]			&	$18.99 \pm 3.80$			& (7)	\\
$M$				&[$M_{\sun}$]			&	$56.1^{+14.5}_{-11.6}$ 		& (6)	\\
$\dot{M}$			&[$M_{\sun}$ yr$^{-1}$] 	&	 $1.9\pm0.2\times10^{-6}$	& (6)	\\
$\varv_{\infty}$ 		&[km~s$^{-1}$] 		&	 $2300\pm100$			& (6)	\\
$\beta$ 			& 					&	 $0.9$					& (6)	\\
$f_\infty$ 			& 					&	 $0.05$					& (6)	\\ \hline
\end{tabular}
\end{center}}
{\footnotesize 
$^{(1)}$Galactic O-Star Catalog: \citet{2014ApJS..211...10S}, $^{(2)}$\citet{2004ApJS..151..103M}, $^{(3)}$\citet{1997MNRAS.284..265H}, $^{(4)}$\citet{2006AstL...32..759G}, $^{(5)}$\citet{1998A&A...331..949M}, $^{(6)}$\citet{2012A&A...544A..67B}, $^{(7)}$\citet{2005A&A...436.1049M}}
\label{tab:Naos_StellarParams}
\end{table}

Standing out from the aforementioned restricted sample of O-type stars is the hot, bright, rapidly-rotating, single, runaway supergiant $\zeta$~Puppis (Table~\ref{tab:Naos_StellarParams}). Owing to its brightness, proximity and status as a single star, $\zeta$~Pup has become a key object for understanding the properties of O-type stars and hot stellar winds. Therefore it is not surprising why several observational studies have been conducted on $\zeta$~Pup across a wide range of wavelength: in the X-ray \citep[e.g.][ and references therein]{2013A&A...551A..83H,2013ApJ...763..143N}, in the UV \citep{1992ApJ...390..266P,1995ApJ...452L..65H}, in the optical \citep{1970MNRAS.150...45D,1996A&A...311..616R,2014MNRAS.445.2878H}, as well as in the radio and sub-millimeter \citep{2003A&A...408..715B}. Amongst these studies, time-resolved spectroscopic and photometric observations have unveiled a number of variabilities on timescales ranging from hours to a couple of days \citep[Table~1 in ][]{2014MNRAS.445.2878H}.

The first variability reported on $\zeta$~Pup came from the monitoring of the central absorption reversal of its H$\alpha$ wind emission line profile which was seen to vary non-sinusoidally with a period of $5.075\pm 0.003$~d \citep{1981ApJ...251..133M}. Subsequent ground-based optical photometric observations in the Str\"{o}mgren $b$ filter \citep{1992MNRAS.254..404B}, as well as UV spectroscopic studies \citep{1995ApJ...452L..65H} revealed periods compatible with this one ($\sim5.2$~d). Up to this time this $5.1$~d period was attributed to the rotation period of the star, which, combined with the measured projected rotational velocity and the radius (Table~\ref{tab:Naos_StellarParams}), implies that the star is seen almost equator on. 

Also, photospheric line-profile variations that were best interpreted as non-radial pulsations (NRPs) with $l=-m=2$ were detected in $\zeta$~Pup, occurring as bumps moving redward across some optical He~{\sc i}, He~{\sc ii}, N~{\sc iv} and C~{\sc iv} lines of the star with a period of $8.54 \pm 0.05$~h \citep{1986ASIC..169..465B,1996A&A...311..616R}. Additionally, a possibility for the presence of low-order oscillatory convection modes with a period of $1.780938 \pm 0.000093$~d and semi-amplitude of $6.9 \pm 0.3$~mmag has been reported from four contiguous years ($2003-2006$) of \emph{Coriolis}/SMEI photometric monitoring of the star \citep{2014MNRAS.445.2878H}.

From the point of view of wind variability, $\zeta$~Pup is often used as a laboratory to study the time-dependent properties of wind structures in O-type stars. Such wind structures are classified in two distinct categories with respect to their spatial extent: small-scale density inhomogeneities dubbed clumps; and large-scale spiral-like structures corotating with the stellar photosphere, the so-called Corotating Interaction Regions (CIRs). Both of these phenomena are now believed to be essentially ubiquitous in O stars \citep{2008A&ARv..16..209P}.

Observational evidence for the presence of clumps in hot luminous star winds was initially detected in stars in the He-burning, He-rich Wolf-Rayet phase \citep{1975IAUS...67..299S,1988ApJ...334.1038M}, a late stage in the evolution of O-type stars \citep[e.g.][]{1991ApJ...368..538L}. In the case of O stars, a phenomenological model consisting of a population of radiatively driven blobs in the wind of $\zeta$ Pup was first proposed to explain its observed X-ray luminosity \citep{1980ApJ...241..300L}. Later $\zeta$ Pup became the first O-type star confirmed to have a clumpy wind:  spectroscopic monitoring of its wind-formed He~{\sc ii}~$\lambda4686$ line profile revealed excess bumps at the $\sim5\%$ level of the line intensity, propagating from the centre towards the edges of the line, which is a signature of the kinematics of clumps randomly appearing in the inner stellar wind \citep{1998ApJ...494..799E}. Subsequent studies confirmed these interpretations and detected similar behaviour in other hot stars \citep{2008AJ....136..548L}.

The physical origin of these clumps still remains to be determined, with two coexistent paradigms to be considered: the first scenario involves line-driving instability that could generate inverse shocks locally compressing wind material to form clumps in a less dense inter-clump medium \citep{1970ApJ...159..879L,1988ApJ...335..914O,1995A&A...299..523F,2003A&A...406L...1D}, whereas the second one involves a contribution from internal gravity waves that could be excited within a subsurface convection zone and induce the formation of clumps near the photosphere-wind interface \citep{2009A&A...499..279C} which could then be subsequently enhanced by the first mechanism further out in the stellar wind.

Besides that, the absorption troughs of unsaturated P Cygni profiles of UV resonance lines of O-type stars are the sites for the appearance and blueward propagation of spectral features called Discrete Absorption Components (DACs), a more general form of features initially called Narrow Absorption Components \citep{1988MNRAS.231P..21P,1989ApJS...69..527H,1992ASPC...22..155H,1995ApJ...452L..65H,1999A&A...344..231K}. These features are best interpreted as the spectroscopic manifestations of the presence of Corotating Interaction Regions (CIRs) in the winds of O stars \citep{1984ApJ...283..303M}, such as those that are also found in the solar wind \citep{1972NASSP.308..393H}. $\zeta$~Pup was one of the three program stars observed during extensive UV spectroscopic monitoring with the \emph{International Ultraviolet Explorer (IUE)} \citep{1995ApJ...452L..53M,1995ApJ...452L..65H}. A mean DAC recurrence period of $19.23\pm0.45$~h was measured from the analysis of their propagation in the Si~{\sc iv}~$\lambda\lambda 1394,1403$ doublet line of the star. With a $5.1$~d rotation period as was believed at that time, $\zeta$~Pup would exhibit on average $5-6$ DACs per rotation cycle \citep[Figure~1 in ][]{1995ApJ...452L..65H}, which is much more than the average of two dominating DACs per rotation cycle observed in other O stars \citep{1999A&A...344..231K}.

As for the origin of these CIRs, the canonical hydrodynamical model in O stars involves photospheric perturbations such as corotating bright spots or non-radial pulsations which initially enhance the local stellar wind speed, generating wind plasma of a different speed that interacts with the ambient wind to produce compressed material shaped as large spiral arms as they corotate with the star \citep{1996ApJ...462..469C}. Velocity plateaus forming ahead of the CIR arms are seen as propagating DACs in the absorption troughs of UV P Cygni resonance lines such as C~{\sc iv}~$\lambda1548$  or Si~{\sc iv}~$\lambda\lambda1394,1403$. Observational evidence for the presence of bright spots on an O star with possible link to its known DACs recurrence period recently emerged from \emph{MOST} photometric monitoring of the single runaway mid-O-type giant $\xi$~Per \citep{2014MNRAS.441..910R}. This discovery raised a motivation to determine if the presence of bright photospheric spots is universal among O stars (like the NAC/DAC phenomenon) and how they are linked to CIR/DAC activities in the stellar wind. Primarily driven by this purpose, we conducted a coordinated optical observing campaign on $\zeta$~Pup spanning $\sim5.5$~months, consisting of space-based photometric monitoring with \emph{BRITE-Constellation} during an observing run on the Vela/Puppis field (December 11, 2014 - June 02, 2015), and contemporaneous multi-site ground-based spectroscopic monitoring of the He~{\sc ii}~$\lambda4686$ line profile of the star. Our \emph{BRITE} observations probe light variations at the level of the stellar photosphere in two passbands, not achievable by any previous space-based mission, while our spectroscopic observations of the He~{\sc ii}~$\lambda4686$ wind emission line allow us to trace any possible activities related to CIRs and clumps in the inner stellar wind.

%%%%%%%%%%%%%%%%%%%%%%%%%%%%%%%%%%%%%%%%%%%%%%%%%%%%%%%%%
\section{Observations}
\label{sec:Naos_Obs}

%%%%%%%%%%%%%%%%%%%%%%%%%%%%%%%%%%%%%%%%%%%%%%%%%%%%%%%%%
\subsection{High-precision space photometry}
\label{subsec:Naos_Obs_BRITE_SMEI}

%%%%%%%%%%%%%%%%%%%%%%%%%%%%%%%%%%%%%%%%%%%%%%%%%%%%%%%%%
\subsubsection{BRITE-Constellation photometry}
\label{subsubsec:Naos_Obs_BRITE}

\begin{table*}
\caption{\emph{BRITE} photometry of $\zeta$~Pup - Journal of Observations. $N_s$ indicates the number of frames stacked onboard. Quantities listed in column 6 through 9 were assessed at post-decorrelation stage: $N_{\rm p, tot}$ is the total number of data points, $N_{\rm p, orb}$ is the median number of points per orbit (along with the range of values), $t_{\rm cont}$  is the median contiguous time per orbit during which observations were performed (along with the range of values), and $\sigma_{\rm rms}$ is the quadratic mean value of mean standard deviations per orbit assessed from linear fits of the flux values within the orbits.}
\centering
{\footnotesize
\begin{center}
\begin{tabular}{l c c c c c c c c }
\hline
\hline
Satellite & Start Date & End Date  & Observing  & $N_{\rm s}$ & $N_{\rm p, tot}$ & $N_{\rm p, orb}$ & $t_{\rm cont}$ & $\sigma_{\rm rms}$ \\
& \multicolumn{2}{c}{[HJD-2451545]} & Mode & & & & [min] & [mmag] \\
\hline
\emph{BRITE-Austria}	
&	$\begin{array}{c} 5458.402 \\ 5527.920 \end{array}$	
&	$\begin{array}{c} 5527.991 \\ 5621.152 \end{array}$ 
&	$\begin{array}{c} \textrm{Stare} \\ \textrm{Chopping} \end{array}$ 
&	$\begin{array}{c} 1 \\  1 \end{array}$ 
& 	$\begin{array}{c} 16198 \\ 22981 \end{array}$
& 	$\begin{array}{c} \textrm{$25~[4-48]$} \\ \textrm{$30~[6-40]$} \end{array}$
& 	$\begin{array}{c} \textrm{$~6.9~[0.8-12.4]$} \\ \textrm{$11.6~[2.3-13.6]$} \end{array}$
& 	$\begin{array}{c} 1.41 \\ 1.19 \end{array}$
\\\\
\emph{BRITE-Toronto}	
&	$\begin{array}{c} 5465.758 \\ 5535.135 \end{array}$
&	$\begin{array}{c} 5535.193 \\ 5624.967 \end{array}$
&	$\begin{array}{c} \textrm{Stare} \\ \textrm{Chopping} \end{array}$ 
&	$\begin{array}{c} 3  \\ 1 \end{array}$
&	$\begin{array}{c} 15744 \\ 54135 \end{array}$
&	$\begin{array}{c} \textrm{$23~[4-30]$} \\ \textrm{$56~[8-60]$} \end{array}$
& 	$\begin{array}{c} \textrm{$16.8~[2.2-22.0]$} \\ \textrm{$15.1~[2.3-15.6]$} \end{array}$
& 	$\begin{array}{c} 1.36 \\ 0.75 \end{array}$
\\\\
\emph{BRITE-Heweliusz}	&	5553.059	&	 5631.313	&	Stare	&	$1$	& $44747$ & $~78~[8-161]$ & $15.9~[2.1-30.1]$ & 1.40
\\
\hline
\end{tabular}
\end{center}}
\label{tab:Naos_Obs_BRITE_Log}
\end{table*}

\begin{figure*}
\includegraphics[width=18cm]{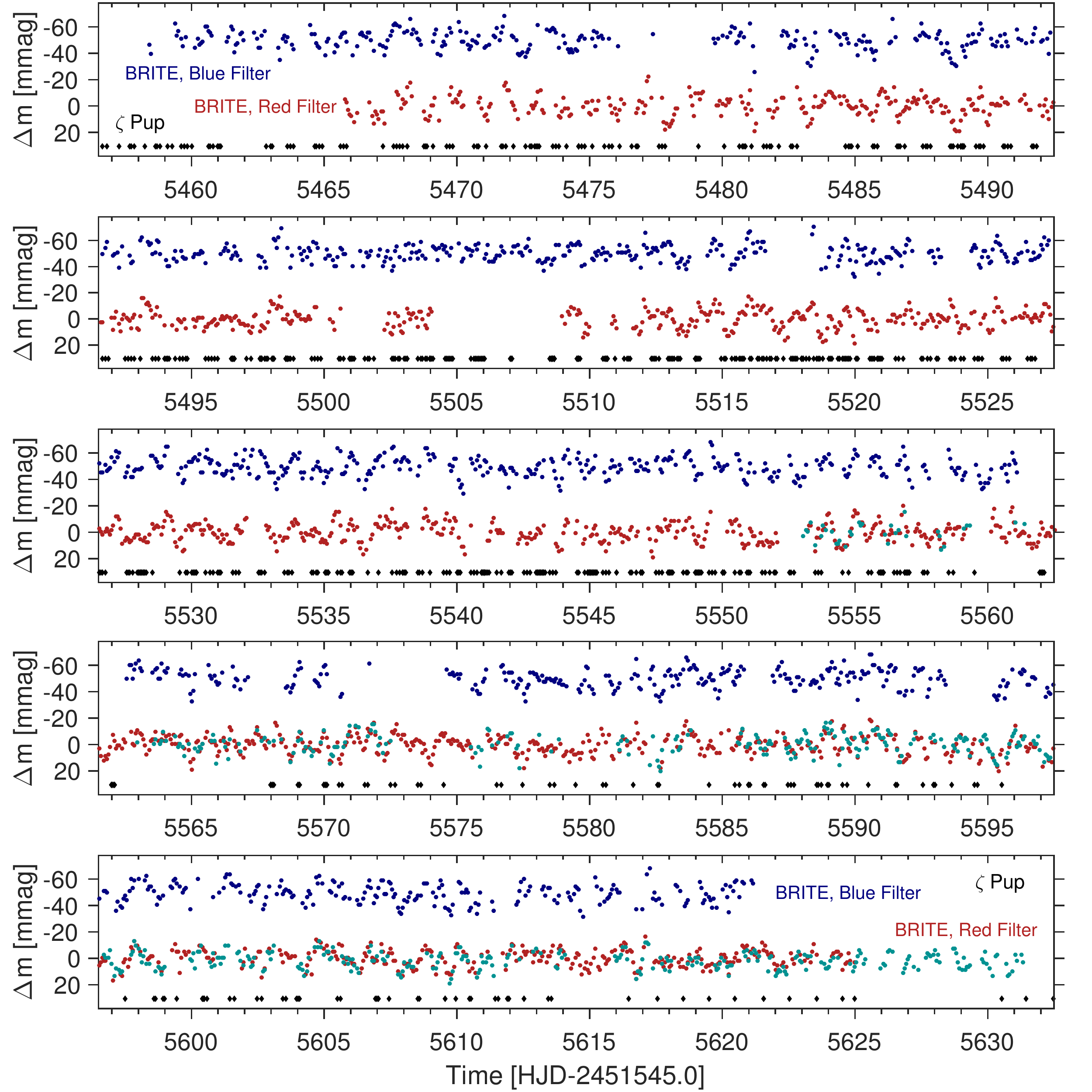}
 \caption{Two-color time series \emph{BRITE} photometry of $\zeta$~Pup binned over each \emph{BRITE} orbit. The observations obtained from \emph{BAb}, \emph{BTr} and \emph{BHr} are in blue, red and dark cyan respectively. For better visibility, an offset of $50$~mmag was added to the light curve in the blue filter. The temporal coverage of the contemporaneous ground-based spectroscopy is also depicted as small diamonds below the \emph{BRITE} light curves.}
  \label{fig:Naos_BRITE_lcs_full}
\end{figure*}

\emph{BRITE-Constellation} \citep{2014PASP..126..573W,2016PASP..128l5001P} as originally planned consists of a network of six nanosatellites each housing a $35$-mm format KAI-11002M CCD photometer fed by a 30-mm diameter f/2.3 telescope through either a blue filter ($390-460$~nm) or a red filter ($545-695$~nm): \emph{BRITE-Austria (BAb)}, \emph{Uni-BRITE (UBr)}, \emph{BRITE-Heweliusz (BHr)}, \emph{BRITE-Lem (BLb)}, \emph{BRITE-Toronto (BTr)} and \emph{BRITE-Montr\'eal (BMb)}, the last letter of the abbreviations denoting the filter type (``b'' for blue and ``r'' for red). With the exception of \emph{BMb} which failed to detach from the upper stage module of its Dnepr launch vehicle, all the satellites were launched into Low Earth Orbits of orbital period of the order of $100$~min, commissioned and now fully operational. With a $\sim24\degr\times20\degr$ effective field of view, each component of \emph{BRITE-Constellation} performs simultaneous monitoring of $15-30$ stars brighter than $V\simeq6$. A given field is observed typically over a $\sim 6$~month time base and, as far as possible, at least two satellites equipped with different filters are set to monitor the field to ensure dual-band observations.

The seventh field monitored by \emph{BRITE} was the Vela/Puppis field, for which photometry of $32$ stars was extracted, $\zeta$~Pup being the sole O-type star (not counting the primary component of the WC+O system $\gamma^2$~Velorum, as the binary is not resolved by \emph{BRITE}, the \emph{BRITE} detector pixel size being $\sim27.3\arcmin\arcmin$). Observations of this field were performed by \emph{BAb}, \emph{BTr} and \emph{BHr} (Table~\ref{tab:Naos_Obs_BRITE_Log}) during $\sim5.5$~months between December 11, 2014 and June 02, 2015 (HJD $2,457,003.40 - 176.31$). Short $1$~s exposures were taken at a median cadence of $15.3$~s during $\sim1-30\%$ of each $\sim100$~min \emph{BRITE} orbit, the remaining time unused due to stray light interferences, blocking by the Earth, and limited data download capacity. Observations were performed in stare mode for all satellites and then switched to chopping mode \citep{2016PASP..128l5001P,2016SPIE.9904E..1RP,2017A&A...605A..26P} for \emph{BAb} and \emph{BTr} for the second half of the observing run. Onboard stacking of three consecutive frames was performed only for the first three setups of the \emph{BTr} observations. Raw light curves were extracted using the reduction pipeline for \emph{BRITE} data which also includes corrections for intra-pixel sensitivity \citep{2017A&A...605A..26P}. Then post-reduction decorrelations with respect to instrumental effects due to CCD temperature variations, centroid position and orbital phase were performed on each observational setup for each satellite according to the method described by \citet{2016A&A...588A..55P}, and flux variations due to changes in point spread function shape as a function of temperature were performed according to the method described by \citet{2017A&A...602A..91B}. In the resulting final decorrelated light curves, we do not notice any obvious variations on timescales shorter than the \emph{BRITE} orbital period ($t \lesssim100$~min) that could be qualified as intrinsic to the star rather than pure instrumental noise. Therefore in order to gain in precision we calculated satellite-orbital mean fluxes to create the final light curves in the two filter bands. We note that we also perform removal of outliers during the decorrelation process, such that it is reasonable to adopt orbital mean fluxes instead of median fluxes or trimmed mean fluxes or mean flux values within 1/4 and 3/4 quartiles. Then, to extract the orbital means, we tested two different methods : (1) a simple average of the flux values taken within an orbit, and (2) an average taken to be the mid-point of a linear fit of the flux values within an orbit. We noticed no significant difference between the resulting root mean square (rms) values of the mean standard deviations obtained from the two methods, the second one being only slightly better. Therefore we adopted the second method to generate the final binned light curves in the two filter bands which we use to extract information on the intrinsic variability of the star (Figure~\ref{fig:Naos_BRITE_lcs_full}).

%%%%%%%%%%%%%%%%%%%%%%%%%%%%%%%%%%%%%%%%%%%%%%%%%%%%%%%%%
\subsubsection{Coriolis/SMEI photometry}
\label{subsubsec:Naos_Obs_BRITE}

We also use the archival light curves of $\zeta$~Pup recorded by \emph{Coriolis}/SMEI during its $2003-2006$ seasonal observing runs, published by \citet{2014MNRAS.445.2878H}. The characteristics of the \emph{Coriolis}/SMEI light curves of $\zeta$~Pup are described in detail in Section~2.1 of \citet{2014MNRAS.445.2878H}.

%%%%%%%%%%%%%%%%%%%%%%%%%%%%%%%%%%%%%%%%%%%%%%%%%%%%%%%%%
\subsection{Ground-based multi-site optical spectroscopy}
\label{subsec:Naos_Obs_Spectro}

\begin{table*}
\caption{Basic characteristics of the spectroscopic observations involved in our campaign. The facilities are ordered according to East longitude, starting from the closest to the Greenwich meridian. Amateur facilities members of the Southern Astro Spectroscopy Email Ring (SASER; http://saser.wholemeal.co.nz) at the time of the campaign are indicated by a star. Estimates of the typical $S/N$ per spectrum were taken in a portion of continuum in the closest vicinity of He~{\small II}~$\lambda4686$.}
\centering
{\footnotesize
\begin{center}
\begin{tabular}{l c c c c c c c c}
\hline
\hline
 Observatory & PI/Observer & East Longitude & Telescope & Instrument & R & $\lambda$ Range (\AA) & $N_{\rm sp}$ & $S/N$ \\
\hline

SAAO	&	T. Ramiaramanantsoa &	$20\degr 48\arcmin37\arcmin\arcmin$&	$1.9$ m  		&	GIRAFFE	&	$39000$ &	$4249 - 6513$ &	$24$ & $370$\\

Shenton Park $^{\star}$	&	P. Luckas &	$115\degr  48\arcmin 54\arcmin\arcmin$&	$0.35$ m 		&	Lhires III	&	$8364$ &	$4579-4729$ &	$257$ & $590$\\

Domain $^{\star}$	&	B. Heathcote	&	$144\degr 59\arcmin  21\arcmin\arcmin$&	$0.28$ m 		&	Lhires III	&	$8529$ &	$4590-4738$ &	$106$ & $700$\\

Latham $^{\star}$	&	J. Powles	&	$149\degr 01\arcmin 41\arcmin\arcmin$&	$0.25$ m 		&	Spectra L200	&	$8774$ &	$4549-4879$ &	$65$ & $500$\\

Mirranook $^{\star}$	&	T. Bohlsen	&	$151\degr 30\arcmin 35\arcmin\arcmin$&	$0.28$ m 		&	Spectra L200	&	$6361$ &	$4484-5022$ &	$1$ & $320$\\

R. F. Joyce $^{\star}$	&	M. Locke	&	$172\degr20\arcmin 59\arcmin\arcmin$&	$0.4$ m 		&	Spectra L200	&	$8432$ &	$4453-4809$ &	$22$ & $725$\\

CTIO	&	T. Ramiaramanantsoa	&	$290\degr48\arcmin23\arcmin\arcmin$ & 	$1.5$ m 	&	CHIRON	&	$24000$ &	$4578-8762$ &	$415$ & $725$\\

Dogsheaven $^{\star}$	&	P. Cacella	&	$312\degr 05\arcmin 20\arcmin\arcmin$&	$0.51$ m 		&	Lhires III	&	$5780$ &	$4550-4777$ &	$164$ & $420$\\

\hline
\end{tabular}
\end{center}}
\label{tab:Naos_Obs_Spectro_Log}
 %\vspace{-0.25cm}
\end{table*}

\begin{figure*}
\includegraphics[width=18cm]{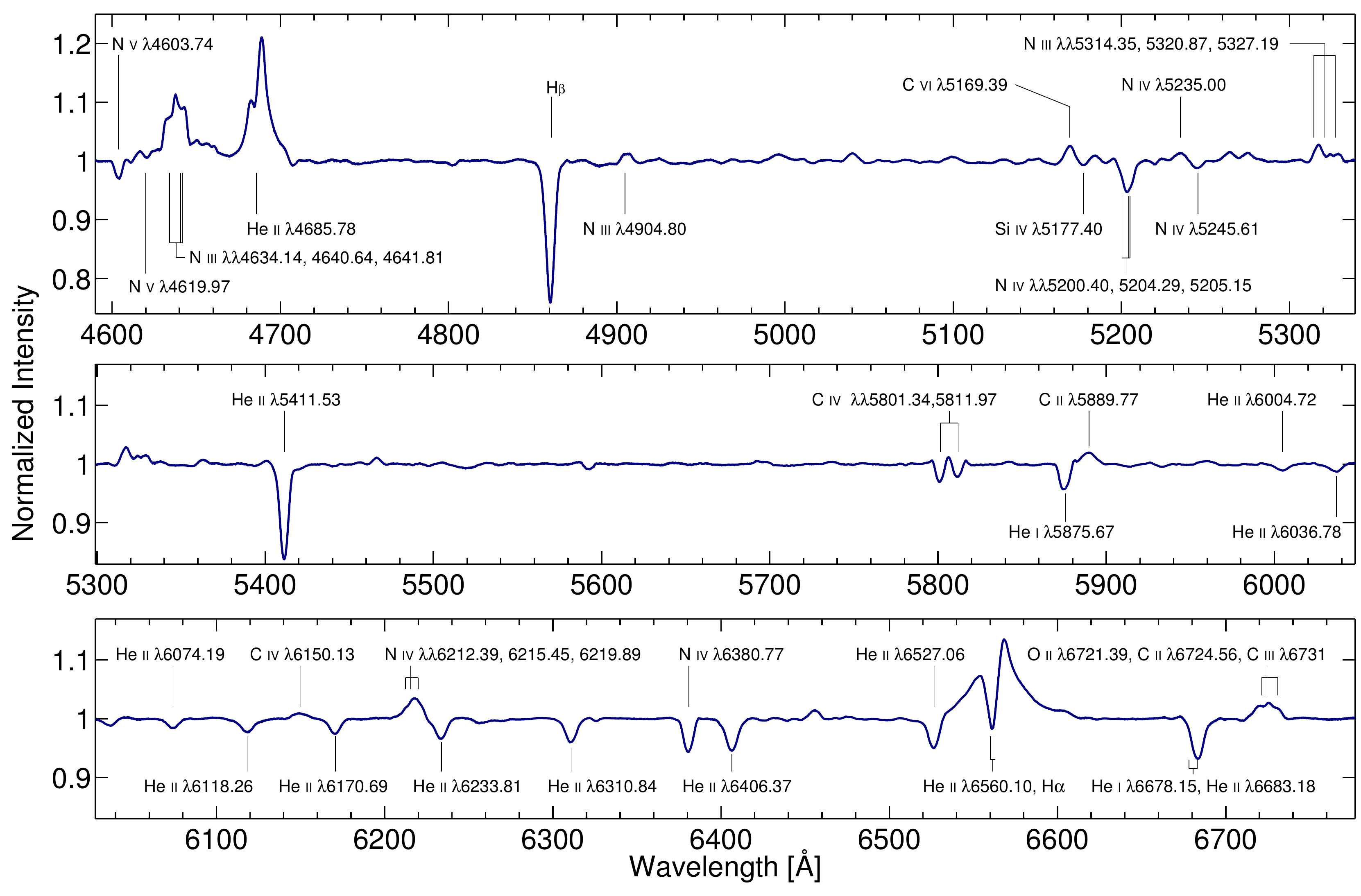}
 %\vspace{-0.5cm}
 \caption{Mean high-resolution ($R\sim80000$) CTIO-SMARTS~1.5m/CHIRON spectrum of $\zeta$~Pup. $S/N\sim2500$ in the continuum close to the vicinity of the He~{\sc ii}~$\lambda4686$ line profile.}
  \label{fig:Naos_CHIRON_spectrum_full}
\end{figure*}

In parallel with the \emph{BRITE} run, we conducted multi-site ground-based spectroscopy of $\zeta$~Pup focusing on the He~{\sc ii}~$\lambda4686$ line in order to monitor possible signatures of CIR/DAC activities in the inner stellar wind. The temporal coverage of the ground-based observations is depicted in Figure~\ref{fig:Naos_BRITE_lcs_full} and the characteristics of the observations are listed in Table~\ref{tab:Naos_Obs_Spectro_Log}.

\subsubsection{CTIO-SMARTS~1.5m/CHIRON}
\label{subsubsec:Naos_Obs_Spectro_CTIO}

We acquired a total of $415$ spectra of $\zeta$~Pup spanning $4578-8762$~\AA~with the CHIRON \'echelle spectrograph \citep{2013PASP..125.1336T} mounted on the Cerro Tololo Inter-American Observatory's (CTIO) 1.5~m telescope operated by the Small and Moderate Aperture Research Telescope System (SMARTS) consortium (NOAO Proposal IDs: 2014B-0082, 2015A-0002; PI: T. Ramiaramanantsoa). We adopted a cadence of $2-6$ visits  per night on $\zeta$~Pup (depending on the observability of the star), between November 28, 2014 and June 20, 2015. Each visit consists of $4-6$ consecutive $15$~s sub-exposures taken in slicer mode ($R\sim80000$), that we stacked and re-binned down to $R\sim24000$ so that one single spectrum has a $S/N\sim725$ per pixel in the continuum close to the He~{\sc ii}~$\lambda4686$ line. In order to correct for the instrumental blaze function, we normalized the extracted one-dimensional wavelength-calibrated spectra of $\zeta$~Pup through a division by cubic spline fits to each of the orders of the spectrum of the B9.5V star HR4468 ($V=4.7$) obtained during the campaign with the same instrument configurations (slicer mode, fixed cross-disperser position). We chose HR4468 for this purpose as it is relatively close to $\zeta$~Pup, relatively bright enough and does not have many strong metal lines. Figure~\ref{fig:Naos_CHIRON_spectrum_full} illustrates the normalized, unbinned ($R\sim80000$) CTIO-SMARTS~1.5m/CHIRON mean spectrum of $\zeta$~Pup over the entire campaign with identification of its strongest spectral lines. 

\subsubsection{SAAO~1.9m/GIRAFFE}
\label{subsubsec:Naos_Obs_Spectro_SAAO}

We collected $24$ optical spectra of $\zeta$~Pup during $14$ nights between January 28, 2015 and February 10, 2015 at the South-African Astronomical Observatory (SAAO) with the Grating Instrument for Radiation Analysis with a Fibre Fed \'Echelle (GIRAFFE) spectrograph hosted in the coud\'e chamber of the $1.9$~m Grubb Parsons telescope (PI: T. Ramiaramanantsoa). Given the two available cross-disperser prisms optimized for the blue domain ($3791-5459$~\AA) and the red domain ($4094-9397$~\AA) with typical $R\sim39000$, we chose to use the red prism in order to better match with the CTIO-SMARTS~1.5m/CHIRON wavelength coverage. Camera flat-fields, fiber flat-fields, bias frames and arc exposures were obtained in the usual way and the data reduced using the python-based pipeline for GIRAFFE data extraction, {\sc indlulamithi}. Exposure times range from $4$~min to $12$~min depending on the weather conditions and the airmass of the star. We adopted a cadence of $2-5$ visits per night well-spread in time, alternating with visits on another primary target ($\gamma^2$~Velorum) and the B9.5V star HR4468 which was used to correct for the instrumental blaze function in the spectra of the two primary targets. Each visit on $\zeta$~Pup consists of $4-6$ consecutive spectra that we later stacked in order to get a $S/N\sim370$ in the continuum near He~{\sc ii}~$\lambda4686$.

\subsubsection{SASER}
\label{subsubsec:Naos_Obs_Spectro_SASER}

Optical spectra covering the He~{\sc ii}~$\lambda4686$ region were obtained by amateur observatories members of the Southern Astro Spectroscopy Email Ring (SASER) located in Australia (Shenton Park observatory, Domain Observatory, Latham Observatory, Mirranook Observatory), New-Zealand (R. F. Joyce Observatory) and Brazil (Dogsheaven Observatory). Data reduction (bias, flat-field, dark and sky background correction) and extraction of one-dimensional wavelength-calibrated unnormalized spectra were performed with the Integrated Spectrographic Innovative Software (ISIS) and the MaxIm DL software. Then, by means of the Image Reduction and Analysis Facility \citep[IRAF\footnote{IRAF is distributed by the National Optical Astronomy Observatories, which are operated by the Association of Universities for Research in Astronomy, Inc., under cooperative agreement with the National Science Foundation}:][]{1986SPIE..627..733T,1993ASPC...52..173T} software, we performed final homogeneous continuum normalization of the $615$ spectra collected by SASER together with the $415$ spectra obtained from CTIO-SMARTS~1.5m/CHIRON and the $24$ SAAO~1.9m/GIRAFFE spectra.

\subsubsection{CFHT~3.6m/Reticon}
\label{subsubsec:Naos_Obs_Spectro_CFHT}

We re-extracted the $50$ archival optical spectra of $\zeta$~Pup taken with the (now decommissioned) spectrograph mounted at the coud\'{e} focus of the $3.6$~m Canada-France-Hawaii Telescope (CFHT) during the nights of December 10/11 and December 12/13, 1995 (PI: Moffat) and published by \citet{1998ApJ...494..799E} in order to compare the properties of the line profile variations (LPVs) that they observed in He~{\sc ii}~$\lambda4686$ to those that we detected during our campaign. Details on the characteristics of these spectra are described in Section~2 of \citet{1998ApJ...494..799E}.

%%%%%%%%%%%%%%%%%%%%%%%%%%%%%%%%%%%%%%%%%%%%%%%%%%%%%%%%%
\section{Photometric variability}
\label{sec:Naos_Results_Photo}

%%%%%%%%%%%%%%%%%%%%%%%%%%%%%%%%%%%%%%%%%%%%%%%%%%%%%%%%%
\subsection{Amplitudes of the observed light variations}
\label{subsec:Naos_Results_Photo_Amp}

\begin{figure}
\includegraphics[width=8.4cm]{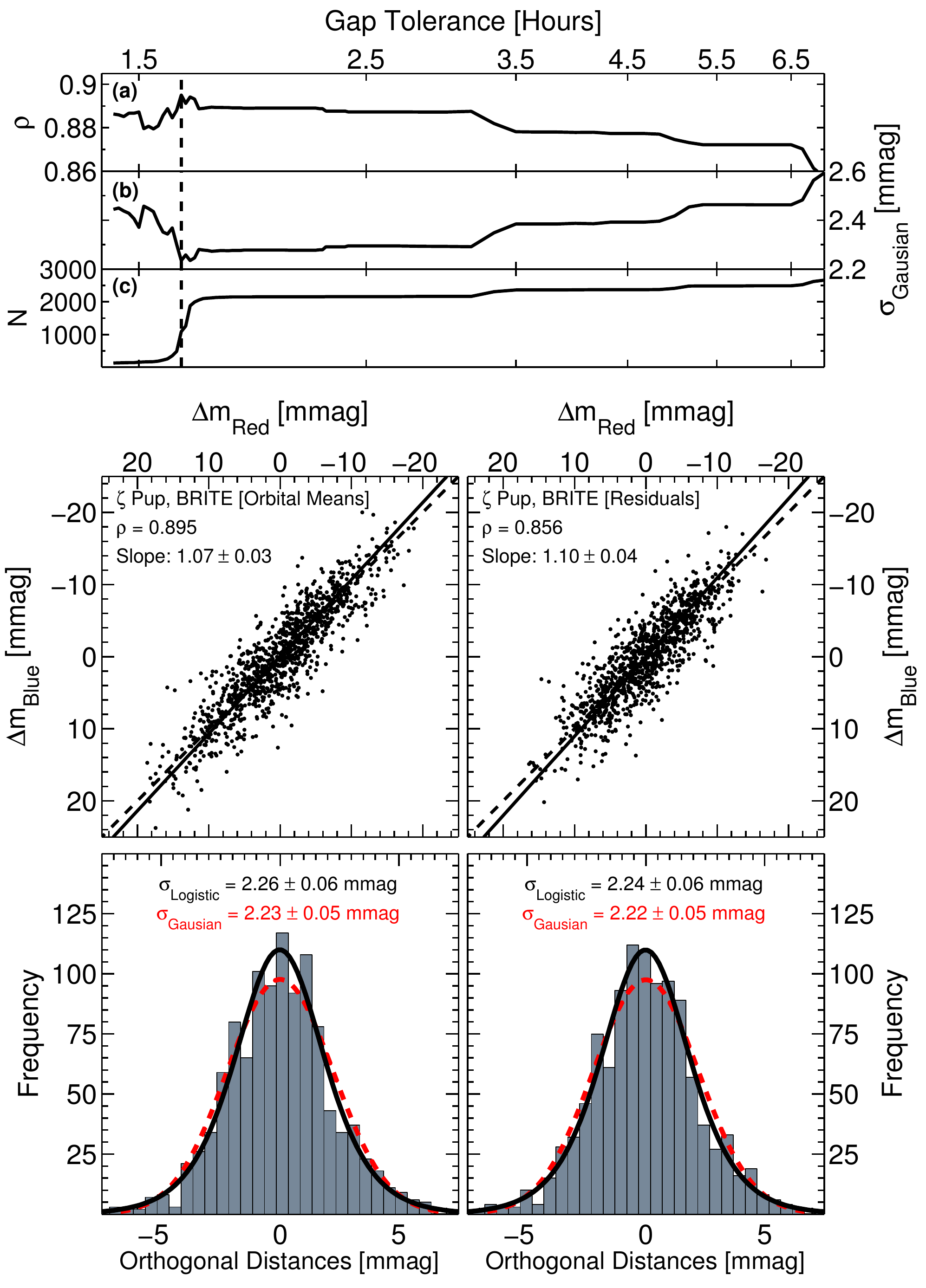}
 %\vspace{1cm}
 \caption{Comparison of the amplitudes of the variations of $\zeta$~Pup observed through the two \emph{BRITE} optical filters. \emph{Top:} Evolution of (a) the Pearson correlation coefficient, (b) the standard deviation of the distribution of orthogonal distances, and (c) the number of points involved, as a function of gap tolerance for the linear interpolation process. The vertical dashed line indicates the location of the optimal gap tolerance value ($1.65$~h). \emph{Middle:} Comparison diagrams for (Left) the \emph{BRITE} light curves presented in Figure~\ref{fig:Naos_BRITE_lcs_full} and (Right) the residual light curves  after removal of the main periodic signal intrinsic to the star (Sections~\ref{subsec:Naos_1.78dperiod}~and~\ref{subsec:Naos_stochastic_variability}). The dashed lines indicate the reference slope $=1$. \emph{Bottom:} Distributions of distances perpendicular to the direction of correlation. The dashed red curves are Gaussian fits to the distributions whereas the continuous black ones are logistic distribution fits.}
  \label{fig:Naos_BRITE_lcs_full_BvsR}
\end{figure}

As depicted in Figure~\ref{fig:Naos_BRITE_lcs_full}, the light variations of $\zeta$~Pup appear to be coherent in the two \emph{BRITE} filters, with qualitatively  the same amplitudes. In order to quantitatively check the validity of this visual impression, we plotted the observed blue versus red amplitude variations (Figure~\ref{fig:Naos_BRITE_lcs_full_BvsR}). Since the measurements from the two filters were not acquired exactly simultaneously, interpolation over the common time sampling grid of the two light curves is required in order to construct such a diagram (which we shall hereafter call a ``\emph{BRITE-b} vs \emph{BRITE-r} diagram''). Therefore a proper treatment of gaps in the time series needs to be considered, since interpolation over a large gap (such as the extreme case of the $\sim5$-day gap during the time interval $[5504-5509]$) would lead to wrong estimates of the correlation coefficient, whereas if the gap tolerance is too short the number of points that remain in the diagram would be too small to allow for the extraction of meaningful slope and correlation coefficient values. The ideal case would be to consider a gap tolerance of the order of a \emph{BRITE} orbit, which is acceptable only if the time samplings in the two light curves are such that a reasonable number of points remain for the interpolation. Thus we scanned a grid of gap tolerance values and measured the Pearson correlation coefficient $\rho$, the standard deviation of the distribution of distances orthogonal to the direction of correlation $\sigma$, and the total number of points $N$ involved in these estimates. The first three upper panels of Figure~\ref{fig:Naos_BRITE_lcs_full_BvsR} trace the evolution of these three quantities over the considered range of gap tolerance values, and show that at a gap tolerance of $1.65$~h (which is of the order of a \emph{BRITE} orbit), $\rho$ reaches its global maximum ($\rho_{\rm max}=0.895$), $\sigma$ its global minimum ($\sigma_{\rm min}=2.23\pm0.05$~mmag) and $N$ is large enough to make these values meaningful ($N=1092$). We established our diagram with this gap tolerance value (middle panels of Figure~\ref{fig:Naos_BRITE_lcs_full_BvsR}). Thus we confirm that the variations observed in the blue filter and the red filter are strongly correlated ($\rho = 0.895$), while the orthogonal linear regression yields a slope indicating that the amplitudes of the variations of $\zeta$~Pup in the blue filter are only $7\pm3~\%$ higher than the amplitudes measured through the red filter (middle-left panel of Figure~\ref{fig:Naos_BRITE_lcs_full_BvsR}). This relatively small difference could be due to the fact that the wavelength coverages of the \emph{BRITE} filters both fall in the domain of validity of the Rayleigh-Jeans approximation for a hot star like $\zeta$~Pup,  where different wavelengths react essentially identically to temperature changes.

Besides these considerations on the amplitudes of the light variations of $\zeta$~Pup measured in the two \emph{BRITE} filters, it is worth noting that the distributions of orthogonal distances depicted in the bottom panels of Figure~\ref{fig:Naos_BRITE_lcs_full_BvsR} are slightly leptokurtic (excess kurtosis $\sim0.7$), such that these orthogonal distances can be better described as logistically distributed rather than normally distributed. More importantly, the value of the standard deviation $\sigma_{\rm Logistic}=2.26\pm0.06$~mmag is indicative of the expected scatter of instrumental origin in the \emph{BRITE} light curves of $\zeta$~Pup. Assuming that the three satellites involved in the observations provide the same data quality, the expected scatter of instrumental origin is $\sigma_{\rm Logistic}/\sqrt2=1.60\pm0.04$~mmag, which is in line with the rms scatter listed in the last column of Table~\ref{tab:Naos_Obs_BRITE_Log}, the rms of these values being $\sim1.25$~mmag.

%%%%%%%%%%%%%%%%%%%%%%%%%%%%%%%%%%%%%%%%%%%%%%%%%%%%%%%%%
\subsection{\emph{BRITE} and \emph{Coriolis}/SMEI probe the stellar photosphere}
\label{subsec:Naos_BRITE_probesphotosphere}

\begin{figure}
\centering
  \includegraphics[width=\columnwidth]{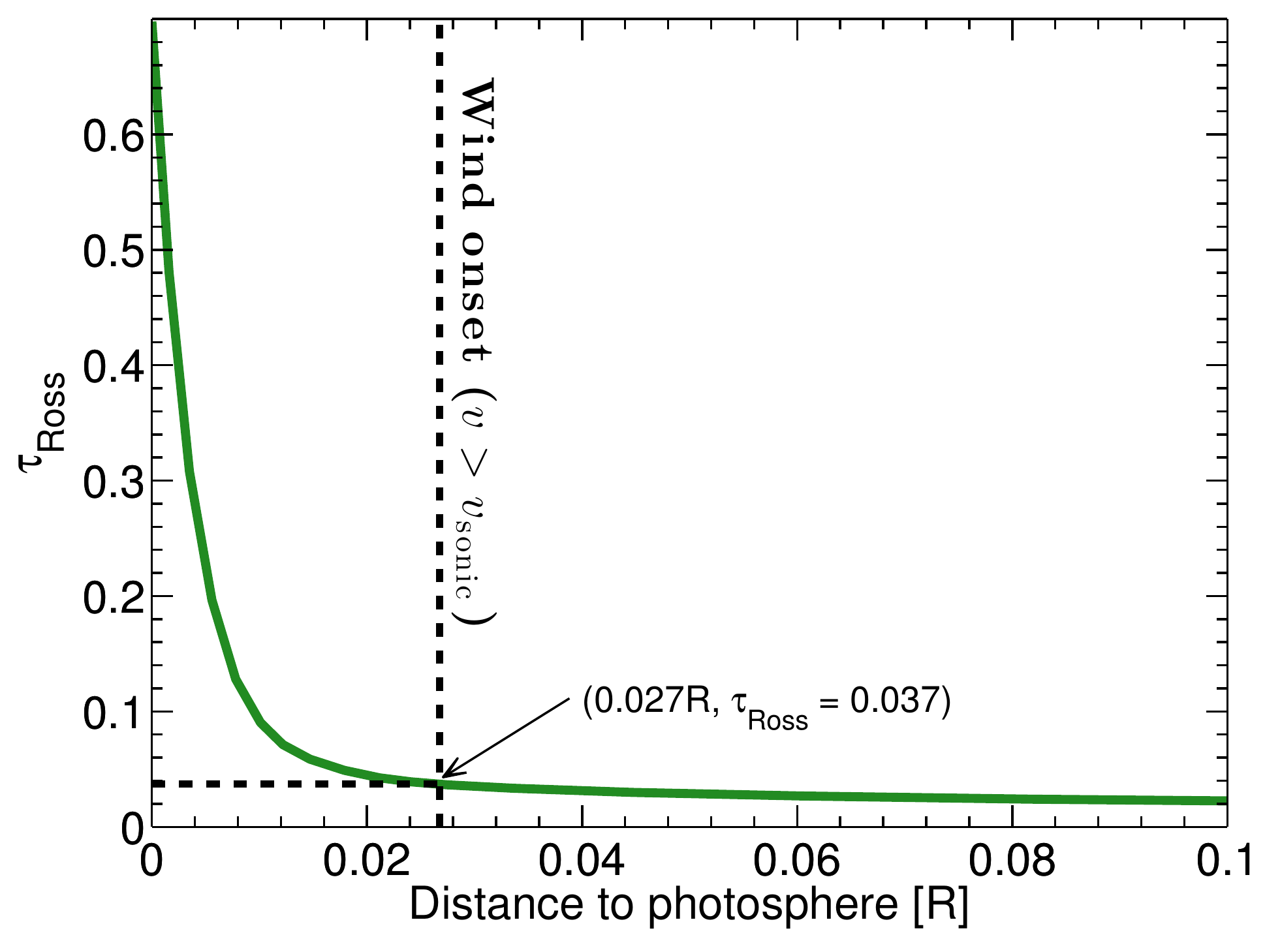}
  \caption{How $\tau_{\rm Ross}$ varies as a function of the radial extent from the photosphere (defined at $\tau_{\rm Ross} = 2/3$), as predicted by the \emph{PoWR} stellar atmosphere code. Radii beyond the vertical dashed line correspond to supersonic expansion velocities, i.e.\ the wind domain.}
\label{tab:Naos_PoWR_tauscale}
\end{figure}

In view of the typical $10-20$~mmag light variations of $\zeta$~Pup as detected in the photometric observations, the next crucial point that we investigate is the stellar domain in which the observed variability originates. In principle, the variability could originate in either the photosphere or the wind of $\zeta$ Pup (or even both), since some fraction of the photospheric continuum photons interact with matter in the wind. To estimate the fraction of scattered photons in the wind, we calculate a model atmosphere for $\zeta$ Pup using the non-local thermodynamic equilibrium (non-LTE) Potsdam Wolf-Rayet (\emph{PoWR}) tool, applicable for any hot star, including OB-type stars \citep[e.g.][]{2011MNRAS.416.1456O,2015ApJ...809..135S,2015A&A...577A..13S}. As described in detail by \citet{2003A&A...410..993H,2004A&A...427..697H} and \citet{2015A&A...579A..75T}, the tool solves the radiative transfer and rate equations in expanding atmospheres under the assumption of spherical symmetry and stationarity of the flow. 

To construct the model, we used the stellar and wind parameters of $\zeta$ Pup listed in Table~\ref{tab:Naos_StellarParams}: $\log(L)$, $T_{\rm eff}$, $\log g$, $M$, $\dot{M}$, $\varv_{\infty}$, $\beta$ and $f_{\infty}$, all derived by \citet{2012A&A...544A..67B}. Figure~\ref{tab:Naos_PoWR_tauscale} depicts the evolution of the Rosseland mean optical depth $\tau_{\rm Ross}$ as a function of the radial distance to the photosphere (defined at $\tau_{\rm Ross} = 2/3$), as predicted by \emph{PoWR}. We also mark the radius at which the stellar wind initiates, which is defined at the point where $\varv (r)$ exceeds $\varv_{\rm sonic}$. It is worth noting that, in the wind domain, $\tau_{\rm Ross}$ is virtually identical to the Thomson optical depth $\tau_{th}$ which originated in the scattering of photons off free electrons. 

It is evident from Figure~\ref{tab:Naos_PoWR_tauscale} that the wind optical depth is $\tau_{\rm Ross, wind} = 0.037$. This corresponds to a scattering of $1 - e^{-\tau} \approx 3.5\%$ of the photospheric continuum light in the wind. This is therefore the maximum amount of variability that is expected to originate from scattering of photospheric photons in the wind. However, since some lines generated primarily in the wind, such as He\,{\sc ii}\,$\lambda 4686$, are variable on the $\sim5-10\%$ level (Section~\ref{subsubsec:Naos_Results_Spectro_CIRs_dynamic_spec}, Figure~\ref{fig:Naos_HeII4686_CIRs}; Section~\ref{subsubsec:Naos_Results_Spectro_Clumps_dynamicspectra}, Figures~\ref{fig:Naos_HeII4686_Clumps_0210}~and~\ref{fig:Naos_HeII4686_Clumps_CFHT95}), the wind can contribute at most $0.35\%$ to the variability in the continuum. This means that in the typical $10-20$~mmag light variations observed by \emph{BRITE} and \emph{Coriolis}/SMEI in $\zeta$~Pup, only $0.035-0.07$~mmag may originate from the wind. We therefore conclude that the variability we observe in the photometric measurements originates primarily in the photosphere of $\zeta$~Pup.

%%%%%%%%%%%%%%%%%%%%%%%%%%%%%%%%%%%%%%%%%%%%%%%%%%%%%%%%%
\subsection{Search for periodic signals}
\label{subsec:Naos_Results_Photo_Fourier}

\begin{figure*}
\includegraphics[width=18cm]{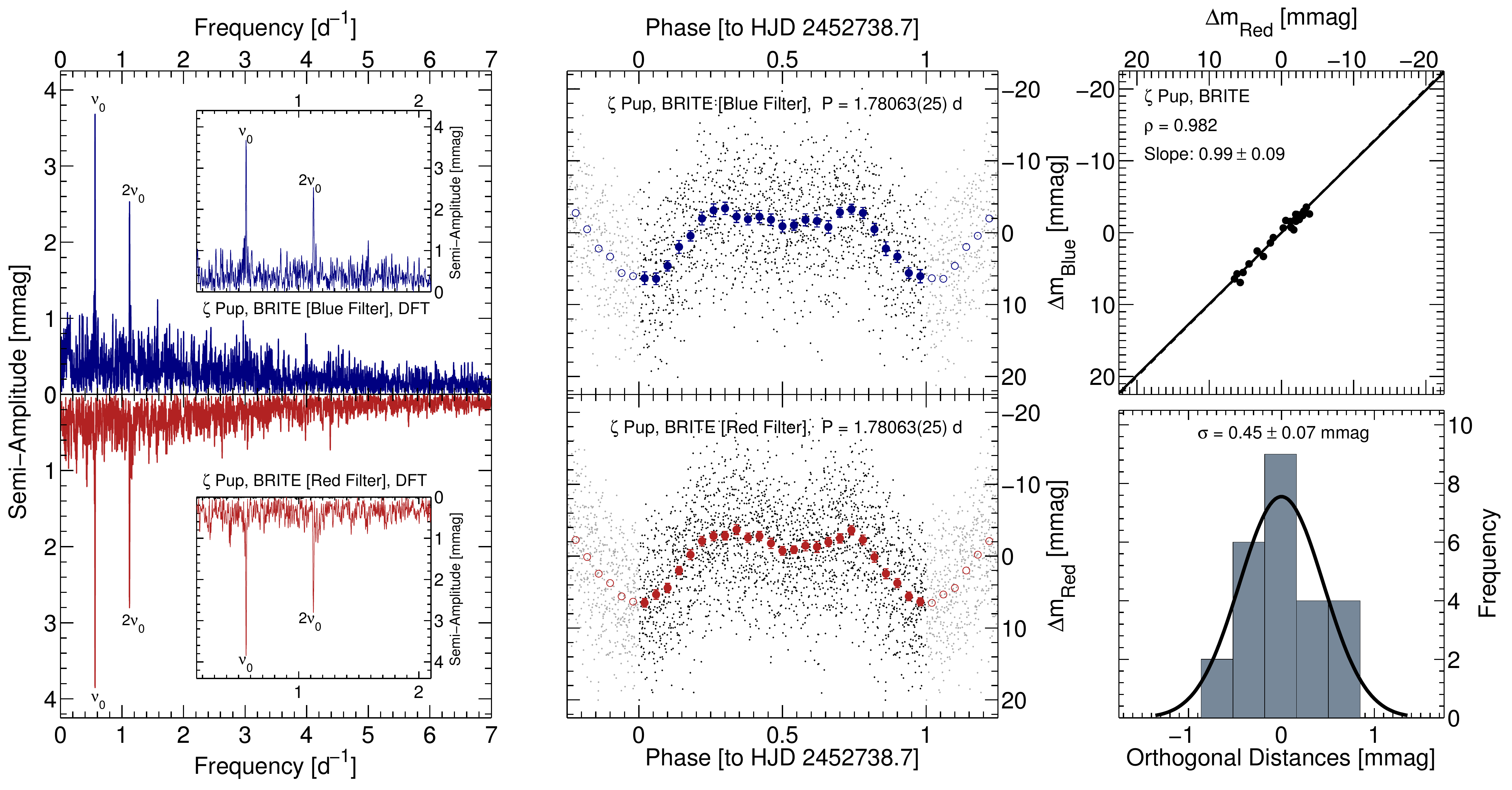}
%\vspace{1cm}
 \caption{\emph{Left:} Discrete Fourier Transform (DFT) of the \emph{BRITE} light curves of $\zeta$~Pup, with the insets showing zooms on the two significant frequency peaks. For ease of direct comparison, the vertical axis for the DFT of the red filter data (bottom panel) is reversed. \emph{Middle:} Rotation of $\zeta$~Pup as observed during the \emph{BRITE} observing run, the value of the rotation period $P = 1.78063(25)$~d assessed as the weighted mean of the values of the $1.78$~d period detected during the \emph{Coriolis}/SMEI and \emph{BRITE} runs (Table~\ref{tab:Naos_SMEI_BRITE_1.8dPeriod}). The original light curves are displayed (small grey points) as well as averages over bins of $0.04$ in phase (large blue/red points) along with their $1\sigma$ uncertainties. The reference epoch HJD~2452738.7 was chosen to fall before the beginning of the \emph{Coriolis}/SMEI 2003--2004 observing run and to allow for an easy monitoring of the evolution of the pattern of two consecutive bumps in the light curves (Section~\ref{subsubsec:Naos_1.78dsig_stability}, Figure~\ref{fig:Naos_BRITE_lcs_dynamic}) and the ``S'' patterns detected in the He~{\sc ii}~$\lambda$4686 wind emission line (Section~\ref{subsubsec:Naos_Results_Spectro_CIRs_dynamic_spec}, Figure~\ref{fig:Naos_HeII4686_CIRs}). \emph{Right:} Comparison diagram for the phased light curves of $\zeta$~Pup observed through the two \emph{BRITE} optical filters (top), together with the distribution of distances orthogonal to the direction of correlation and its Gaussian fit (bottom).}
  \label{fig:Naos_BRITE_DFT_lc178d_BvsR}
\end{figure*}

\begin{table*}
\caption{Fourier analysis of the \emph{Coriolis}/SMEI and \emph{BRITE} light curves of $\zeta$~Pup. The second column indicates the frequency resolution for each observing run. Formal $1\sigma$ uncertainties on the parameters were estimated from $100$ Monte Carlo simulations. $S/N$ values were estimated within intervals spanning $2$~d$^{-1}$ centered on each frequency. For \emph{BRITE}, all frequencies that are detected in both the red and the blue filters are listed even if their $S/N<4$.}
\centering
{\footnotesize
\begin{center}
\begin{tabular}{c c c c c c c}
\hline
\hline
 Observing Run	&	$\nu_{\rm res}$ 	&	Frequency			&	Period				&	Semi-amplitude	& 	$S/N$ 	&	Denotation		\\
				&	[d$^{-1}$]		&	[d$^{-1}$]			&	[d]					&	[mmag]			& 			&					\\
\hline

\emph{Coriolis}/SMEI&					&						&						&					&			&			\\
2003 -- 2004		&	0.00246			&	$0.56116\pm0.00011$	&	$1.78202\pm0.00035$	&	$6.64\pm0.48$	&	$7.3$	&	$\nu_0$	\\
2004 -- 2005		&	0.00473			&	$0.56148\pm0.00019$	&	$1.78101\pm0.00060$	&	$7.38\pm0.50$	&	$8.3$	&	$\nu_0$	\\
2005 -- 2006		&	0.00569			&	$0.56181\pm0.00031$	&	$1.77996\pm0.00098$	&	$6.27\pm0.52$ 	&	$6.1$	&	$\nu_0$	\\		
2003 --- 2006		&	0.00093			&	$0.56150\pm0.00002$	&	$1.78094\pm0.00006$	&	$6.61\pm0.23$	&	$12.2$	&	$\nu_0$	\\\\

\emph{BRITE} 2014 -- 2015		&	&	&	&	&	&	\\

%%%%% Blue Filter:
Blue Filter		&	0.00615	
&	$\begin{array}{c} \textrm{$0.56249\pm0.00024$} \\	\textrm{$1.12087\pm0.00102$}		\\	\textrm{$1.12694\pm0.00120$}				\\	\textrm{$1.14385\pm0.00100$} \\	\textrm{$1.28213\pm0.00121$} \\	\textrm{$1.72701\pm0.00236$} \\	\textrm{$1.77456\pm0.00094$}	\end{array}$
&	$\begin{array}{c} \textrm{$1.77781\pm0.00076$} \\	\textrm{$0.89216\pm0.00081$}		\\	\textrm{$0.88736\pm0.00094$}				\\	\textrm{$0.87424\pm0.00076$} \\	\textrm{$0.77995\pm0.00074$} \\	\textrm{$0.57904\pm0.00079$} \\	\textrm{$0.56352\pm0.00030$}	\end{array}$
&	$\begin{array}{c} \textrm{$3.85\pm0.03$}	   \\	\textrm{$2.33\pm0.16$}				\\	\textrm{$1.86\pm0.17$}						\\	\textrm{$1.14\pm0.03$}		 \\	\textrm{$0.89\pm0.03$}		 \\	\textrm{$0.99\pm0.03$}		 \\	\textrm{$0.89\pm0.03$}			\end{array}$
&	$\begin{array}{c} \textrm{$8.7$}		   	   \\   \textrm{$6.7$}						\\	\textrm{$4.1$}								\\	\textrm{$3.4$}				 \\	\textrm{$2.7$}				 \\	\textrm{$2.9$}				 \\	\textrm{$2.8$}					\end{array}$
&	$\begin{array}{c} \textrm{$\nu_{0}$}		   \\   \textrm{$\nu^{\prime}_{0}\sim2\nu_{0}$}				\\	\textrm{$\nu^{\prime\prime}_{0}\sim2\nu_{0}$}				 \\  \textrm{$\nu_{1}$}			 \\ \textrm{$\nu_{2}$} 		 	 \\ \textrm{$\nu_{3}$} 		 	 \\ \textrm{$\nu_{4}$} 				\end{array}$
\\\\

%%%%% Red Filter:
Red Filter		&	0.00604
&	$\begin{array}{c} \textrm{$0.56250\pm0.00017$} \\	\textrm{$1.12151\pm0.00213$}		\\	\textrm{$1.12813\pm0.00264$}				\\	\textrm{$1.14242\pm0.00087$} \\	\textrm{$1.28509\pm0.00085$} \\	\textrm{$1.72947\pm0.00170$} \\	\textrm{$1.76946\pm0.00101$} \\	\textrm{$4.38282\pm0.00768$}	\end{array}$
&	$\begin{array}{c} \textrm{$1.77778\pm0.00054$} \\	\textrm{$0.89166\pm0.00169$}		\\	\textrm{$0.88642\pm0.00207$}				\\	\textrm{$0.87533\pm0.00067$} \\	\textrm{$0.77816\pm0.00051$} \\	\textrm{$0.57821\pm0.00057$} \\	\textrm{$0.56514\pm0.00032$} \\	\textrm{$0.22816\pm0.00040$}	\end{array}$
&	$\begin{array}{c} \textrm{$3.88\pm0.22$}	   \\	\textrm{$2.78\pm0.33$}				\\	\textrm{$1.10\pm0.23$}						\\	\textrm{$1.03\pm0.25$}		 \\	\textrm{$0.84\pm0.18$}		 \\	\textrm{$0.86\pm0.20$}		 \\	\textrm{$0.76\pm0.18$}		 \\	\textrm{$0.71\pm0.20$}			\end{array}$
&	$\begin{array}{c} \textrm{$9.6$}			   \\	\textrm{$7.8$}						\\	\textrm{$3.4$}								\\	\textrm{$3.0$}				 \\	\textrm{$2.5$}				 \\	\textrm{$2.6$}				 \\	\textrm{$2.6$}				 \\	\textrm{$4.3$}					\end{array}$
&	$\begin{array}{c} \textrm{$\nu_{0}$}		   \\   \textrm{$\nu^{\prime}_{0}\sim2\nu_{0}$}				\\	\textrm{$\nu^{\prime\prime}_{0}\sim2\nu_{0}$}				 \\  \textrm{$\nu_{1}$}		 	 \\ \textrm{$\nu_{2}$} 		 	 \\ \textrm{$\nu_{3}$} 		 	 \\ \textrm{$\nu_{4}$} \\	\textrm{$\nu_{5}$} 				\end{array}$
\\\\

%%%%% Combined Blue+Red:
$\begin{array}{c} \textrm{Two Filters} \\ \textrm{Combined} \end{array}$		&	0.00579	
&	$\begin{array}{c} \textrm{$0.56241\pm0.00015$} \\	\textrm{$1.12111\pm0.00047$} 		\\	\textrm{$1.12714\pm0.00078$} 				\\	\textrm{$1.14280\pm0.00065$} \\	\textrm{$1.28323\pm0.00678$} \\	\textrm{$1.72788\pm0.00172$} \\	\textrm{$1.77440\pm0.00090$} \\	\textrm{$4.38313\pm0.00101$}	\end{array}$
&	$\begin{array}{c} \textrm{$1.77806\pm0.00047$} \\	\textrm{$0.89197\pm0.00037$} 		\\	\textrm{$0.88720\pm0.00061$} 				\\	\textrm{$0.87504\pm0.00050$} \\	\textrm{$0.77928\pm0.00412$} \\	\textrm{$0.57874\pm0.00058$} \\	\textrm{$0.56357\pm0.00029$} \\	\textrm{$0.22815\pm0.00005$}	\end{array}$
&	$\begin{array}{c} \textrm{$3.86\pm0.16$}	   \\	\textrm{$2.46\pm0.36$}		 		\\	\textrm{$1.50\pm0.34$}		 				\\	\textrm{$1.05\pm0.16$}		 \\	\textrm{$0.80\pm0.15$}		 \\	\textrm{$0.77\pm0.34$}		 \\	\textrm{$0.76\pm0.17$}		 \\	\textrm{$0.58\pm0.15$}			\end{array}$
&	$\begin{array}{c} \textrm{$10.1$}			   \\	\textrm{$8.0$}				 		\\	\textrm{$4.3$}				 				\\	\textrm{$3.1$}				 \\	\textrm{$2.6$}				 \\	\textrm{$2.7$}				 \\	\textrm{$2.7$}				 \\	\textrm{$4.0$}					\end{array}$
&	$\begin{array}{c} \textrm{$\nu_{0}$}		   \\   \textrm{$\nu^{\prime}_{0}\sim2\nu_{0}$}				\\	\textrm{$\nu^{\prime\prime}_{0}\sim2\nu_{0}$}				 \\  \textrm{$\nu_{1}$}		 	 \\ \textrm{$\nu_{2}$} 		 	 \\ \textrm{$\nu_{3}$} 		 	 \\ \textrm{$\nu_{4}$} \\	\textrm{$\nu_{5}$} 				\end{array}$
\\
\hline
\end{tabular}
\end{center}}
\label{tab:Naos_SMEI_BRITE_1.8dPeriod}
\end{table*}

We performed a Fourier analysis of the \emph{BRITE} light curves of $\zeta$~Pup using the discrete Fourier transform-based software package Period04 \citep{2005CoAst.146...53L}, suitable for the extraction of individual sinusoidal components in multiperiodic unevenly sampled time series through a prewhitening procedure, with the possibility to perform iterative Monte Carlo simulations to assess formal uncertainties on the extracted parameters. A global view of the raw amplitude spectra of the \emph{BRITE} light curves of $\zeta$~Pup (Figure~\ref{fig:Naos_BRITE_DFT_lc178d_BvsR}, left panel) shows a gradual increase in power towards lower frequencies, suggesting the presence of a red noise component that happens to have roughly the same amplitudes in the two filters. Therefore, at each stage of the prewhitening we computed the $S/N$ for the detected frequency peak $\nu_i$ within an interval spanning $[\nu_i-1;\nu_i+1]$~d$^{-1}$ which is a good compromise for a reasonable local estimate of the mean noise level while taking into account the red noise trend. Then we only consider that the frequency peak is significant if its $S/N>4.0$, a threshold that was validated both empirically \citep{1993A&A...271..482B,1999A&A...349..225B} and from numerical simulations \citep[Figure~4b in ][]{1997A&A...328..544K}. 

As conspicuously depicted in the amplitude spectra of the \emph{BRITE} light curves of $\zeta$~Pup, only two significant frequency peaks are simultaneously detected in the two filters, corresponding to the $1.78$~day period previously found in \emph{Coriolis}/SMEI observations of $\zeta$~Pup, but this time its first harmonic is also prominent. This visual impression is quantified in Table~\ref{tab:Naos_SMEI_BRITE_1.8dPeriod} in which we report all the Fourier components that are either detected simultaneously in the observations through the two \emph{BRITE} filters regardless of their statistical significance, or are significant but detected only in one filter. Only the $1.78$~day component (denoted by $\nu_0$ in Table~\ref{tab:Naos_SMEI_BRITE_1.8dPeriod}) and its first harmonic have $S/N>4$ \emph{and} are detected in both filters. We also point out in Table~\ref{tab:Naos_SMEI_BRITE_1.8dPeriod} that a pair of frequencies, $\{\nu^{\prime}_{0};\nu^{\prime\prime}_{0}\}$, is closely spread around the true value of the first harmonic. It is known that a new frequency extracted at one stage of the prewhitening procedure could be very close to a component that was found during one of the previous stages \citep[e.g.][]{2011A&A...533A...4B}. That behaviour is not surprising and is easily conceivable if for instance the physical phenomenon responsible for that signal is subject to a phase shift during the time of the observations. In all cases, one must adopt a resolution criterion in order to decide if a frequency peak is truly unique. In the present case, we use the criterion of \citet{1978Ap&SS..56..285L} stating that two frequencies must be separated by at least $1.5$ times the resolution frequency to be considered unique. The resolution frequency for each set of observations is also provided in Table~\ref{tab:Naos_SMEI_BRITE_1.8dPeriod}. Using this criterion, we conclude that the pair $\{\nu^{\prime}_{0};\nu^{\prime\prime}_{0}\}$ represents the same frequency, namely the first harmonic of the fundamental frequency $\nu_0$.

We also note that one frequency peak, $\nu_{5}=4.38313(101)$~d$^{-1}$ [$P=5.47560(126)$~h], is detected with a $S/N=4.3$ in the observations in the red filter \emph{but undetected in the blue filter}. This periodicity has never been found in previous observational campaigns on $\zeta$~Pup. At this point caution must be exercised as for the physical interpretation of this frequency. Future observations will confirm whether this signal is intrinsic to the star or an artifact of instrumental origin in our set of observations in the red filter. 

Finally, we detect a set of frequencies, labelled $\{\nu_{1}; \nu_{2}; \nu_{3}; \nu_{4}\}$ in Table~\ref{tab:Naos_SMEI_BRITE_1.8dPeriod}, corresponding to periods in the range $\sim13.5-21$~h, that are present in the two filters but with a rather low significance ($S/N\sim2.5-3.5$).

Since the $1.78$-day signal was discovered for the first time in \emph{Coriolis}/SMEI observations of $\zeta$~Pup spanning $2003 - 2006$ \citep{2014MNRAS.445.2878H}, for comparison purposes and in order to explore the stability of this signal (Section~\ref{subsubsec:Naos_1.78dsig_stability}) we also revisited the \emph{Coriolis}/SMEI light curves and report in Table~\ref{tab:Naos_SMEI_BRITE_1.8dPeriod} the values of all frequencies detected with $S/N>4$ from the Fourier analysis of the three seasonal observing runs taken separately ($2003 - 2004$, $2004 - 2005$, $2005 - 2006$), and the ensemble of the observations spanning $2003 - 2006$. We confirm that the only significant periodicity that is present in the \emph{Coriolis}/SMEI light curves of $\zeta$~Pup is the $1.78$-day period, and that the first harmonic of that period is not detected with a high significance level when taking the three seasonal observing runs separately nor when considering the entire dataset as a single light curve as did \citet{2014MNRAS.445.2878H}. In the latter case, the values that we obtained for the fundamental frequency and its amplitude as well as their formal uncertainties are in full agreement with the values reported by \citet{2014MNRAS.445.2878H} even if $(1)$ they used a date-compensated discrete Fourier transform algorithm \citep[DCDFT: ][]{1981AJ.....86..619F} instead of the classical discrete Fourier transform that we adopted in our analyses, and $(2)$ their formal $1\sigma$ uncertainties were assessed from $10000$ iterations of Monte-Carlo simulations of synthetic data whereas ours were extracted from only $100$ iterations. 

%%%%%%%%%%%%%%%%%%%%%%%%%%%%%%%%%%%%%%%%%%%%%%%%%%%%%%%%%
\subsection{The $1.78$-day period}
\label{subsec:Naos_1.78dperiod}

Here we investigate the properties of the $1.78$-day monoperiodic signal in $\zeta$~Pup by first determining its origin (Section~\ref{subsubsec:Naos_pulsations_vs_rotational_modulation}), then tracking its evolution during the \emph{Coriolis}/SMEI and \emph{BRITE} observing runs (Section~\ref{subsubsec:Naos_1.78dsig_stability}), and finally modeling the photospheric source causing it (Section~\ref{subsubsec:Naos_LI}).

%%%%%%%%%%%%%%%%%%%%%%%%%%%%%%%%%%%%%%%%%%%%%%%%%%%%%%%%%
\subsubsection{Stellar pulsations versus rotational modulation}
\label{subsubsec:Naos_pulsations_vs_rotational_modulation}

As already mentioned in Section~\ref{sec:Naos_Intro}, \citet{2014MNRAS.445.2878H} proposed an interpretation of the $1.78$-day periodic signal detected by \emph{Coriolis}/SMEI in $\zeta$~Pup as possibly due to low-order $(0 < l \le  2)$ oscillatory convection (non-adiabatic g$^{-}$) modes. The discovery of oscillatory convection modes with high-order azimuthal degree ($l \gtrsim 10$) in theoretical pulsation models for stars with $\log L/L_{\odot}=5.0$ and $3.65 \leq \log T_{\rm eff} \leq 4.0$ was first reported by \citet{1981PASJ...33..427S}. Later the \emph{theoretical} work of \citet{2011MNRAS.412.1814S} unveiled the possibility for the excitation of low-order $(0 < l \le  2)$ oscillatory convection modes in luminous stars with higher effective temperatures ($\log L/L_{\sun}\gtrsim4.5$, $\log T_{\rm eff} \gtrsim4.0$), necessarily associated with a subsurface convection zone caused by the opacity peak due to the partial ionization of iron-group elements near $\log T \sim 5.3$. In their thorough analysis of the \emph{Coriolis}/SMEI light curves of $\zeta$~Pup, \citet{2014MNRAS.445.2878H} encountered the classical dilemma ``pulsations versus rotational modulation'' for the interpretation of the newly discovered $1.78$-day signal. Such a situation is not uncommon and, depending on the amount of information available it may not be straightforward to draw a conclusion \citep[e.g.][]{2010A&A...519A..38D,2011A&A...536A..82D,2013A&A...557A.114A,2014MNRAS.441..910R}. In the case of $\zeta$~Pup, \citet{2014MNRAS.445.2878H} concluded that the $1.78$-day monoperiodic signal could arise from stellar oscillations rather than rotational modulation for the following reasons: 

\begin{enumerate}[labelindent=4.0pt,leftmargin=*]
\renewcommand\labelenumi{\textbf{[\roman{enumi}]}}
\item the period was found to be ``marginally consistent with the shortest possible rotation period''
\item there was only a ``minor periodogram peak'' at the first harmonic of the $1.78$~day period, such that the resulting phase-folded light curve was ``only slightly non-sinusoidal'' showing only one dominant bump \citep[Figure~2 in][]{2014MNRAS.445.2878H}
\item the observed pulsation constant ($Q=P\sqrt{\bar{\rho}/\bar{\rho}_\odot} \simeq P\sqrt{M/M_\odot/(R/R_\odot)^3}$, $\bar{\rho}$ and $\bar{\rho}_\odot$ being the mean stellar and solar densities) implied by the $1.78$~d period is of the order of $0.1-0.2$~d, which is consistent with the predictions in the models of Saio (2011) for low-order oscillatory convection modes ($Q \sim 0.2-0.3$~d).
\end{enumerate}

\emph{However}, now the \emph{BRITE} observations provide us with additional information, the most remarkable of which being the change of shape of the phased light curves (Figure~\ref{fig:Naos_BRITE_DFT_lc178d_BvsR}, middle panel; see also Figure~\ref{fig:Naos_BRITE_lcs_dynamic}, Section \ref{subsubsec:Naos_1.78dsig_stability}) with respect to the single-bumped modulation observed during the \emph{Coriolis}/SMEI observing run (Figure~2 in \citeauthor{2014MNRAS.445.2878H}~\citeyear{2014MNRAS.445.2878H}; see also Figures~\ref{fig:Naos_SMEI20052006_lcs_dynamic},~\ref{fig:Naos_SMEI20042005_lcs_dynamic},~\ref{fig:Naos_SMEI20032004_lcs_dynamic}): the \emph{BRITE} light curves phase-folded on the $1.78$~d period are highly non-sinusoidal, characterized by a pattern of two consecutive bumps separated by $\Delta\phi\sim0.4$ which explains why the first harmonic is prominent in the Fourier spectra. The highly non-sinusoidal nature of this monoperiodic signal, coupled with the change of shape of the phased light curves during the two observing runs are strong indications that it cannot arise from pulsations but rather a signature of rotational modulation. Note that the monoperiodic and non-sinusoidal nature of the signal alone are not sufficient for drawing any conclusion on whether it comes from stellar oscillations or from rotational modulation, as some radial pulsators are known to exhibit highly non-sinusoidal modulation of their light curves (e.g. \citeauthor{1952AJ.....57..158A}~\citeyear{1952AJ.....57..158A}, \citeauthor{2008ApJ...674L..93N}~\citeyear{2008ApJ...674L..93N}, \citeauthor{2010MNRAS.409.1244S}~\citeyear{2010MNRAS.409.1244S}). Also, it is worth noting that the shape-changing property of the light curve alone could be qualified as grossly similar to the shape-changing nature of the observed light variations in RV Tau variables \citep[e.g. ][]{1996MNRAS.279..949P}, which are pulsators showing the period doubling effect \citep{1990ApJ...355..590M}. To date, all the pulsators known to exhibit this effect are radial pulsators, although in theory the pulsation modes involved do not necessarily need to be radial but need to satisfy the half-integer resonance criterion, i.e. two modes A and B satisfying $(2n+1)\nu_A=2\nu_B$ with $n$ integer \citep{1990ApJ...355..590M}. In that case, due to resonant phase locking, mode B is not observed as an independent frequency: in the amplitude spectrum, mode A manifests at its frequency $\nu_A$ along with its harmonics, whereas mode B manifests through sub-harmonic frequencies of $\nu_A$. Therefore, if the $1.78$-day variability in $\zeta$~Pup were to be explained in terms of the period doubling effect, the main pulsation mode A has to be the one oscillating at the $\sim21.4$~h period, while mode B would manifest as the $1.78$~d period. But then, in that situation, the fact that there are no signs of any sub-harmonics of the $\sim21.4$~h period (except the $1.78$~d period) in any of the observing runs remains unexplained, and more importantly the $\sim21.4$~h period itself is quasi-absent during the $\sim4$ years of \emph{Coriolis}/SMEI observing run (Figure~2 in \citeauthor{2014MNRAS.445.2878H}~\citeyear{2014MNRAS.445.2878H}; Figures~\ref{fig:Naos_SMEI20052006_lcs_dynamic},~\ref{fig:Naos_SMEI20042005_lcs_dynamic},~\ref{fig:Naos_SMEI20032004_lcs_dynamic}). Based on all these considerations, we conclude that the highly non-sinusoidal nature of this monoperiodic shape-changing $1.78$~d signal in $\zeta$~Pup is incompatible with stellar oscillations but can be naturally explained as arising from rotational modulation due to the presence of bright spots appearing and disappearing at different locations on the stellar surface.

The pattern of two consecutive bumps observed in the phase domain during the \emph{BRITE} run is noteworthy. Patterns of two consecutive dips are common in the non-sinusoidally modulated light curves of cool stars having two dominant dark surface spots (e.g. \citeauthor{2013MNRAS.432.1203M}~\citeyear{2013MNRAS.432.1203M}, \citeauthor{2015ApJ...806..212D}~\citeyear{2015ApJ...806..212D}), whereas patterns of two consecutive bumps are common in the phased light curves of stars exhibiting bright surface chemical inhomogeneities \citep{2009MNRAS.396.1189B,2015A&A...581A.138B,2015AN....336..981B,2016A&A...588A..54W}. Over the past decade, an increasing number of light curves of massive OB stars have been seen to show low-frequency non-sinusoidal signal best explained as arising from rotational modulation rather than pulsations as the corresponding phased light curves show a pattern of two consecutive bumps and the periodograms show prominent peaks at the harmonics of the fundamental frequency \citep{2002A&A...393..965D,2010A&A...519A..38D,2011A&A...536A..82D,2016MNRAS.457.3724B}. In the case of $\zeta$~Pup, one noticeable property in the frequency domain is the prominence of the fundamental frequency (\emph{Coriolis}/SMEI and \emph{BRITE} observing runs) and the first harmonic (\emph{BRITE} observing run), while higher harmonics are not detected. This behaviour is reminiscent of the findings of  \citet{2016MNRAS.457.3724B} who scanned a sample of OB stars observed during the K2 mission and found that $54\%$ of them show rotational modulation rather than pulsations: $39\%$ of these rotational variables show a single dominant low frequency peak in their periodogram (no harmonics) and $61\%$ have both the fundamental frequency and its first harmonic prominent in their periodogram while higher harmonics are absent. This particular behaviour reconciles with the theoretical investigations of \citet{2003A&A...407.1029C} who found that the Fourier series decomposition of stellar light variations due to surface spots is characterized by significant contributions from the fundamental and the first harmonic, with the strengths of the higher harmonics rapidly vanishing as the order increases.

\begin{figure}
\includegraphics[width=8.4cm]{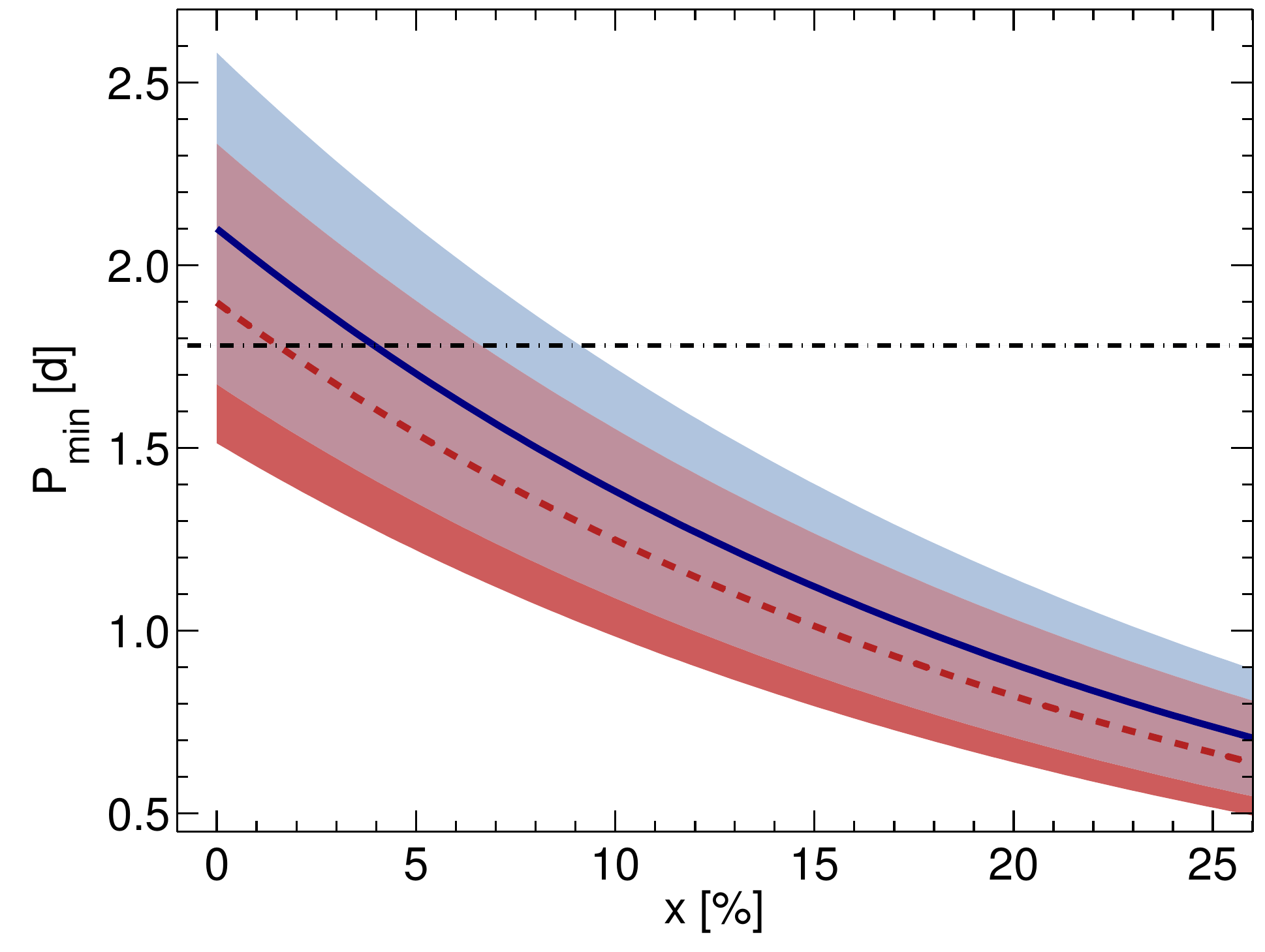}
% \vspace{-0.25cm}
 \caption{Minimum rotation period of $\zeta$~Pup (Equation~\ref{eq:Naos_Protmin}) as a function of the dimensionless fraction $x = (\log g_{\rm p} / 3.64) -1$ which represents the percentage of deviation of the polar gravity from the value of the effective surface gravity $\log g~(\textrm{cm~s}^{-2}) = 3.64\pm 0.1$ derived by \citet{2012A&A...544A..67B}. Two cases are considered: $R_{\rm e}=\sqrt{1.5}R$ (Solid Blue) and $R_{\rm e}=R$ (Dashed Red). In each case, the shaded area delimits the upper and lower bounds imposed by the uncertainties on the stellar radius and a $2.75\%$ uncertainty on $\log g$. The dash-dot horizontal line is the period $P=1.78063(25)$~d (its $1\sigma$ uncertainty is of the order of the thickness of the line itself).}
  \label{fig:Naos_Protmin}
\end{figure}

\begin{table}
 \caption{Estimates of the minimum possible rotation period $P_{\rm min}$, the equatorial velocity $\varv_{\rm e}$, and the inclination angle $i$ of $\zeta$~Pup, taking $P = 1.78063(25)$~d for the value of the rotation period and considering the two cases where $R_{\rm e} = \sqrt{1.5}R$ and $R_{\rm e} = R$. The values of $P_{\rm min}$ were calculated for $g_{\rm p} = g$.}
 \label{tab:Naos_Protmin_Vc_Ve_i}
 {\normalsize
\begin{center}
 \begin{tabular}{l r c c}
  \hline
  \hline
  Parameter & & $R_{\rm e} = \sqrt{1.5}R$ & $R_{\rm e} = R$ \\
  \hline
 $P_{\rm min}$ & [d] & $2.10_{-0.43}^{+0.48}$ & $1.90_{-0.39}^{+0.43}$ \\[7pt] 
$\varv_{\rm e}$ 		& [km~s$^{-1}$]	& 	$661\pm132$	& 	$539\pm108$	 \\[7pt] 
$i$ 			& [$\degr$]		& 	$19.4_{-4.7}^{+7.3}$		&	$24.0_{-5.9}^{+9.4}$\\[5pt]
  \hline
 \end{tabular}
\end{center}}
\end{table}

Under these considerations, now that our analyses indicate that the $1.78$ signal in $\zeta$~Pup is related to rotational modulation rather than pulsations, we revisit the three points \textbf{[i]}, \textbf{[ii]} and \textbf{[iii]} mentioned above. We note that since the calculation of the pulsation constant mentioned in point \textbf{[iii]} involves the stellar mass and radius, which have large uncertainties in the case of $\zeta$~Pup due to the large uncertainty in its distance, the last point \textbf{[iii]} can only be confirmed or ruled out once a refinement on the value of the distance is available, e.g. with the \emph{Gaia} mission \citep{2012A&A...538A..78L,2014SPIE.9143E..0YM}. Also, as previously discussed in detail, point \textbf{[ii]} is an integral part of the key property that the shape of the light curve changes due to the appearance of spots at different locations on the stellar surface. Finally, point \textbf{[i]} also depends on the stellar radius which is affected by the uncertainty in the distance. As pointed out by \citet{2014MNRAS.445.2878H}, in order to maintain a positive equatorial effective gravity in a Roche model, the stellar rotation period must not be shorter than 
\begin{equation}
P_{\rm min} = 3\pi\sqrt{\frac{R_{\rm e}}{g_{\rm p}}},
\label{eq:Naos_Protmin}
\end{equation}
where $R_{\rm e}$ is the equatorial radius and $g_{\rm p}$ the gravity at the pole. Note that to date the available values of $R$ and $g$, which are listed in Table~\ref{tab:Naos_StellarParams}, were all derived from spherical models \citep{2005A&A...436.1049M,2012A&A...544A..67B}. The values of $P_{\rm min}$ for $g_{\rm p}=g$ and considering the two cases where $R_{\rm e}=\sqrt{1.5}R$ and $R_{\rm e}=R$ are summarized in Table~\ref{tab:Naos_Protmin_Vc_Ve_i}, indicating that the $1.78$-day period is close to the minimum possible value of the stellar rotation period, but not entirely excluded by virtue of the estimated uncertainties. Also, $\zeta$~Pup being a fast rotator, if effects of gravity darkening are non-negligible, the value for $P_{\rm min}$ could be lower as the polar gravity would be higher than the value of the effective surface gravity $\log g~(\textrm{cm~s}^{-2}) = 3.64\pm0.1$ listed in Table~\ref{tab:Naos_StellarParams}. This is clearly illustrated on Figure~\ref{fig:Naos_Protmin}: for instance in the case $R_{\rm e}=\sqrt{1.5}R$, if $\log g_{\rm p}$ is $10\%$ higher than $\log g = 3.64$, then the minimum rotation period drops to $1.38_{-0.29}^{+0.34}$~d. From all these considerations we conclude that the three points \textbf{[i]}, \textbf{[ii]} and \textbf{[iii]} are compatible with the fact that the $1.78$~d signal in $\zeta$~Pup comes from rotational modulation.

Given the values of the $1.78$~d period yielded by the Fourier analyses of the different data sets of $\zeta$~Pup (Table~\ref{tab:Naos_SMEI_BRITE_1.8dPeriod}), we adopt a unique value of the rotation period throughout our investigation by assessing the weighted average of the values of the $1.78$~d period detected in the three \emph{Coriolis}/SMEI seasonal observing runs and the combined \emph{BRITE} light curve. We use the inverse square of the $1\sigma$ uncertainties derived from the Monte-Carlo simulations as weights on the periods, yielding a final weighted average of $P = 1.78063(25)$~d. Such a relatively short rotation period for $\zeta$~Pup is quite surprising at first sight, but is more consistent with the suspected rotational evolution of the star involving past interactions within a massive binary (\citeauthor{1998ASSL..232.....V}~\citeyear{1998ASSL..232.....V}; see also Section~\ref{sec:Naos_Discussion} and Figure~\ref{fig:Naos_EvolutionaryScenario}) or even a multiple system (\citeauthor{2012ASPC..465..342V}~\citeyear{2012ASPC..465..342V}). In all cases, the relatively short rotation period has strong implication on both the value of the stellar inclination angle $i$ with respect to the line of sight and the equatorial velocity $\varv_{\rm e}$. Given the available estimates of the stellar radius $R=18.99\pm3.80$~$R_{\sun}$ and the projected rotational velocity $\varv_{\rm e} \sin i = 219 \pm 18$~km~s$^{-1}$ (Table~\ref{tab:Naos_StellarParams}), the stellar inclination angle can be directly assessed as:
\begin{equation}
i = \arcsin\left(\frac{P \varv_{e} \sin i}{2 \pi R_{\rm e}}\right),
\label{eq:Naos_inclination}
\end{equation}
while the equatorial velocity is: 
\begin{equation}
\varv_{\rm e} =  \frac{2 \pi R_{\rm e}}{P}.
\label{eq:Naos_equatorialvelocity}
\end{equation}

The values of these two parameters are also reported in Table~\ref{tab:Naos_Protmin_Vc_Ve_i} for the two cases $R_{\rm e} = \sqrt{1.5}R$ and $R_{\rm e} = R$. As clearly depicted by the numbers in Table~\ref{tab:Naos_StellarParams}, we can conclude that the short rotation period implies that the star is seen at a very low inclination angle, but bearing in mind that all these values strongly depend on the stellar radius which remains highly uncertain as a result of the large uncertainty in the distance to the star. 

%%%%%%%%%%%%%%%%%%%%%%%%%%%%%%%%%%%%%%%%%%%%%%%%%%%%%%%%%
\subsubsection{Evolution of the $1.78$-day signal}
\label{subsubsec:Naos_1.78dsig_stability}

\begin{figure*}
\includegraphics[width=18cm]{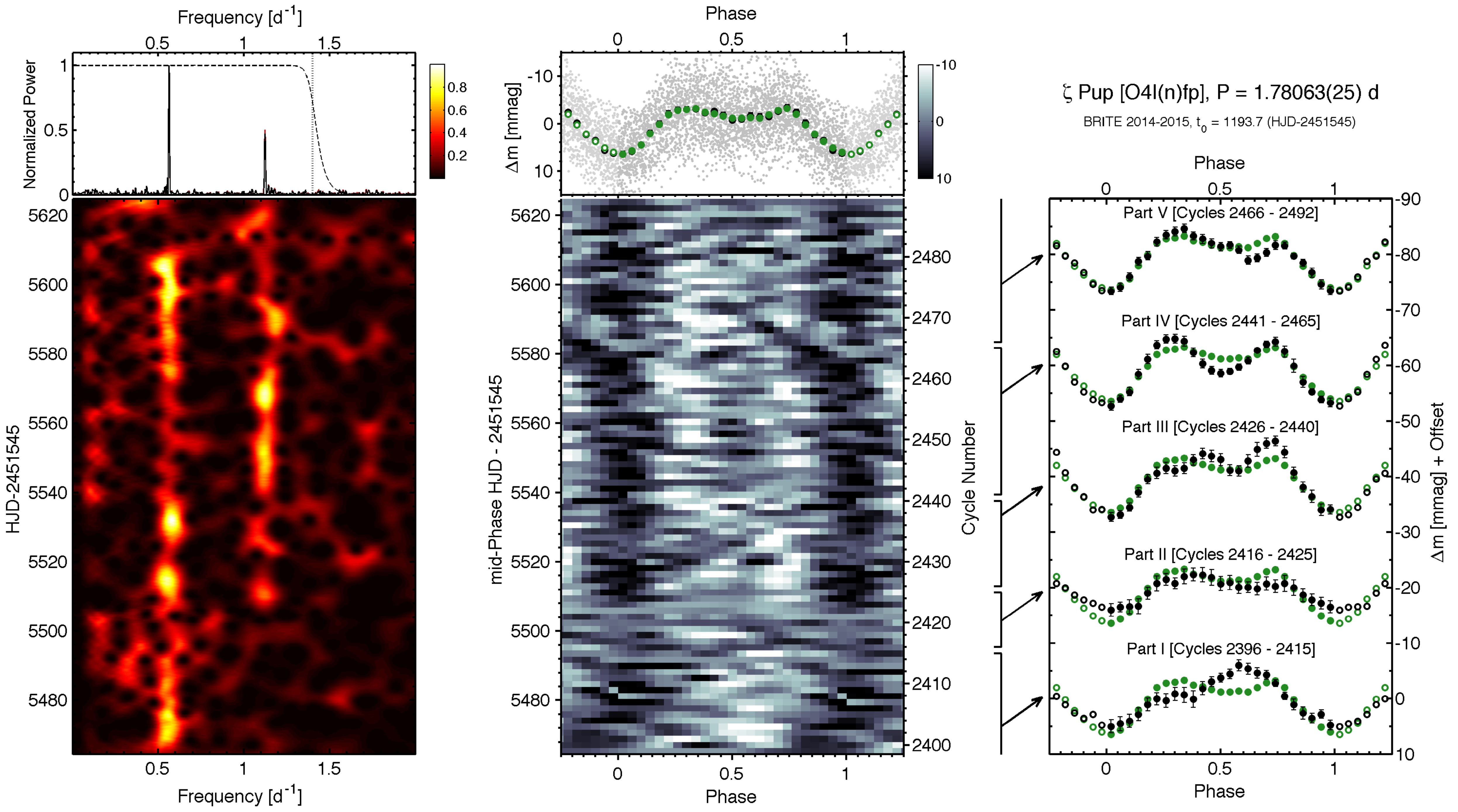}
 %\vspace{-0.5cm}
 \caption{Investigation of the evolution of the $1.78$~d signal in $\zeta$~Pup during the \emph{BRITE} observing run $(2014-2015)$. Time increases upwards. \emph{Left:} Time-frequency diagram of the combined filtered \emph{BRITE} light curves constructed from a $12$~d sliding window Fourier transform. The upper panel shows the power spectrum of the entire unfiltered light curve (red line) compared to that of the entire filtered light curve (black line), along with the magnitude response of the low-pass filter (dashed line) and the location of its cutoff frequency (vertical dotted line). \emph{Middle:} Continuous dynamic phased light curves of $\zeta$~Pup. The upper panel shows the phased light curve for the entire unfiltered data (black filled circles) compared to that from the entire filtered light curve (green filled circles). \emph{Right:} Average shapes of the phased light curve of $\zeta$~Pup (black diamonds) over the first $20$ rotational cycles (Part~I, see text), the next $10$ rotational cycles at the beginning of the transition phase (Part~II), the last $15$ rotational cycles during the transition phase (Part~III), the next $25$ rotational cycles (Part~IV) and the last $27$ rotational cycles (Part~V). These shapes are compared to the overall shape of the phased light curve (green filled circles).}
  \label{fig:Naos_BRITE_lcs_dynamic}
\end{figure*} 

As noted in the previous section, the key property leading to the conclusion that the $1.78$~d signal in $\zeta$~Pup is related to rotational modulation rather than stellar oscillations is the change of shape of the phased light curve from the epoch of the \emph{Coriolis}/SMEI observations to the epoch of the \emph{BRITE} run. This behaviour led us to also investigate the evolution of the amplitude of the $1.78$~d signal in $\zeta$~Pup through the \emph{BRITE} observing run and through each \emph{Coriolis}/SMEI seasonal run. In order to reduce the effects of high-frequency stochastic variability intrinsic to the star (that we shall characterize in Section~\ref{subsec:Naos_stochastic_variability}), we applied an infinite impulse response (IIR) low-pass $32$\textsuperscript{nd}-order Butterworth filter to the light curves, cutting off at $\nu_{\rm c} = 1.4$~d$^{-1}$. It is known that Butterworth filters have the advantage of having a monotonic magnitude response that is optimally flat (no ripples) in the passband, which is the prime reason why we chose it over other types of filters, at the cost of a less steep rolloff that is mitigated by increasing the order of the filter. We inspected the reliability of our low-pass Butterworth filter in studying the $1.78$~d signal in $\zeta$~Pup by comparing the power spectra of the original light curve and the filtered light curve (Figure~\ref{fig:Naos_BRITE_lcs_dynamic}, upper-left panel). There is globally no significant difference between the two power spectra in the filter bandpass, the loss of \emph{power} being at the $\sim2\%$ level at the fundamental frequency peak $\nu_0$ and $\sim8\%$ at the first harmonic $2\nu_0$. Besides, in the time domain, the rms scatter of the residuals between the phased light curve before and after applying the low-pass filter is $0.26$~mmag, which is of the order of half the rms value of the $1\sigma$ uncertainties of the mean points in the original phased light curve (Figure~\ref{fig:Naos_BRITE_lcs_dynamic}, upper-middle panel). From these results we conclude that the low-pass filtering process is reliable to study the evolution of the $1.78$~d signal while reducing the contribution of the high-frequency stochastic variations intrinsic to the star which shall be discussed in detail in Section~\ref{subsec:Naos_stochastic_variability}. Thus we performed all of the analyses related to the $1.78$~d signal on the filtered light curves.

In the frequency domain, we conducted a Fourier analysis of the \emph{BRITE} light curve in sliding windows over restricted time intervals spanning $\Delta T = 12$~d (with a step of $\Delta T \times 1\%$), thus covering at least $\sim6$~stellar rotational cycles at each step. The lower-left panel of Figure~\ref{fig:Naos_BRITE_lcs_dynamic} depicts the resulting time-frequency diagram, clearly showing that the fundamental frequency is always present from the beginning of the observing run until almost the end (around day $5610.0$), while the first harmonic only emerges from day $\sim5510.0$ and remains visible until the end of the observing run. Between days $\sim 5540.0$ and $\sim 5575.0$, the fundamental frequency is still present but its power is partly absorbed by the first harmonic which is very prominent during that time interval. 

Besides the time-frequency analysis, we also tracked the evolution of the $1.78$-day signal by splitting the light curve into consecutive rotational cycles and stacking them on top of each other to form the grayscale diagram in the middle panel of Figure~\ref{fig:Naos_BRITE_lcs_dynamic} as a function of rotational phase and cycle number. Such a diagram is not common practice for analyzing periodic/cyclic signals in photometric observations of stars as it requires a large number of points per cycle, but could be useful in detecting and characterizing subtle variations (e.g. \citeauthor{2012AJ....143..137G}~\citeyear{2012AJ....143..137G}, \citeauthor{2015ApJ...806..212D}~\citeyear{2015ApJ...806..212D}). In our case, the combined filtered \emph{BRITE} light curve of $\zeta$~Pup provides us with an increased number of points per rotational cycle and increased S/N so that the diagram turns out to be a very useful complement to the time-frequency analysis as it helps us track events in the light curve that explain the behaviour that we found in the time-frequency diagram. More specifically, before day $\sim 5490$ the phased light curves show one large bump centered around phase $0.6$, which is the reason why we only see the fundamental frequency and not the first harmonic during that epoch of the observations. This behaviour could be due to the presence of one relatively large bright spot or a group of smaller bright spots at the photosphere. Then the time between day $\sim 5490$ and day $\sim 5538$ is characterized by a transition that eventually leads to the formation of two bumps separated by $\Delta \phi \sim 0.44$ in the phased light curve at the end of the transition. These two bumps are clearly defined between days $\sim 5540.0$ and $\sim 5575.0$, causing the prominence of the first harmonic of the $1.78$~d signal during that time interval in the time-frequency diagram. The presence of these two bumps could be the manifestation of two dominant bright spots at the stellar surface, separated by $360\degr\times\Delta \phi \simeq158\degr$ in longitude. The separation of these two bumps in phase seems to become larger towards the end of the observing run, a sign that the surface spots experience a slight migration in longitude, which is then an appropriate explanation for the detection of the pair of frequencies $\{\nu^{\prime}_{0};\nu^{\prime\prime}_{0}\}$ closely spread around the true value of the first harmonic during the prewhitening procedure as pointed out in Section~\ref{subsec:Naos_Results_Photo_Fourier}. The right panel of Figure~\ref{fig:Naos_BRITE_lcs_dynamic} shows the mean phased light curves over several consecutive rotational cycles, thus summarizing the above descriptions:

\begin{enumerate}[labelindent=4.0pt,leftmargin=*]
\renewcommand\labelenumi{\textbf{[\roman{enumi}]}}
\item the first $20$ rotational cycles observed before the transition phase, showing one large bump centered around phase $0.6$ (Cycles $2396-2415$, which we will hereafter refer to as Part~I),
\item the next first $10$ rotational cycles during the transition phase (Cycles $2416-2425$; Part~II),
\item the last $15$ rotational cycles during the transition phase (Cycles $2426-2440$; Part~III),
\item the $25$ rotational cycles right after the transition phase, during which the two bumps separated by $\Delta \phi \sim 0.44$ are clearly visible (Cycles $2441-2465$; Part~IV),
\item the last $27$ rotational cycles of the observing run (Cycles $2466-2492$; Part~V).
\end{enumerate}

We note that the behaviour of the star during the transition (Part~II and Part~III) is not clearly defined at this point, the two most pertinent scenarios being either that the large spot/spot group in Part~I disintegrates and the resulting most dominant parts \emph{migrate} to form the two dominant spots that we see in Part~IV at locations on the surface separated by $\sim158\degr$ in longitude, or the large spot/spot group in Part~I gradually disappears while two \emph{new} dominant spots take birth. 

\citet{2014MNRAS.445.2878H} already investigated the stability of the $1.78$~d signal during the \emph{Coriolis}/SMEI observing run by performing a DCDFT on the seasonal subsets of the light curve and on $50$-d subsets with $50\%$ overlaps. They noticed that the fundamental frequency of the signal does not change, but its amplitude varies by a factor 2 during the observing run, while it experiences a modest phase excursion \citep[Figure~3 in ][]{2014MNRAS.445.2878H}. In order to investigate the stability of the $1.78$~d signal on the $12$~yr timescale covered by the \emph{Coriolis}/SMEI and the \emph{BRITE} observing runs, we established the same dynamic plots as Figure~ \ref{fig:Naos_BRITE_lcs_dynamic} for the \emph{Coriolis}/SMEI seasonal observing runs (Figures~\ref{fig:Naos_SMEI20052006_lcs_dynamic},~\ref{fig:Naos_SMEI20042005_lcs_dynamic}~and~\ref{fig:Naos_SMEI20032004_lcs_dynamic}). The time-frequency plots show that the fundamental frequency is always detected except at some points where it is so weak that its power is comparable with the noise level (e.g. between HJD$-2451545 \sim 1450-1500$). Furthermore, the dynamic phased light curves show that the minima wander within the interval $0.4-0.6$, confirming the behaviour reported by \citet{2014MNRAS.445.2878H}. More interestingly, the phased light variations observed during the epoch of the \emph{Coriolis}/SMEI observing run present only one relatively large bump located around phase zero, which could indicate that  the surface spots seen during the \emph{Coriolis}/SMEI observing run appear at different longitudes compared to those observed during the epoch of the \emph{BRITE} observations. However, at this point a strict interpretation of the position of the maxima in the phased light curve cannot be ascertained since the exact value of the rotation period remains unknown, and the value of its error bar $\sigma = 0.00025$~d could lead to an apparent phase shift of $\sim0.03$ in one year, or $\sim0.35$ over the $\sim12$~years covered by the \emph{Coriolis}/SMEI and the \emph{BRITE} observing runs. Nevertheless, even considering such a large phase shift, the clear shape-changing nature of the light variations over these twelve years, characterized by the presence of one large bump during the \emph{Coriolis}/SMEI observing run and the presence of two dominating bumps during the epoch of the \emph{BRITE} observations, is already a strong indication that the spots seen in $2003-2006$ are not the same as the ones seen in $2014-2015$.

%%%%%%%%%%%%%%%%%%%%%%%%%%%%%%%%%%%%%%%%%%%%%%%%%%%%%%%%%
\subsubsection{Mapping the spotted stellar surface}
\label{subsubsec:Naos_LI}

Approaches to the modeling of light curves showing effects of rotational modulation associated with the presence of surface spots can be classified into two distinct categories depending on whether the problem is solved in a direct manner or in an inverse manner. The direct method consists in calculating the emerging stellar light variations produced by \emph{a given number of spots} (often taken as a fixed parameter), provided some \emph{a priori} assumptions on their \emph{shape} (often assumed to be circular), their \emph{decay law}, their \emph{size}, and their \emph{location} as well as the \emph{time of their first appearance} on the stellar surface. Then these last three free parameters are refined for each spot until the calculated light curve reasonably fits the observed light curve. Obviously, in that method a convergence towards a reasonable fit to the observed light curve strongly depends on the assumed number of spots. In practice, one could gradually increase the assumed number of spots until the smallest number that produces a good fit is reached, and then argue that in some sense the simplest acceptable spot distribution has been found. But then the other five parameters for each of the spots are still being assumed. This direct method, also known as the analytical approach, has been used in the modeling of light curves of cool low-mass stars exhibiting active surface regions  (e.g. \citeauthor{2012MNRAS.427.2487K}~\citeyear{2012MNRAS.427.2487K}, \citeauthor{2014ApJ...788....1B}~\citeyear{2014ApJ...788....1B}, \citeauthor{2015ApJ...806..212D}~\citeyear{2015ApJ...806..212D}) and a few cases of hotter stars \citep{2010A&A...509A..43L,2011A&A...536A..82D, 2014MNRAS.441..910R}, offering the advantage of being fast in terms of computation time but suffering from the disadvantage that the number of spots, their shapes and their decay law have to be assumed a priori along with good estimates of starting points for the other three free parameters for each spot.  Alternatively, the inverse approach \citep{2000AJ....120.3274H,2011AJ....141..138R,2013ApJ...767...60R,2016ApJ...832..207R,2007A&A...470.1089K,2013ARep...57..548K,2014AstBu..69..179K,2015AdSpR..55..808K} consists in assessing the distribution of specific intensity at the stellar surface that reasonably fits the observed light curve, with no a priori assumptions concerning the six parameters mentioned previously, the sole assumption being that the stellar surface contains spot-like features. This conceptually constitutes an inverse ill-posed problem and has been proven to require more computation time than the direct approach (thus its designation as the ``numerical approach''), keeping in mind its great advantage of being free from any prior assumption on the five spot parameters enumerated previously.

In order to model the bright spot-induced light variations of $\zeta$~Pup as observed by \emph{BRITE} and \emph{Coriolis}/SMEI, we adopted the numerical approach by using a constrained non-linear light curve inversion algorithm \citep[LI\footnote{Initially denominated ``matrix light curve inversion (MLI)'' in \citet{2000AJ....120.3274H} as it was inspired from a matrix-based algorithm for the inversion of light curves of planetary objects to assess their surface albedo distribution \citep{1989PASP..101..844W,1991ApJ...368..622W}. However, the actual implementation of the algorithm no longer involves matrices; thus, we adopt its common current designation: light curve inversion (LI).}: ][]{2000AJ....120.3274H}. This algorithm for mapping the surface of spotted stars was initially formulated and tested on simulated data by \citet{2000AJ....120.3274H}, and applied to the analyses of the evolution of dark spots on the K2IV primary component of II Pegasi \citep{2011AJ....141..138R} and cool K-type \emph{Kepler} targets (KIC 5110407: \citeauthor{2013ApJ...767...60R}~\citeyear{2013ApJ...767...60R}; KOI-1003: \citeauthor{2016ApJ...832..207R}~\citeyear{2016ApJ...832..207R}). Here we apply the algorithm for the first time to map the locations and evolution of bright spots at the surface of a hot massive star.

%%%%%%%%%%%%%%%%%%%%%%%%%%%%%%%%%%%%%%%%%%%%%%%%%%%%%%%%%
\paragraph{Light curve inversion - The algorithm\\\\}
\label{paragraph:Naos_LI_algorithm}

Let us represent the observed light curve as the time series $\mathbf{I} = \left\{I(t_{k})\right\}_{1\leq k\leq K}$, thus containing $K$ measurements. The algorithm as it stands assumes rigid rotation, and thus currently does not explicitly allow for surface differential rotation. Then the basis of LI is to partition the stellar surface into $N$ latitude bands of equal angular size ($\Delta\vartheta=\pi/N$), each $n$\textsuperscript{th} latitude band containing $P_n$ spherical rectangles having exactly equal area. In addition, the partitioning is made such that all the patches on the stellar surface are of nearly equal area, which is achieved by having $P_n$ proportional to the cosine of the latitude of the center of the band, subject to the constraint that $P_n$ must be an integer. Thus, at a given time $t_{k}$, if say $J_{np}$ is the contribution of the specific intensity along the outward normal for patch $(n;p)$ that subtends a solid angle $\Omega_{np}(t_{k})$ as seen from the observer, and $\mathcal{L}_{np}(t_{k})$ the value of the limb-darkening at the location of that patch, then to the limit of a large total number of patches, the intensity from the star at that time takes the discretized form
\begin{equation}
I(t_{k}) = \sum_{n=1}^{N} \sum_{p=1}^{P_n} \Omega_{np}(t_{k}) \mathcal{L}_{np}(t_{k}) J_{np},
\label{eq:Naos_LI_specificintensity}
\end{equation}
in which the outer summation needs only to be performed up to $n_s \leq N$, which is the index of the southernmost latitude band that is visible by the observer as a consequence of the inclination of the star (thus $n_s = N$ if the star is seen equator on). Then the idea is to find the set of relative patch intensities $\mathbf{\hat{J}}=\{\hat{J}_{np}\}_{1\leq n \leq n_s; 1\leq p \leq P_n}$ that produces a reconstructed light curve $\mathbf{\hat{I}} = \{\hat{I}(t_{k})\}_{1\leq k\leq K}$ such that the rms deviation (expressed in magnitudes) between $\mathbf{\hat{I}}$ and the observed light curve $\mathbf{I}$ is equal to the estimated rms noise in $\mathbf{I}$ (also expressed in magnitudes). The ill-posed nature of the problem clearly appears here, the key component that is involved being the noise in the observations that shows as high-frequency ripples that would be indistinguishable from the effect of a multitude of small bright and dark spots randomly distributed over the stellar surface. Simply trying to find the distribution of patch intensities that best fits the observed light curve, i.e. through a standard minimization of a goodness-of-fit criterion, would inevitably yield a granulated surface dominated by noise artifacts. As in any ill-posed problem, a regularization process must be involved in order to mitigate noise sensitivity and ensure convergence towards a unique solution \citep{Twomey1977,1986ipag.book.....C}. In our case, regularization is achieved by solving for the distribution of patch intensities $\mathbf{\hat{J}}$ through a Lagrange constrained minimization of an objective function defined as a Lagrangian of the form
\begin{equation}
E(\mathbf{\hat{J}},\mathbf{I},\lambda,B) = G(\mathbf{\hat{J}},\mathbf{I}) + \lambda S(\mathbf{\hat{J}},B),
\label{eq:Naos_LI_objectivefunc}
\end{equation}
in which $G(\mathbf{\hat{J}},\mathbf{I})$ is the term that measures the goodness-of-fit between the reconstructed light curve and the observed light curve, while the second term $\lambda S(\mathbf{\hat{J}},B)$ controls the smoothness of the reconstructed stellar surface. The function $S(\mathbf{\hat{J}},B)$ is therefore called \emph{smoothing function}, such that the lower value it takes the smoother the surface is. Thus in this context the Lagrange multiplier $\lambda$ acts as a \emph{smoothing parameter}, a property that can be easily understood when considering the two asymptotic cases where $\lambda \rightarrow 0$ and $\lambda \rightarrow +\infty$: the first case amounts to minimizing $E(\mathbf{\hat{J}},\mathbf{I},\lambda,B) \simeq G(\mathbf{\hat{J}},\mathbf{I})$ which would yield solutions dominated by noise artifacts as discussed above, while the second case would yield very smooth solutions that poorly fit to the observed light curve as the term $\lambda S(\mathbf{\hat{J}},B)$ would dominate over $G(\mathbf{\hat{J}},\mathbf{I})$. The optimal solution of the inverse problem corresponds to an intermediate value of $\lambda$ that is the best trade-off between these two limit cases, yielding a reconstructed light curve $\mathbf{\hat{I}}$ that reasonably fits the observed light curve $\mathbf{I}$ while mitigating the tendency of overfitting.

The subtle role of the smoothing function $S(\mathbf{\hat{J}},B)$ can be understood from its closed-form expression which is a generalized Tikhonov regularizer:
\begin{equation}
S(\mathbf{\hat{J}},B) =  \sum_{n=1}^{n_s} \sum_{p=1}^{P_n} w_{n} c_{np} \left[ \hat{J}_{np} - \langle \hat{J} \rangle \right]^2 ,
\label{eq:Naos_LI_penaltyfunction}
\end{equation}
where $\langle \hat{J} \rangle$ is the average patch intensity, $w_{n}$ are latitude-dependent weighting factors, and the coefficients $c_{np}$ incorporate the constraint that the surface exhibits confined discernible spots on a uniform background. This is achieved through the definition:
\begin{equation}
c_{np} = \left\{\begin{array}{l c r l}
        1 & & \textrm{if} & \hat{J}_{np} < \langle \hat{J} \rangle \\
        B & & \textrm{if} & \hat{J}_{np} \geq \langle \hat{J} \rangle
        \end{array}\right.,
\label{eq:Naos_LI_biasparameter}
\end{equation}
with $B$ a strictly positive real constant. Keeping in mind that $S(\mathbf{\hat{J}},B)$ is involved in the objective function $E(\mathbf{\hat{J}},\mathbf{I},\lambda,B)$ to be minimized, two distinct cases arise from this definition of $c_{np}$: $B>1$ and $0<B<1$. The first case would mean that a patch that tends to be brighter than $\langle \hat{J} \rangle$ by a given amount rather than being dimmer than $\langle \hat{J} \rangle$ by the same amount will increase $S(\mathbf{\hat{J}},B)$ by a factor of $B>1$, thus inflicting a penalty in the minimization of $E(\mathbf{\hat{J}},\mathbf{I},\lambda,B)$. Conversely, if $0<B<1$ the penalty becomes a gain in the minimization of $E(\mathbf{\hat{J}},\mathbf{I},\lambda,B)$ as the contribution of the patch to $S(\mathbf{\hat{J}},B)$ decreases by a factor of $1/B>1$. In other words, the information on whether we want to model dark or bright spots on a uniform background is incorporated in the coefficients $c_{np}$: the modeling of dark spots requires $B>1$ as was adopted in the work of \citet{2011AJ....141..138R,2013ApJ...767...60R,2016ApJ...832..207R}, while in this investigation we constrain $0<B<1$ to model the bright spots on $\zeta$~Pup. Note that for these reasons, $S(\mathbf{\hat{J}},B)$ is also called \emph{the penalty function} and $B$ the \emph{bias parameter} since it biases the solution in favor of either dark or bright spots.

Also, the presence of the latitude-dependent weighting factors $w_{n}$ is noteworthy and comes from the necessity to take into account the fact that the maximum projected area of a patch increases the closer to the sub-Earth latitude it is located. Hence the amount of light modulation induced by any given spot is equivalent to that of a smaller spot located closer to the sub-Earth latitude. This degeneracy would lead the algorithm to a systematic convergence towards solutions with spots lying near the sub-Earth point, as a result of the fact that small spots incur small values of the penalty function $S(\mathbf{\hat{J}},B)$. To reduce that systematic behaviour, the $n$\textsuperscript{th} latitude band is weighted by a coefficient $w_{n}$ that is proportional to the difference between the maximum and minimum values of the product of the projected area and the limb darkening for patches in the band, namely:
\begin{equation}
w_{n} = \frac{1}{\mathcal{W}} \times \left| \max\limits_{1 \leq p \leq P_n} \{\mathcal{A}_{np}\} - \min\limits_{1 \leq p \leq P_n} \{\mathcal{A}_{np}\} \right|,
\label{eq:Naos_LI_latitudeWeight}
\end{equation}
in which $\mathcal{W} = \max\limits_{1 \leq n \leq n_s} \{w_{n}\}$ is a normalization factor, while $\mathcal{A}_{np}$ is the product of the projected area and the limb-darkening value for the patch $(n;p)$:
\begin{eqnarray}
\mathcal{A}_{np} &=& \frac{1}{2}\left[ \sin i \left( \cos\varphi_1 - \cos\varphi_2 \right) \left( \frac{\pi}{N} + \frac{1}{2} \left( \sin2\vartheta_1 - \sin2\vartheta_2 \right) \right) \right. \nonumber \\  
&&\left. + \frac{2\pi}{P_n} \cos i \left( \cos^2\vartheta_1- \cos^2\vartheta_2 \right) \right] \times \mathcal{L}_{np},
\label{eq:Naos_LI_projectedArea}
\end{eqnarray}
in which $i$ is the stellar spin axis inclination angle with respect to the line-of-sight, $\vartheta_1$ and $\vartheta_2$ are the colatitudes of the northern and southern edges of the latitude band, whereas $\varphi_1$ and $\varphi_2$ are the longitudes of the western and eastern edges of the patch. As a matter of fact, $\max\limits_{1 \leq p \leq P_n} \{\mathcal{A}_{np}\}$ is reached at the sub-Earth meridian, while $\min\limits_{1 \leq p \leq P_n} \{\mathcal{A}_{np}\}$ is reached at the anti-Earth meridian. Although it was originally found that the introduction of the latitude weighting coefficients $w_n$ generally resulted in a modest improvement on the latitude resolution and an increased computation time \citep{2000AJ....120.3274H}, here we decided to keep them in order to optimize the output of LI. 

Finally, the goodness-of-fit term $G(\mathbf{\hat{J}},\mathbf{I})$ is defined to be the rms deviation in magnitudes of the reconstructed light curve from the observed light curve. Evaluated in units of the estimated noise variance in magnitudes $\sigma^2$ in the observed light curve, the closed-form expression for $G(\mathbf{\hat{J}},\mathbf{I})$ is:

\begin{equation}
G(\mathbf{\hat{J}},\mathbf{I}) = \frac{\left( 2.5\log e \right)^2 }{K \sigma^2} \sum_{k=1}^{K} \left( \frac{I(t_{k}) - \hat{I}(t_{k})}{I(t_{k})} \right)^2.
\label{eq:Naos_LI_goodnessofFit}
\end{equation}

As mentioned earlier, the goal of LI is to find the optimal distribution of patch intensities $\mathbf{\hat{J}}$ that yield a reconstructed light curve $\mathbf{\hat{I}}$ such that the rms variance between $\mathbf{\hat{I}}$ and the observed light curve $\mathbf{I}$ reaches $\sigma^2$. In other words, an inversion consists in finding the optimal values of the Lagrange multiplier $\lambda$ and the bias parameter $B$, for which we have:

\begin{equation}
G(\mathbf{\hat{J}}(\lambda;B),\mathbf{I}) = 1.
\label{eq:Naos_LI_f}
\end{equation}

That is achieved through a two-step process involving a Van Wijngaarden-Dekker-Brent root-finding algorithm \citep{Press:2007:NRE:1403886} to first solve Equation~\ref{eq:Naos_LI_f} for $\lambda$ for a set of values of the bias parameter $B$ constrained within the interval $[0;1]$, yielding $\lambda_{\rm opt}(B)$ which is in turn used to find $B_{\rm opt}$ as the root of the function
\begin{equation}
g(B) = \frac{\min \left\{ \hat{J}_{np}(\lambda_{\rm opt}(B); B) \right\} }{\langle \hat{J}_{np}(\lambda_{\rm opt}(B); B) \rangle} - \frac{J^{\rm (s)}}{J^{\rm (p)}},
\label{eq:Naos_LI_g}
\end{equation}
where $J^{\rm (s)}$ and $J^{\rm (p)}$ are respectively the spot and photosphere specific intensities, two parameters that we determine by evaluating the Planck function at the central wavelength of the filter passband for the spot and photosphere temperatures. In principle a better way to obtain a proxy for these two parameters is to integrate the Planck function over the filter passband for the spot and photosphere temperatures, the most ideal way being a numerical integration of the stellar spectral energy distribution (SED) over the filter passband. However, given the typical uncertainty in spot temperatures, the method that we adopted is enough for achieving a nearly optimal precision.\\

%%%%%%%%%%%%%%%%%%%%%%%%%%%%%%%%%%%%%%%%%%%%%%%%%%%%%%%%%
\paragraph{Spot temperatures\\\\}
\label{paragraph:Naos_Tspot}

At this point, particular consideration has to be given to spot temperatures: since our \emph{BRITE} observations were performed in two different filters, any amplitude difference between observations through the two filters could be exploited in order to extract potential information on the typical temperatures of the spots. This can be easily conceived by assuming to first order that all the spots that are present on the surface at a given time $t_{k}$ are all at the same temperature $T^{\rm (s)}$ (a reasonable approximation if the spots roughly probe the same stellar atmosphere layer), and assuming that their combined area at that snapshot in time is equivalent to a single spot occupying a fractional area $f$ on the stellar surface subtending a limb angle $\theta$ as seen from the center of the star and emitting $I^{\rm (s)}$ per unit area, while the unperturbed stellar photosphere emits $I^{\star}$ per unit area in that filter. This situation amounts to the configuration considered by \citet{2003A&A...407.1029C} (see his Figure~1), so that the observed photometric signal takes exactly the same closed-form expression as his Equation~1:
\begin{equation}
F^{}_j(\theta) = \mathcal{K} \left[ I_j^{\star} + f \left( I_j^{\rm (s)} - I_j^{\star}  \right) \cos \theta \right],
\label{eq:Naos_Tspot_ObservedFlux}
\end{equation}
in which we have the additional subscript $j$ to indicate observations through filter $j$ ($j =$~``$b$'' or ``$r$'' in our case). The factor $\cos\theta$ takes into account the foreshortening of the projected area of the spot according to the angle $\theta$, whereas the factor $\mathcal{K}$ is a constant that depends on the stellar radius and the distance to Earth. Furthermore, Equation~\ref{eq:Naos_Tspot_ObservedFlux} can be expressed in terms of magnitudes and approximated to first order as:  
\begin{equation}
m^{}_j - m_j^{\star} \simeq  -2.5 (\log e) \left( \frac{I_j^{\rm (s)} - I_j^{\star}}{I_j^{\star}}  \right) f \cos \theta.
\label{eq:Naos_Tspot_ObservedMag_1stOrder}
\end{equation}
with $m_j^{\star} = -2.5\log\left(\mathcal{K}I_j^{\star}\right)$ the stellar magnitude in the absence of spots. Therefore the ratio of the observed variations in the blue filter to the observed variations in the red filter is:

\begin{equation}
\alpha = \frac{\Delta m^{}_{b}}{\Delta m^{}_{r}}  = \frac{m^{}_{b} - m_{b}^{\star}}{m^{}_{r} - m_{r}^{\star}} \simeq \frac{I_{b}^{\star} - I_{b}^{\rm (s)}}{I_{r}^{\star} - I_{r}^{\rm (s)}} \times \frac{I_{r}^{\star}}{I_{b}^{\star}}.
\label{eq:Naos_Tspot_ObservedMag_1stOrder_ratioBR}
\end{equation}

But by definition, this ratio $\alpha$ turns out to be exactly the slope in our \emph{BRITE-b} vs \emph{BRITE-r} diagrams! In terms of color indices, Equation~\ref{eq:Naos_Tspot_ObservedMag_1stOrder_ratioBR} yields:
\begin{equation}
(b-r)_{\rm d} = (b-r)_{\star} -2.5\log \alpha,
\label{eq:Naos_Tspot_ObservedMag_1stOrder_ratioBR_colorindices}
\end{equation}
where $(b-r)_{\rm d}$ is the color index of the difference between the spot and the photosphere, while $(b-r)_{\star}$ is the color index for the unperturbed star. In other words, Equation~\ref{eq:Naos_Tspot_ObservedMag_1stOrder_ratioBR_colorindices} remarkably tells us that a simple measurement of the slope in the \emph{BRITE-b} vs \emph{BRITE-r} diagram can give us an estimate of spot temperatures by means of calibrations between $(b-r)$ and effective temperatures. Also, Equation~\ref{eq:Naos_Tspot_ObservedMag_1stOrder_ratioBR_colorindices} can be applied to any dual-band photometric time series observations of any star when appropriate calibrations of color indices versus effective temperatures are available. Unfortunately in our situation we measure a slope $\alpha=0.99\pm0.09$ in our \emph{BRITE-b} vs \emph{BRITE-r} diagram for the $1.78$~d signal in $\zeta$~Pup (right panel of Figure~\ref{fig:Naos_BRITE_DFT_lc178d_BvsR}), which does not allow us to extract any information on spot temperatures. This insensivity could be due to the fact that the two \emph{BRITE} red and blue filters fall in the domain of validity of the Rayleigh-Jeans approximation for a hot star like $\zeta$~Pup. Thus, given the fact that there is no amplitude difference in the observations in the two filters, we performed our inversions on the combined \emph{BRITE} light curve of $\zeta$~Pup, allowing us to have decreased errors in the mean points and reduced gaps in rotational phase coverage, even if it has been demonstrated by \citet{2000AJ....120.3274H} that the inversion of light curves in multiple filters can improve the spot latitude resolution.

\begin{figure*}
\includegraphics[width=18cm]{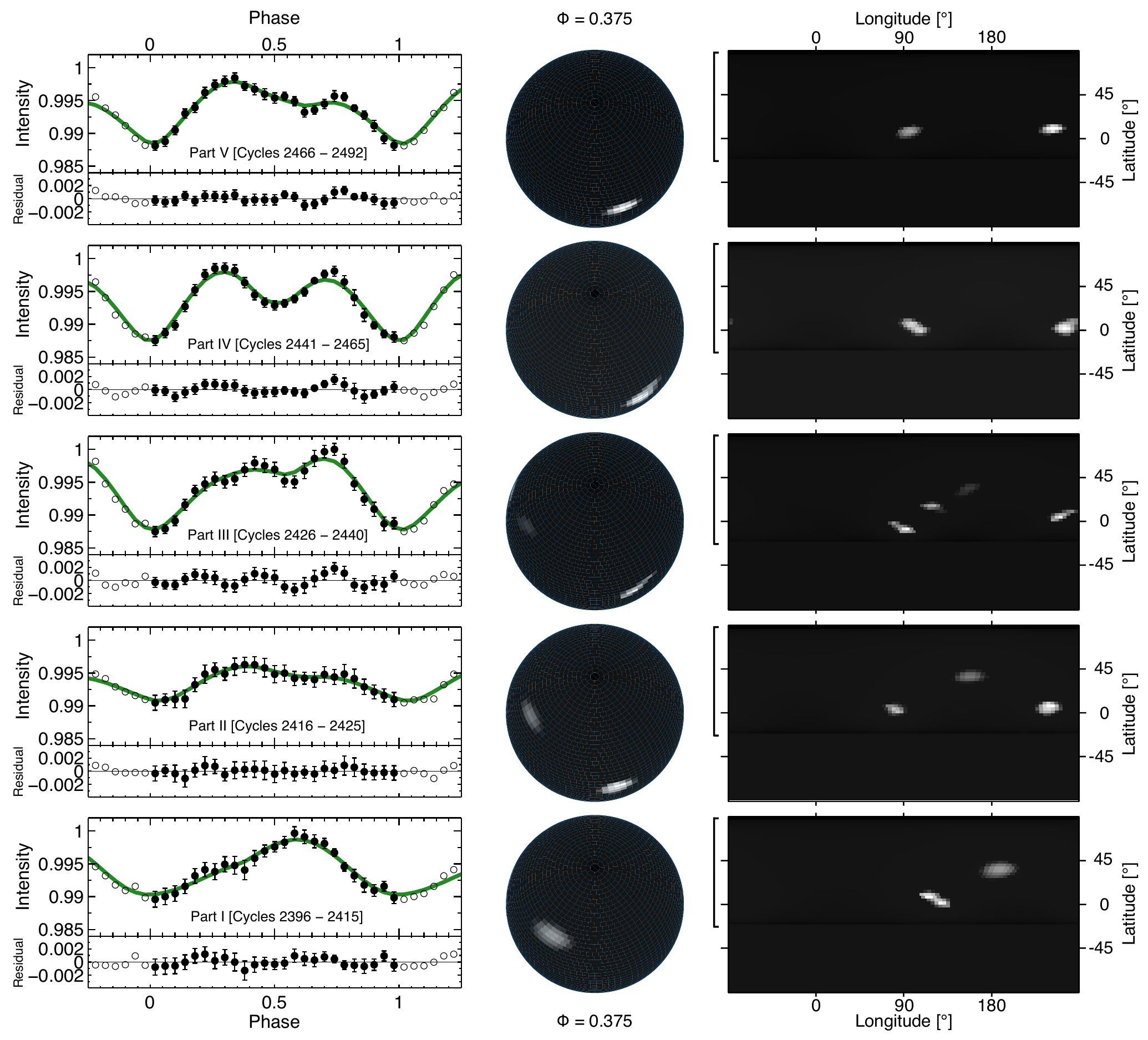}
 %\vspace{-0.5cm}
 \caption{Light curve inversion: mapping the photosphere of $\zeta$~Pup as observed by \emph{BRITE} in 2014-2015, for $T^{\rm (s)}=42.5$~kK and a stellar inclination angle $i=24\degr$. Time increases upwards. The left panel illustrates the observed light curve (filled circles) during Part~I...V of the \emph{BRITE} observing run as defined on Figure~\ref{fig:Naos_BRITE_lcs_dynamic}, Section~\ref{subsubsec:Naos_1.78dsig_stability}, along with the reconstructed light curve (green line), with the residuals plotted below the light curves. Then follows a view of the star at rotational phase $0.375$ (\emph{Middle panel}) and the pseudo-Mercator projection of the stellar surface (\emph{Right panel}). The vertical open brackets on the left of the pseudo-Mercator projections indicate the range of latitudes visible by the observer. The sub-Earth point is at longitude $0\degr$ at rotational phase zero.}
  \label{fig:Naos_BRITE1415_LI_maps_inc24}
\end{figure*}

\begin{figure*}
 \includegraphics[width=18cm]{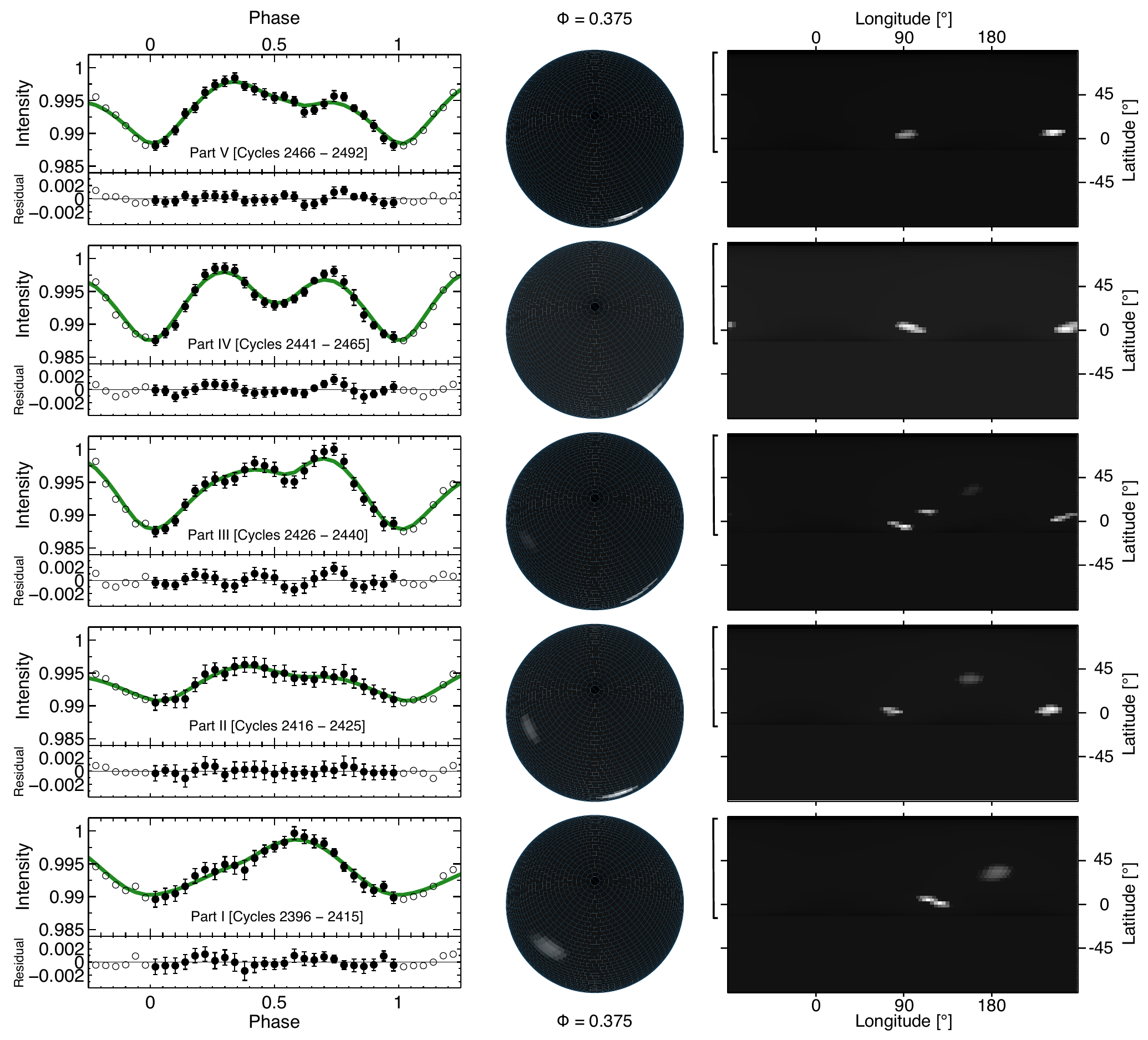}
 %\vspace{-0.5cm}
 \caption{Same as in Figure~\ref{fig:Naos_BRITE1415_LI_maps_inc24} but for $i=i_{\rm min}=15\degr$. The locations of the detected spots remain the same as for $i=24\degr$ (Figure~\ref{fig:Naos_BRITE1415_LI_maps_inc24}), while their shapes appear to be more stretched in longitude.}
  \label{fig:Naos_BRITE1415_LI_maps_inc15}
\end{figure*}

\begin{figure*}
 \includegraphics[width=18cm]{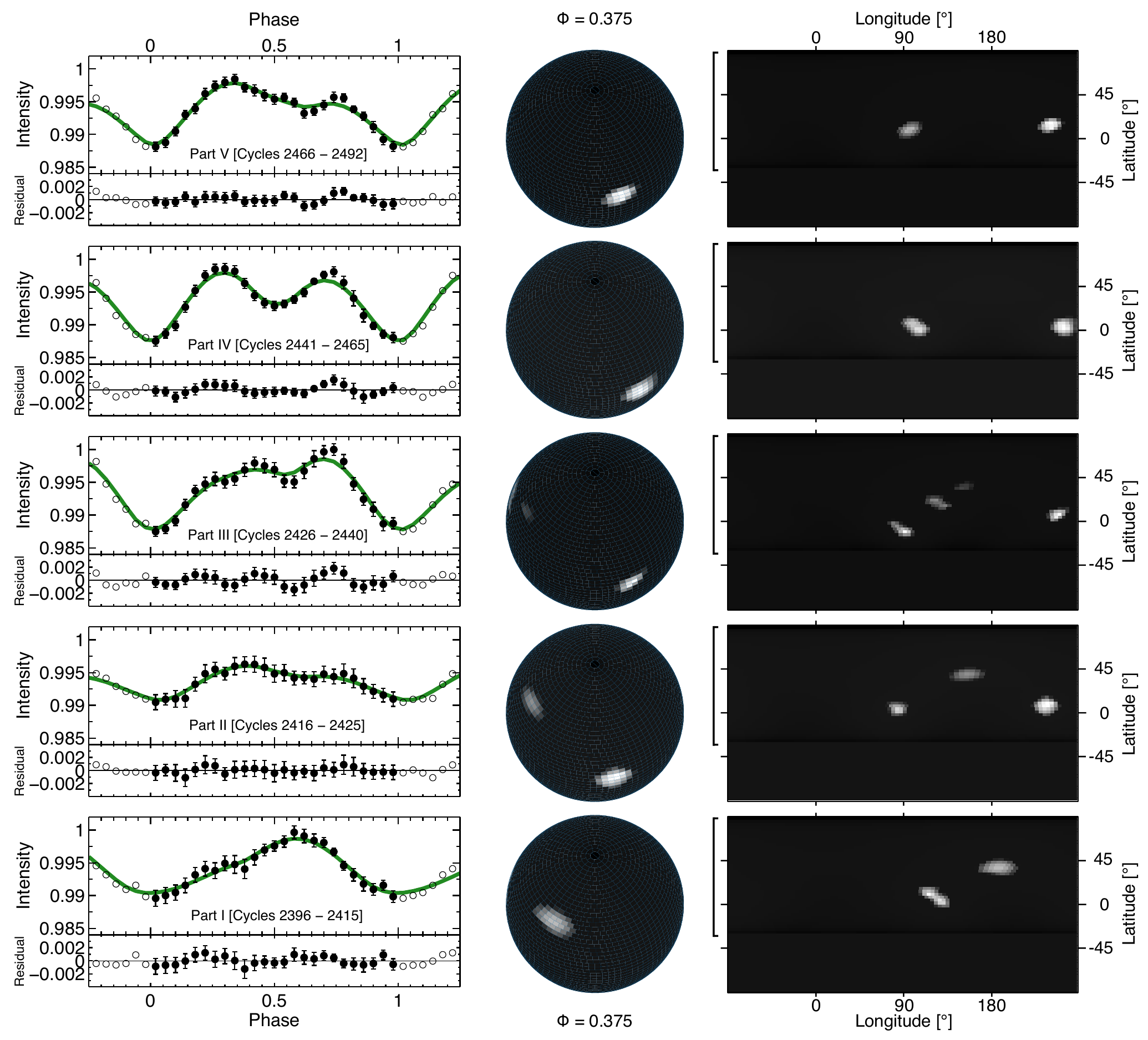}
%\vspace{-0.5cm}
 \caption{Same as in Figure~\ref{fig:Naos_BRITE1415_LI_maps_inc24} but for $i=i_{\rm max}=33\degr$. The locations of the detected spots remain the same as for $i=24\degr$ (Figure~\ref{fig:Naos_BRITE1415_LI_maps_inc24}), while their shape appear to be rounder.}
  \label{fig:Naos_BRITE1415_LI_maps_inc33}
\end{figure*}

%%%%%%%%%%%%%%%%%%%%%%%%%%%%%%%%%%%%%%%%%%%%%%%%%%%%%%%%%
\paragraph{Mapping the surface of $\zeta$~Pup during the \emph{BRITE} observing run\\\\}
\label{paragraph:Naos_LI_BRITE20142015}

Under all the considerations in \S~\ref{paragraph:Naos_LI_algorithm} and \S~\ref{paragraph:Naos_Tspot}, we performed inversions of the \emph{BRITE} light curves of $\zeta$~Pup by partitioning the stellar surface into $N=60$ latitude bands with the first four equatorial bands containing $90$ patches, and we adopt as input parameters: 

\begin{enumerate}[labelindent=4.0pt,leftmargin=*]
\renewcommand\labelenumi{\textbf{[\roman{enumi}]}}
\item the estimated rms noise $\sigma$ for the obverved light curves split into different parts (Section~ \ref{subsubsec:Naos_1.78dsig_stability}). As mentioned earlier, the goal of LI is to find the ``smoothest'' solution until the rms residual between the reconstructed and observed light curves reaches the level of $\sigma$. It has been demonstrated that in practice \citep{2000AJ....120.3274H,2011AJ....141..138R,2013ApJ...767...60R,2016ApJ...832..207R} the reconstructed surface will show obvious noise artifacts when the assumed noise level becomes too low. Our criterion for choosing the best solution is thus defined by the effective noise level at which this behaviour starts to occur. Therefore, we only assessed the rms noise in our observed light curves in order to choose the appropriate range of assumed values of $\sigma$ that need to be considered. We measured $\sigma$ in the range $1.2-0.7$~mmag in the phased light curves of $\zeta$~Pup observed by \emph{BRITE}, hence we scanned the following range of $\sigma$ in our inversions: $1.5$, $1.2$, $0.9$, $0.8$, $0.7$, $0.6$, $0.5$ and $0.4$~mmag.
\item the inclination angle: we performed inversions for $i=24^\circ$, which is the mean value defined by the limits that we found in section~\ref{subsubsec:Naos_pulsations_vs_rotational_modulation} (Table~\ref{tab:Naos_Protmin_Vc_Ve_i}), and we also performed inversions for the lower limit $i_{\rm min}=15^\circ$ and the upper limit $i_{\rm max}=33^\circ$.  
\item the photosphere temperature (which we take to be $T_{\rm phot}=T_{\rm eff} = 40.0$~kK: Table~\ref{tab:Naos_StellarParams}), and a range of discrete values for the spot temperature: $T^{\rm (s)}$ = $41.5$~kK, $42.5$~kK, $45$~kK, $47.5$~kK and $50$~kK.
\item limb-darkening coefficients: we adopted a quadratic limb-darkening law and extracted the appropriate coefficients for the stellar parameters of $\zeta$~Pup in the Bessel V filter (the closest to the middle of BRITE combined filters) from the grid of limb darkening coefficients calculated by \citet{2016MNRAS.456.1294R} using non-LTE, line-blanketed TLUSTY model atmospheres. 
\end{enumerate}

\begin{figure*}
\includegraphics[width=18cm]{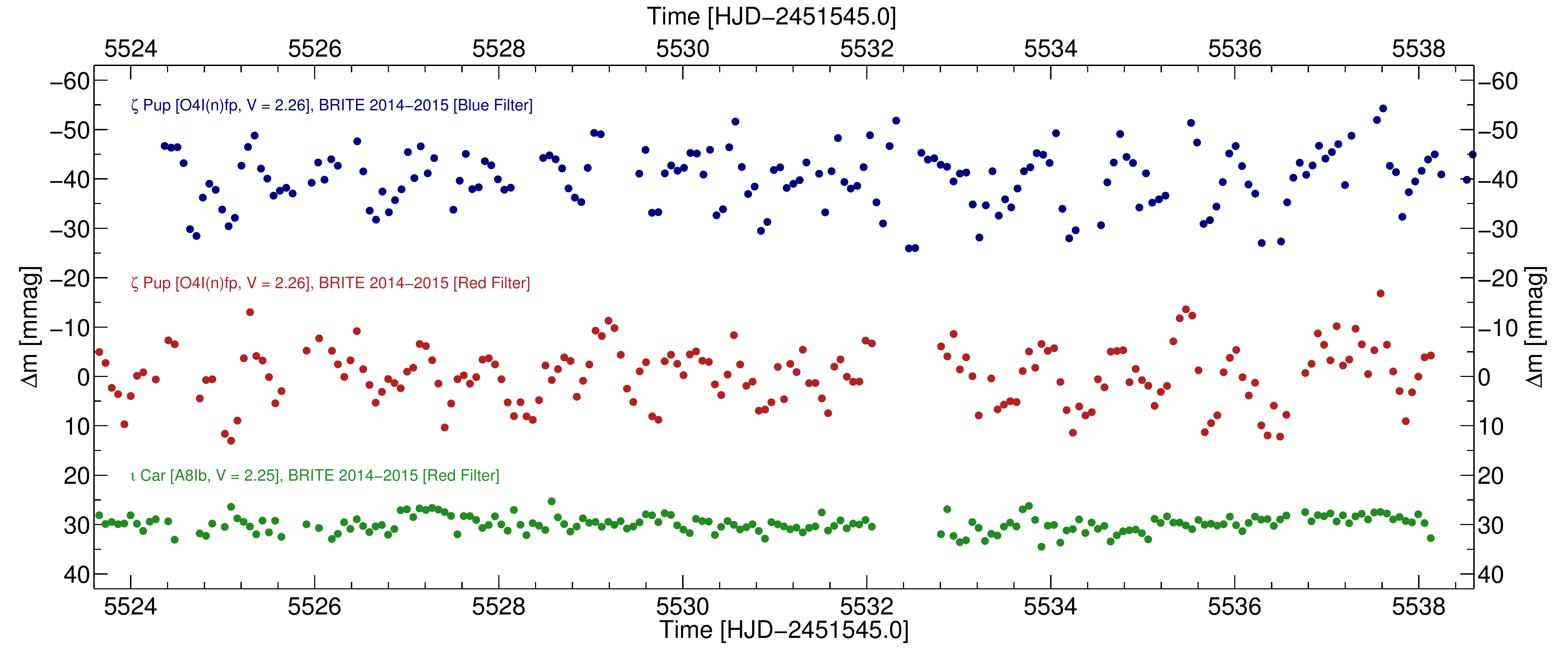}
 %\vspace{-0.5cm}
 \caption{A $15$~day subset of the residual \emph{BRITE} (2014-2015) light curve of $\zeta$~Pup after removal of the $1.78$-d periodicity related to rotational modulation, along with the simultaneously observed \emph{BRITE} (2014-2015) red light curve of $\iota$~Car (HD 80404, A8Ib, V=2.25) used as a comparison star.}
  \label{fig:Naos_BRITE_residuals}
\end{figure*}

The first important point that we noticed was that the inversions for the assumed range of spot temperatures only slightly differ in the resulting spot-to-photosphere contrast: higher spot temperatures resulted in slightly brighter spots (which is expected for the reasonable values of spot temperature that we considered), but negligibly affected the locations and the shapes of the spots found by the algorithm. Figures~\ref{fig:Naos_BRITE1415_LI_maps_inc24},~\ref{fig:Naos_BRITE1415_LI_maps_inc15} and \ref{fig:Naos_BRITE1415_LI_maps_inc33} illustrate the outcome of our inversions of the \emph{BRITE} light curve of $\zeta$~Pup for $T^{\rm (s)}=42.5$~kK and for $i=24\degr$, $i=i_{\rm min}=15\degr$  and $i=i_{\rm max}=33\degr$ respectively. We clearly note that the range within which the stellar inclination angle varies does not influence the locations of the detected spots on the stellar surface. However, we notice that, as the inclination angle gets lower, the shapes of the spots tend to be more stretched in longitude. In other words, the spots are rounder for $i=33\degr$ than for $i=15\degr$. The equatorial spots appear to be the most affected by this behavior, which could then be explained as follows. Given a certain value of the stellar inclination angle, consider an equatorial spot having a certain shape that provides the right amount of light modulation at a given rotational phase: if the stellar inclination is decreased, the lower part of the spot would become less visible (its overall projected area onto the plane of the sky decreases), such that the spot has no other choice than to flatten in order to compensate for that behavior and provide the same amount of light modulation as in the case of the higher inclination angle. Also, in general it is expected that there should be a tendency for the spots to shift to higher latitudes as the assumed inclination angle decreases, since the sub-Earth latitude moves northward in that case. However, for \emph{small} equatorial spots, which is the case for most of the spots detected by the algorithm here, that tendency will not be as pronounced because a small equatorial spot will be visible for half the rotation period regardless of the value of the inclination angle. This explains why the range within which the stellar inclination angle varies here does not affect the locations of the spots detected by the algorithm. In all cases, our inversions of the \emph{BRITE} light curve of $\zeta$~Pup converge towards the following behaviour: 

\begin{enumerate}[labelindent=4.0pt,leftmargin=*]
\renewcommand\labelenumi{\textbf{[\roman{enumi}]}}
\item Part~I: one spot $\mathcal{S}_1$ is detected lying close to the equator, along with a slightly larger spot $\mathcal{S}_2$ located at mid-latitude ($\sim55.5\degr$). Their separation is $\sim66\degr$ in longitude.
\item Part~II: the two spots $\mathcal{S}_1$ and $\mathcal{S}_2$ seem to be subject to a shift in longitude, while a third spot ($\mathcal{S}_3$) appears close to the equator, separated from the previous equatorial spot $\mathcal{S}_1$ by $\sim158\degr$ in longitude.
\item Part~III: spot $\mathcal{S}_2$ has almost faded away, while the equatorial spots $\mathcal{S}_1$ and $\mathcal{S}_3$ remain there. A fourth spot $\mathcal{S}_4$ seems to be located mid-way between $\mathcal{S}_1$ and the previous location of $\mathcal{S}_2$. We identify three possible scenarios to explain the presence of $\mathcal{S}_4$: either it is exactly $\mathcal{S}_2$ that is migrating towards $\mathcal{S}_1$, or part of the disintegration of $\mathcal{S}_2$ migrating towards $\mathcal{S}_1$, or a completely new spot. 
\item Part~IV: only two dominant equatorial spots, $\mathcal{S}_1$ and $\mathcal{S}_3$, are visible separated by $\sim158\degr$ in longitude.
\item Part~V: the two dominant equatorial spots at $\mathcal{S}_1$ and $\mathcal{S}_3$ remain visible. Both spots seem to have slightly shifted in longitude (by $\sim6\degr$ for $\mathcal{S}_1$ and $\sim16\degr$ for $\mathcal{S}_3$), and their separation becomes $\sim148\degr$ in longitude. 
\end{enumerate}

Thus, our inversion of the \emph{BRITE} light curve of $\zeta$~Pup has enabled us to map the surface of the star, find the locations of the surface bright spots that best explain the observed variations, and track the evolution of these surface inhomogeneities. We also performed inversions of the light curves of $\zeta$~Pup obtained during the three \emph{Coriolis}/SMEI seasonal runs. The resulting surface maps for $T^{\rm (s)}=42.5$~kK and $i=24\degr$ are illustrated on Figures~\ref{fig:Naos_SMEI0506_LI_maps},~\ref{fig:Naos_SMEI0405_LI_maps}~and~\ref{fig:Naos_SMEI0304_LI_maps}. We detect on average $\sim2-3$ spots at the equator and at mid-latitudes during the epoch of the \emph{Coriolis}/SMEI observations. The spots clearly appear at different longitudes compared to the \emph{BRITE} observing run, but whenever there are two equatorial spots, their separation seems to be $\sim148\degr -180\degr$ in longitude.   

%%%%%%%%%%%%%%%%%%%%%%%%%%%%%%%%%%%%%%%%%%%%%%%%%%%%%%%%%
\subsection{Stochastic variability}
\label{subsec:Naos_stochastic_variability}

As mentioned in Section~\ref{subsec:Naos_Results_Photo_Fourier}, the sole significant period that emerged from our Fourier analysis of the \emph{BRITE} and \emph{Coriolis}/SMEI light curves of $\zeta$~Pup was the $1.78$~d signal due to rotational modulation (with the first harmonic of the fundamental period also prominent in the amplitude spectrum of the \emph{BRITE} light curves). In order to check for any signs of other types of variability in the light curves, we therefore removed this $1.78$~d periodic signal. Since the shape of this signal changes in time, the standard removal by prewhitening is the least reliable method as it assumes a constant amplitude and phase. Instead of calculating the residual light curve by prewhitening the $1.78$~d signal, we performed its removal in the phase domain by subtracting templates of the phased light curves from the different parts of the observing runs: in the case of the \emph{BRITE} observations, we adopted the five different parts of the observing run (Part~I...V; left panel of Figure~\ref{fig:Naos_BRITE_lcs_dynamic}), while in the case of the \emph{Coriolis}/SMEI observations we adopted the different parts defined on the left panels of Figures~\ref{fig:Naos_SMEI20052006_lcs_dynamic},~\ref{fig:Naos_SMEI20042005_lcs_dynamic},~and~\ref{fig:Naos_SMEI20032004_lcs_dynamic}. Then we recomputed the amplitude spectra of the residual light curves to confirm that there is no sign of the $1.78$~d signal left and no new significant periodicity introduced during the removal process, thus confirming our previous findings in Section~\ref{subsec:Naos_Results_Photo_Fourier}. Now a close inspection of the residual \emph{BRITE} light curves in the two filters reveals variations reaching $\sim20$~mmag in peak-to-peak amplitudes, stochastically generated, but coherent for several hours (Figure~\ref{fig:Naos_BRITE_residuals}). The first hint indicating that this stochastic signal is intrinsic to the star is the fact that it behaves the same way in the observations in the two \emph{BRITE} filters as clearly seen in Figure~\ref{fig:Naos_BRITE_residuals}. This is further confirmed by the behaviour of the \emph{BRITE-b} vs \emph{BRITE-r} diagram for the residual light curves in which we measure a strong correlation ($\rho=0.856$) and a slope $\alpha=1.10\pm0.04$ (right panel of Figure~\ref{fig:Naos_BRITE_lcs_full_BvsR}). Also as already shown in Figure~\ref{fig:Naos_BRITE_residuals}, our ultimate check that confirms that this stochastic signal is not of instrumental origin nor due to an artifact introduced during the decorrelation process is a comparison with the \emph{BRITE} light curve of $\iota$~Car (HD 80404, A8Ib, V=2.25), which is of similar magnitude as $\zeta$~Pup, was also observed by \emph{BRITE} as part of the Vela/Pup observing run, has data that were decorrelated with the same decorrelation routines, and shows no obvious signs of periodic variability from our Fourier analysis. From all these considerations we conclude that this stochastic signal is intrinsic to the star. We emphasize here the fact that the stochastic nature of this signal describes the way each feature in the signal appears to be generated, while a given feature seems to be well organized (or coherent) for several hours after its generation. Also from these considerations, we can take the liberty to call this signal ``noise'', as long as we keep in mind that it is a noise intrinsic to the star and not of instrumental origin. It is thus worth noting that this stochastic behaviour is reminiscent of the findings of \citet{1992MNRAS.254..404B} who came to the conclusion that $\zeta$~Pup is an irregular microvariable (with amplitudes $\sim20$~mmag) from ground-based Str\"{o}mgren $b$ filter photometric monitoring of the star.

\begin{figure*}
\includegraphics[width=18cm]{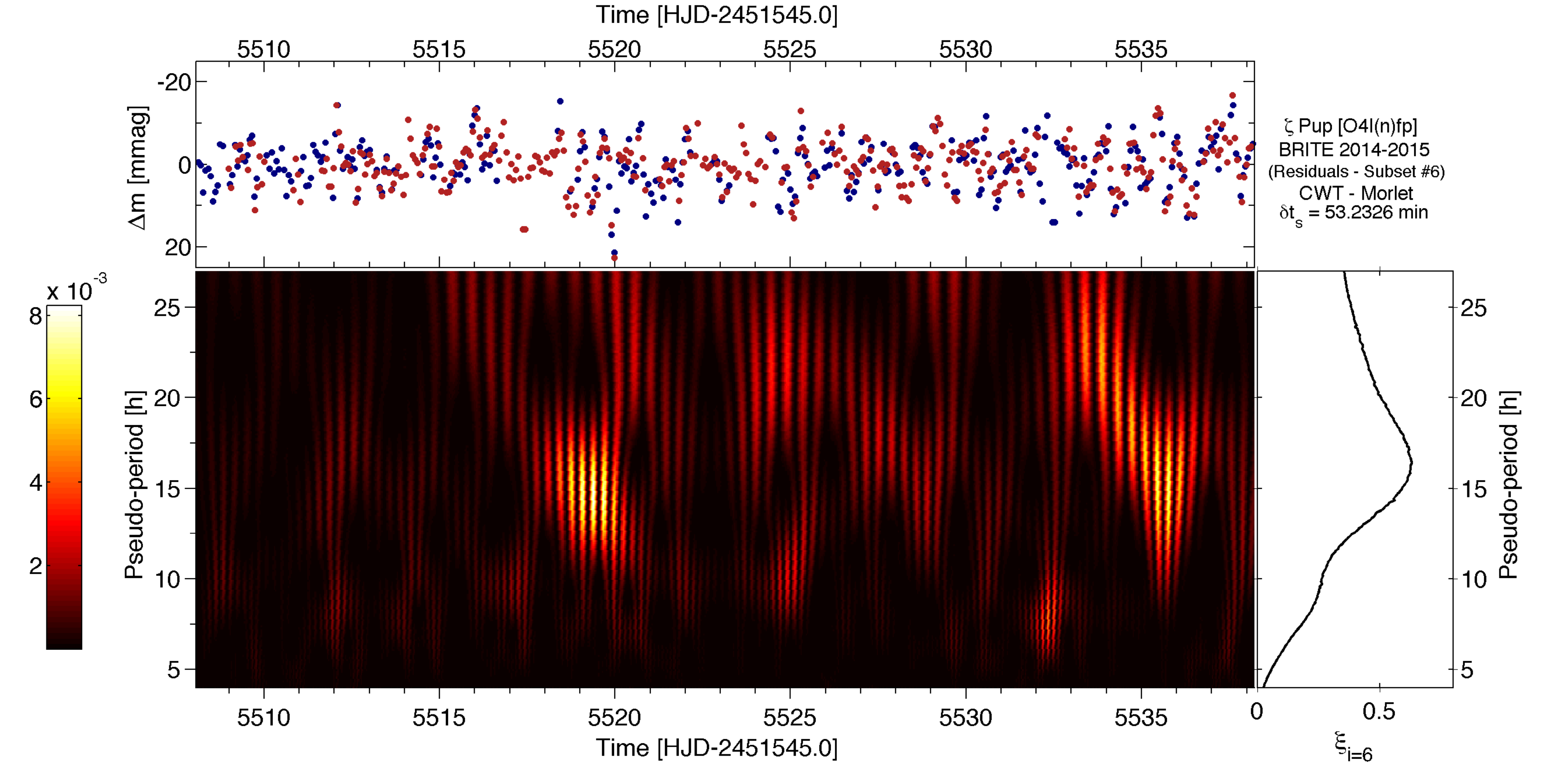}
 %\vspace{-0.5cm}
 \caption{Wavelet analysis of a $30$-day subset of the combined residual \emph{BRITE} light curve of $\zeta$~Pup. The \emph{main panel} depicts the scalogram (percentage of energy for each wavelet coefficient) obtained from a Morlet wavelet family-based continuous wavelet transform (CWT) of the analyzed $30$-day portion of light curve illustrated on the \emph{top panel}. The \emph{right panel} shows the sum of the percentages of energy per wavelet coefficient over time, thus representing the overall energy distribution of pseudo-periods detected during this subset of the observations.}
  \label{fig:Naos_BRITE_residuals_CWT}
\end{figure*}

The origin of this stochastic signal remains unknown, with potential candidates being unresolved randomly excited p-mode oscillations similar to those discovered in the O-type primary component of the massive binary system HD 46149 \citep{2010A&A...519A..38D}, or internal gravity waves (IGWs) generated from either a subsurface convection zone  \citep{2009A&A...499..279C} or from the convective core \citep{2015ApJ...806L..33A}, the latter case having been proposed to be to date the best interpretation of the red noise components in the \emph{CoRoT} observations of the three O-type dwarfs HD 46223, HD 46150 and HD 46966 in the young open cluster NGC 2244 \citep{2011A&A...533A...4B}. Also, perhaps the set of frequencies $\{\nu_{1};\nu_{2}\nu_{3};\nu_{4}\}$ ($P\sim13.5-21$~h) detected by our Fourier analysis in both the observations in the red and blue filters but with low significance ($S/N\sim2.5-3.5$; Section~\ref{subsec:Naos_Results_Photo_Fourier}, Table~\ref{tab:Naos_SMEI_BRITE_1.8dPeriod}) is part of this stochastic signal, which would explain their low significance as they would be present/excited only occasionally and last for a limited time. In all cases, regardless of its origin, in view of its typical amplitudes the quantification of this stochastic variability in terms of its amplitudes and timescales is as important as the characterization of the $1.78$~d periodic rotationally-induced modulation itself.

\subsubsection{Stochastic variability - Amplitudes}
\label{subsubsec:Naos_stochastic_variability_amplitudes}

One striking property of this stochastic variability in $\zeta$~Pup is that it reaches $\sim20$~mmag in peak-to-peak amplitude during the epoch of the \emph{BRITE} observations (Figure~\ref{fig:Naos_BRITE_residuals} and right panel of Figure~\ref{fig:Naos_BRITE_lcs_full_BvsR}), while we should recall that the periodic $1.78$~d signal does not exceed $\sim15$~mmag peak-to-peak (Figure~\ref{fig:Naos_BRITE_DFT_lc178d_BvsR}). The overall standard deviation in the amplitudes of the residual \emph{BRITE} light curves of $\zeta$~Pup is $\sigma_{\textnormal{\scriptsize \textsc{brite}}}\simeq6$~mmag. But as assessed in Section~\ref{subsec:Naos_Results_Photo_Amp}, the scatter of instrumental origin is $\sigma_{\textnormal{\scriptsize \textsc{brite}, \tiny i}}=1.60\pm0.04$~mmag. Since that instrumental scatter is much smaller than the overall scatter in the residual light curve, we can infer that the standard deviation of the amplitudes associated with the stochastic signal intrinsic to the star during the epoch of the \emph{BRITE} observations is of the order of $6$~mmag.

Regarding the \emph{Coriolis}/SMEI observations of $\zeta$~Pup, the overall standard deviation in the amplitudes of the residual \emph{Coriolis}/SMEI light curve of the star is higher: $\sigma_{\textnormal{\scriptsize \textsc{smei}}}\simeq16.8$~mmag. However the contribution of instrumental effects to this scatter remains ambiguous. Indeed, in view of the \emph{Coriolis}/SMEI observations of the late O-type subgiant $\zeta$~Oph (O9.2IVnn; $V=2.6$) having an rms dispersion of $\sim20$~mmag dominated by instrumental effects while contemporaneous \emph{MOST} observations of the star has a dispersion of $\sim6$~mmag, most of which is intrinsic to the star \citep{2014MNRAS.440.1674H}, it is impossible to constrain the nature of the stochastic component of the intrinsic variability of $\zeta$~Pup during the epoch of the \emph{Coriolis}/SMEI observations.

\subsubsection{Stochastic variability - Timescales}
\label{subsubsec:Naos_stochastic_variability_timescales}

As mentioned earlier, the features constituting the intrinsic stochastic signal in $\zeta$~Pup seem to be coherent for several hours. In order to characterize the timescales of these transient features, given the random nature of their generation we performed a continuous wavelet analysis of the residual combined \emph{BRITE} light curves. We tested two cases:~1)~application of the $1.10$ factor on the amplitudes of the observations in the red filter prior to the generation of the combined light curves, and~2)~no amplitude correction prior to the combination of the light curves. We noticed no significant difference in the resulting continuous wavelet transforms, thus we chose to work on the combined light curves without any amplitude correction. Also it is important to note that in general, the continuous formalism can only be applied on well-sampled and equally spaced data. Our overall combined \emph{BRITE} light curve of $\zeta$~Pup is well-sampled enough, but obviously not regularly sampled. However, by choosing appropriate subsets of the combined light curve that are densely sampled and resampling these subsets with one single sampling interval, it is reasonable to adopt the continuous formalism. Therefore our approach is to subdivide the combined light curve into subsets that contain no gaps greater than $5$~h ($39$ subsets), then refining the selection by keeping only subsets spanning at least $2.5$~d, so that we ended up with $11$ subsets with time bases between $\sim3-52$~d and mean sampling intervals in the range $53.2-75.4$~min. Lastly, through a shape-preserving piecewise cubic interpolation \citep{Sprague:1990:SPC:918286} we resampled all the $11$ subsets with one single sampling interval taken to be the minimum of the mean sampling intervals of the subsets ($\delta t_s = 53.2326$~min). Once the subsets are defined, we come to the choice of an appropriate wavelet family that will help us optimally and robustly characterize the timescales of the transient features composing the stochastic signal intrinsic to $\zeta$~Pup. Given the facts that the observed variations in the light curve seem to equally have bumps and dips between $\pm10$~mmag (Figures~\ref{fig:Naos_BRITE_residuals}~and~\ref{fig:Naos_BRITE_lcs_full_BvsR}), and also since we want to have the most reliable conversion from wavelet scales to pseudo-periods, the wavelet family that is the most appropriate to our case is the family of real-valued Morlet wavelets, for which the mother wavelet is a sinusoid modulated by a Gaussian envelope. Furthermore, this definition itself clearly indicates that this family of wavelets is the most suitable one for the characterization of sinusoidal signals with finite lifetimes, which could be the case of the observed stochastic variations here if it is related to unresolved stochastically excited pulsations with finite lifetimes.

\begin{figure}
\includegraphics[width=8.4cm]{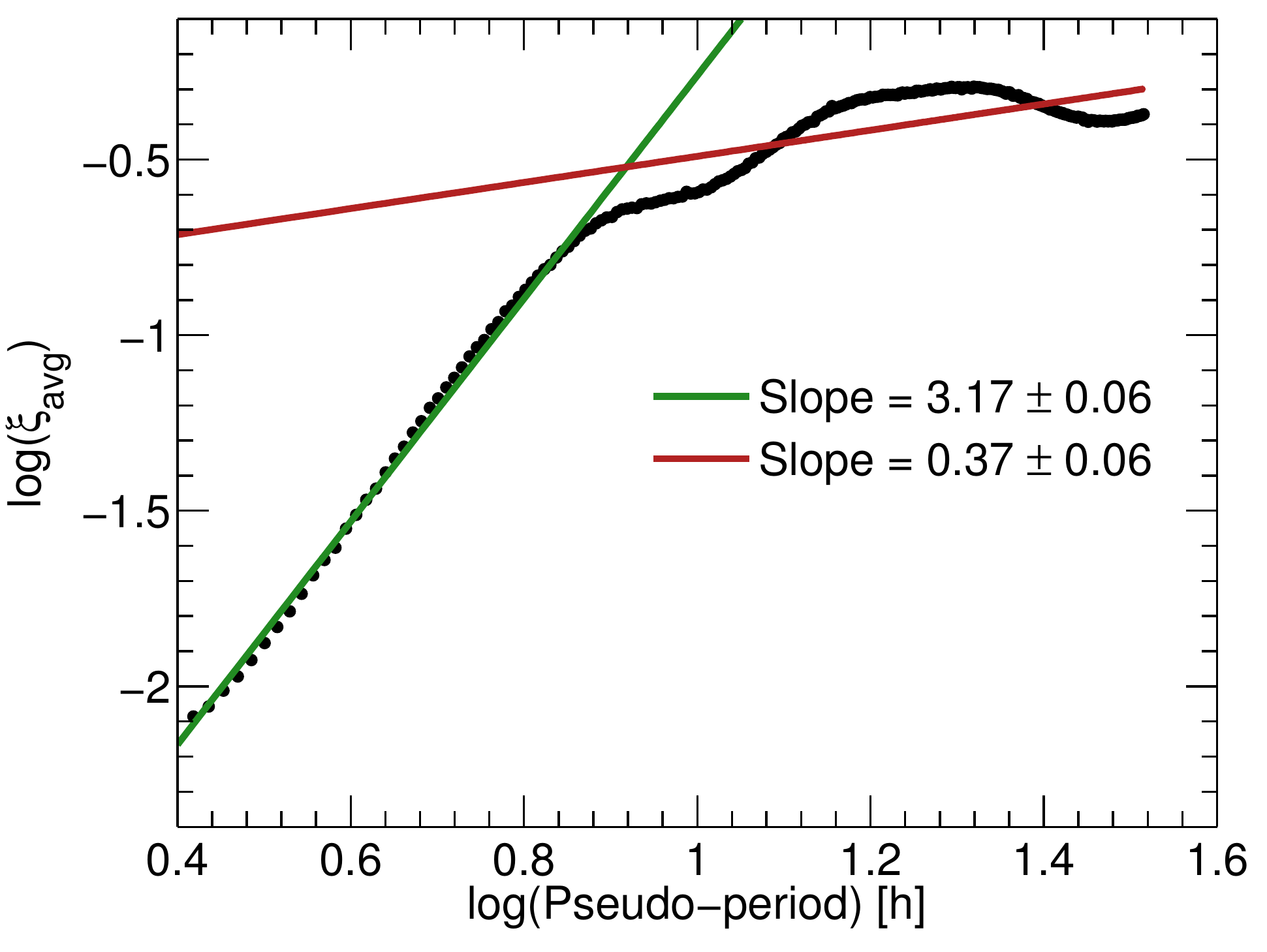}
 \caption{Weighted average energy distribution of pseudo-periods resulting from the CWT of the $11$ subsets in the combined \emph{BRITE} light curve of $\zeta$~Pup (taking as weights the time bases of the subsets), showing a change of slope at a pseudo-period of $\sim8$~h. The continuous lines are linear fits for pseudo-periods less than $8$~h (Green) and pseudo-periods beyond $8$~h (Red).}
   \label{fig:Naos_BRITE_residuals_CWT_sumpow}
\end{figure}

Figure~\ref{fig:Naos_BRITE_residuals_CWT} shows an example of the outcome of our continuous wavelet transform (CWT) of a $30$~d subset of the combined residual \emph{BRITE} light curve of $\zeta$~Pup, indicating a gradual increase in power towards longer timescales, followed by a drop in power for timescales longer than $\sim17$~h. The bump at a pseudo-period of $\sim16.5$~h might correspond to the $16.67\pm0.81$~h and $16.90 \pm 0.48$~h periods reported by \citet{1996A&A...306..899B} from an $11$~d simultaneous X-ray and H$\alpha$ monitoring of $\zeta$~Pup. These ``periods'' have never been found again since then, and \citet{2013ApJ...763..143N} invoked the possible transient nature of the signal associated with them. In our present investigations, the scalogram on Figure~\ref{fig:Naos_BRITE_residuals_CWT} clearly shows that the $\sim16.5$~h pseudo-period is only sporadically present during the \emph{BRITE} observing run, most prominently around time $\sim 5519.5$ and $\sim 5535.0$, and lasts about $\sim2.5-4$~d when it is excited. Also given the width of the bump at the $\sim16.5$~h pseudo-period, it is not relevant to assign an exact value for it, as for instance a Gaussian fit to the bump yields $16.42\pm9.5$~h. In view of the scalogram on Figure~\ref{fig:Naos_BRITE_residuals_CWT}, we notice a pseudo-period drift from long timescales ($\sim22$~h) to shorter timescales ($\sim12$~h), mostly visible between times $\sim 5517.5 - 5521.0$ during which there is a pseudo-period drift from $\sim16$~h to $\sim11$~h, and between times $\sim 5533.0 - 5536.0$ during which pseudo-periods drift from $\sim22$~h to $\sim16$~h, a behaviour that remains unexplained but could be the cause for the large spread of power around the bump at the $\sim16.5$~h pseudo-period. Finally, we also note a minor pseudo-period bump at $\sim8$~h, which also seems to be only intermittently excited during the observing run, but has a shorter lifetime ($\lesssim 1$~d) than the $\sim16.5$~h pseudo-period. From all these considerations, it could also be that these two bumps are just part of the spectrum of pseudo-periods constituting the stochastic variability of the star. Therefore, from the CWT of all the $11$ subsets that we analyzed, we calculated the weighted mean energy distribution of the timescales (with the lengths of the subsets as weights), best fit with a power-law of index $3.17\pm0.06$ for timescales less than $8$~h while timescales greater than $8$~h follow a less steep power-law of index $0.37\pm0.06$ (Figure~\ref{fig:Naos_BRITE_residuals_CWT_sumpow}).

%%%%%%%%%%%%%%%%%%%%%%%%%%%%%%%%%%%%%%%%%%%%%%%%%%%%%%%%%
\section{Spectroscopic variability}
\label{sec:Naos_Results_Spectro}

\begin{figure}
\includegraphics[width=8.4cm]{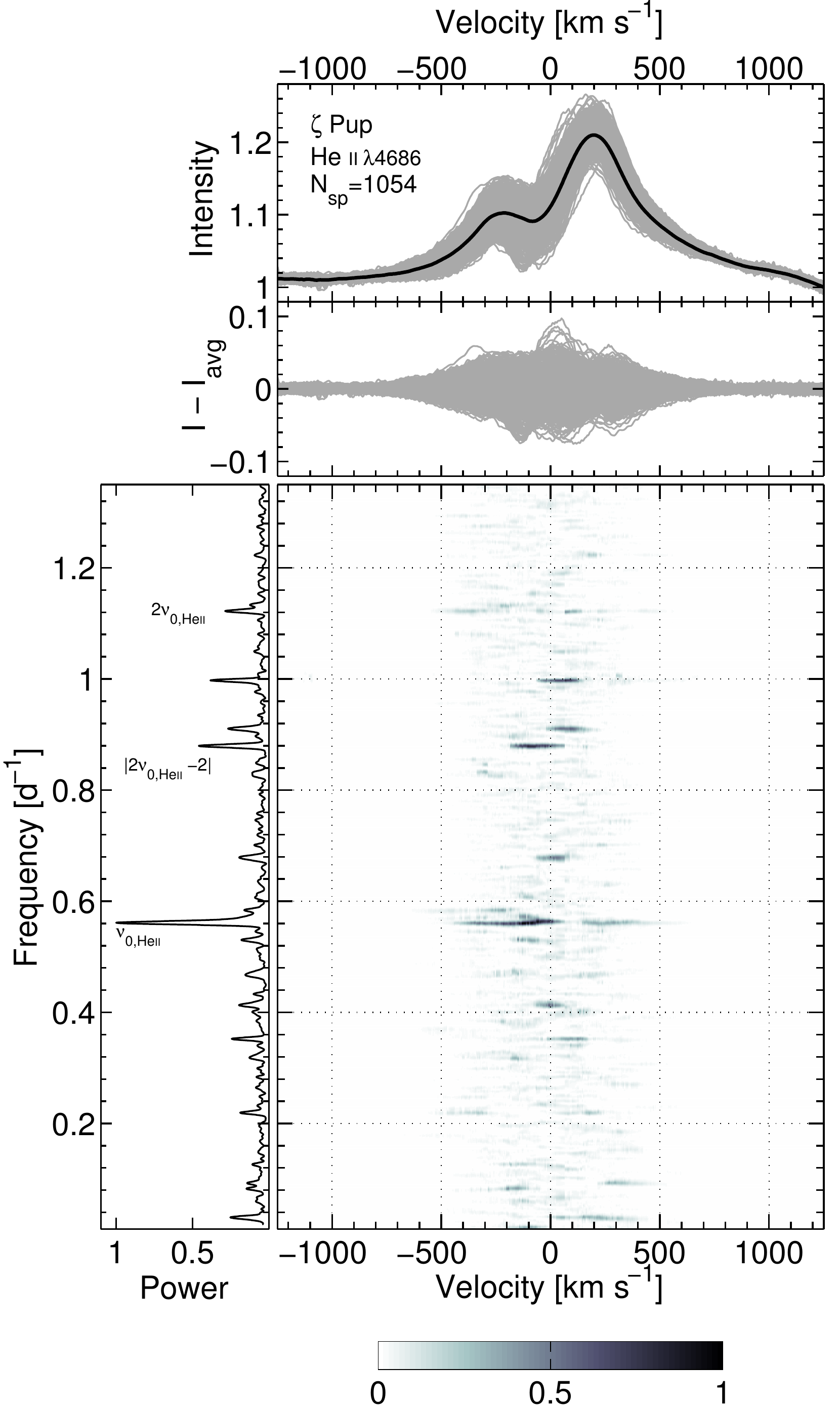}
 %\vspace{-0.5cm}
 \caption{Global time series analysis of the He~{\sc ii}~$\lambda4686$ wind emission line profile of $\zeta$~Pup. \emph{Top:} The $1054$ profiles of the line as recorded throughout the entire campaign (grey) along with their average (black). \emph{Middle:} Difference profiles with respect to the average profile. \emph{Main:} CLEAN periodogram per Doppler velocity within $\pm1275$~km~s$^{-1}$ for the time series of difference spectra, obtained after $10$~CLEAN iterations. The side panel depicts the normalized integrated periodogram power within Doppler velocities $\pm500$~km~s$^{-1}$ where most of the line profile variations are detected.}
  \label{fig:Naos_HeII4686_CIRs_DynSpec}
\end{figure}   

\begin{figure*}
\includegraphics[width=18cm]{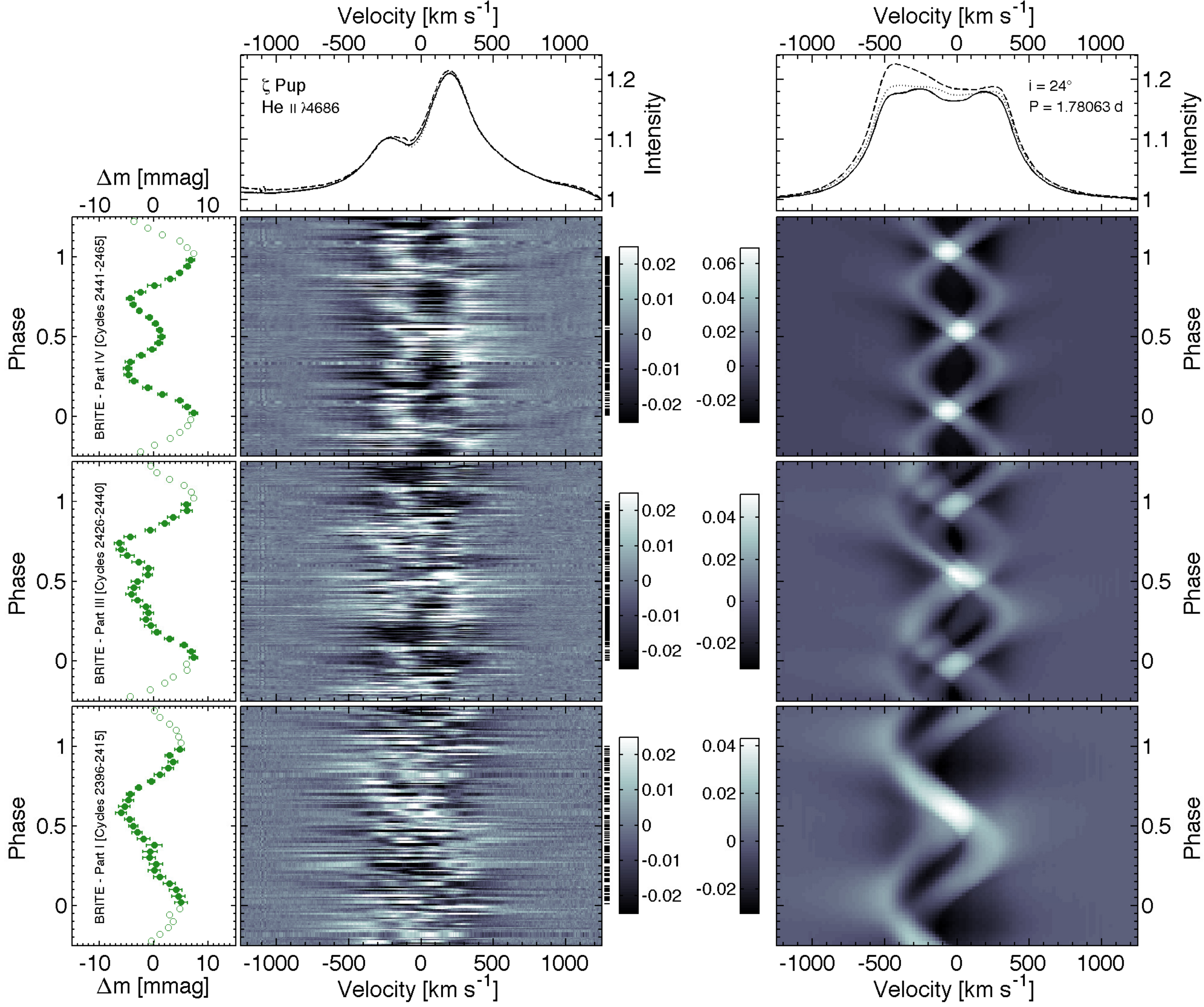}
 %\vspace{-0.5cm}
 \caption{Arms of Corotating Interaction Regions in the wind of $\zeta$~Pup as traced in its He~{\sc ii}~$\lambda4686$ emission line profile in parallel with the \emph{BRITE} 2014-2015 observing run. Time increases upwards. \emph{Left:} Observed dynamic difference spectra of $\zeta$~Pup phase-folded on the rotation period $P = 1.78063(25)$~d, during Part~I (Bottom), Part~III (Second from the bottom), and Part~IV (Third from the bottom), along with the corresponding light variations observed by \emph{BRITE} illustrated on the side panels. Small horizontal ticks on the right side of each dynamic diagram indicate the phase sampling. The average profiles of the He~{\sc ii}~$\lambda4686$ line during each of these three parts of the observing campaign are depicted in the top panel (dashed: Part~I; dotted:  Part~III; solid: Part~IV). \emph{Right:} Models of CIRs corresponding to each of the three parts of the observing campaign, with an assumed stellar inclination angle $i=24\degr$ and taking the locations of the bright spots detected by the light curve inversion algorithm (Section ~\ref{subsubsec:Naos_LI}; Figure~\ref{fig:Naos_BRITE1415_LI_maps_inc24}) as input locations of the photospheric footprints of the CIRs. The top panel depicts the average modelled profiles corresponding to the three parts of the observing run (dashed: Part~I; dotted:  Part~III; solid: Part~IV), not accounting for the resonance zone effect as explained in Section~\ref{paragraph:Naos_Results_Spectro_CIRs_models_Results}.}
  \label{fig:Naos_HeII4686_CIRs}
\end{figure*}

%%%%%%%%%%%%%%%%%%%%%%%%%%%%%%%%%%%%%%%%%%%%%%%%%%%%%%%%%
\subsection{Large-scale wind structures}
\label{subsec:Naos_Results_Spectro_CIRs}

%%%%%%%%%%%%%%%%%%%%%%%%%%%%%%%%%%%%%%%%%%%%%%%%%%%%%%%%%
\subsubsection{Searching for periodicity in the He~{\sc ii}~$\lambda4686$ line}
\label{subsubsec:Naos_Results_Spectro_CIRs_dynamic_spec}

As a first step in our investigations of the variability in the He~{\sc ii}~$\lambda4686$ wind emission line of $\zeta$~Pup, we searched for periodicity in the global time series of difference profiles of the line (with respect to the global average profile) by performing a CLEAN Fourier analysis \citep{1987AJ.....93..968R} for each Doppler velocity in the range $\pm1275$~km~s$^{-1}$ (Figure~\ref{fig:Naos_HeII4686_CIRs_DynSpec}). The resulting power spectrum unveils a dominant peak at $\nu_{0, \textrm{\scriptsize He{\sc ii}}} = 0.5615(1)$~d$^{-1}$ [$P_{0, \textrm{\scriptsize He{\sc ii}}} = 1.7809(3)$~d], for which we assessed the uncertainty by extracting the HWHM of a Gaussian fit to the peak. This frequency peak, detected across $\pm500$~km~s$^{-1}$ in Doppler velocity and mostly prominent on the blue side of the line profile, corresponds to the rotation frequency of the star that was also detected in the \emph{BRITE} and \emph{Coriolis}/SMEI observations (Section~\ref{subsec:Naos_1.78dperiod}; Table~\ref{tab:Naos_SMEI_BRITE_1.8dPeriod}; Figures~\ref{fig:Naos_BRITE_DFT_lc178d_BvsR}~and~\ref{fig:Naos_BRITE_lcs_dynamic}). Moreover, the first harmonic $2\nu_{0, \textrm{\scriptsize He{\sc ii}}}$ is visible in the power spectrum but with a lower strength. Also some of the aliases of these two peaks are still visible in the power spectrum, such as the peak at  $\mid2\nu_{0, \textrm{\scriptsize He{\sc ii}}}-2\mid\sim0.877$~d$^{-1}$ as well as the $1$~d$^{-1}$ alias peak, all due to the unavoidable sampling gaps from the ground-based observations. Apart from the rotation period and its first harmonic we do not detect any other significant periodicity in the He~{\sc ii}~$\lambda4686$ line during our campaign. The absence of any sign of the $5.1-5.2$~d period through the entire width of the line profile is noteworthy. Furthermore, we do not detect any signs of that period in $H\alpha$ which was only covered by our \emph{CTIO-SMARTS 1.5m/CHIRON} observations. That period was previously detected in $H\alpha$ \citep{1981ApJ...251..133M} and in UV observations \citep{1995ApJ...452L..65H}. However, in view of our findings that the true stellar rotation period is the $1.78$~d period and not the $5.1-5.2$~d period (Section~\ref{subsec:Naos_1.78dperiod}), the previous interpretations of the latter needs to be revised.

In view of the detection of the stellar rotation period in the variations of the He~{\sc ii}~$\lambda4686$ line profile, we inspected the behaviour of the difference spectra phase-folded on the rotation period, adopting the value $P = 1.78063(25)$~d that we derived from the \emph{Coriolis}/SMEI and \emph{BRITE} photometric observations (Section~\ref{subsubsec:Naos_pulsations_vs_rotational_modulation}). As we know that the bright spots inducing this signal in the light curve evolve during the observing campaign (Sections~\ref{subsubsec:Naos_1.78dsig_stability}~and~\ref{subsubsec:Naos_LI}), we subdivided the spectroscopic observations according to the five different parts of the \emph{BRITE} observing run, Part~I...V, and looked at the phase-folded dynamic difference spectra for each of these parts. Due to insufficient phase sampling, the phased dynamic difference spectra during the first part of the transition (Part~II) and towards the end of the campaign (Part~V) did not show any obvious pattern that could be used to characterize the nature of the line profile variations (LPVs) related to the stellar rotation in the He~{\sc ii}~$\lambda4686$ line. The phased dynamic spectra for Part~I (before the transition), Part~III (end of the transition) and Part~IV are illustrated on the left panel of Figure~\ref{fig:Naos_HeII4686_CIRs} in which we note the following behaviour:

\begin{enumerate}[labelindent=4.0pt,leftmargin=*]
\renewcommand\labelenumi{\textbf{[\roman{enumi}]}}
\item Part~I: an ``S'' pattern in excess emission is mildly visible, reaching maximum redshift around phase $0.35$ and maximum blueshift around phase $0.85$.  
\item Part~III: the behaviour of the LPVs are fuzzy, but the clearest one is the bump on the red side of the line profile (between $0-400$~km~s$^{-1}$) before phase $\sim0.5$, which appears to be part of the ``S'' pattern seen in Part~I.    
\item Part~IV: a double ``S'' bumped pattern is strikingly visible, wandering between $\sim \pm400$~km~s$^{-1}$. Note that this corresponds to the part during which the two bumps in the \emph{BRITE} light curves are clearly visible, caused by the two dominant equatorial surface spots separated by $\sim158\degr$ in longitude.
\end{enumerate}

The double ``S'' pattern that we clearly see during Part~IV is a typical sign of the presence of two CIR arms induced by two equatorial perturbations \citep{2002A&A...395..209D}, which, according to our findings in Section~\ref{subsubsec:Naos_LI}, are the equatorial bright spots separated by $\sim158\degr$ in longitude observed by \emph{BRITE} during this part of the observing campaign (Figure~\ref{fig:Naos_BRITE1415_LI_maps_inc24}). This situation confirms the theoretical predictions of \citet{2002A&A...395..209D}, as not only did they use the stellar parameters of $\zeta$~Pup in all their CIR models, but more importantly our findings for the behaviour in Part~IV are remarkably similar to the behaviour that they found in their Model C \citep[Figure 3 in][]{2002A&A...395..209D} for which the assumed photospheric perturbations were two equatorial bright spots diametrically opposed on the stellar surface and the assumed stellar inclination angle was $20\degr$ (the actual inclination angle of $\zeta$~Pup is $i\sim24\degr$ according to our findings: Section~\ref{subsubsec:Naos_pulsations_vs_rotational_modulation}), with the assumed line emission region (LER) extending from $2R$ to $4R$.

%%%%%%%%%%%%%%%%%%%%%%%%%%%%%%%%%%%%%%%%%%%%%%%%%%%%%%%%%
\subsubsection{CIR model}
\label{subsubsec:Naos_Results_Spectro_CIRs_models}

Following the detection of the patterns of manifestations of CIR arms in the phased dynamic difference profiles of the He~{\sc ii}~$\lambda4686$ line, most prominently the double ``S'' pattern that we see during Part~IV of the observing run, we further investigated the consistency of these observed LPVs with models of CIR arms that would be generated by the bright photospheric spots observed by \emph{BRITE} and mapped by LI. To this end, we adopted an analytical approach to determine the behaviour of LPVs induced by a given number of CIR arms in a hot-star wind emission line profile. In our approach, the input parameters consist of:

\begin{enumerate}[labelindent=4.0pt,leftmargin=*]
\renewcommand\labelenumi{\textbf{[\roman{enumi}]}}
\item the inclination angle $i$ of the star, its rotation period $P$, and its radius $R$.
\item the radius $r_{\textnormal{\scriptsize \textsc{i}}}$ at which the stellar wind initiates, as well as the terminal wind speed $\varv_{\infty}$.
\item two radial functions, one for localizing the LER, and another one for accounting for ionization gradients through the wind.
\item the number of CIR arms, each of them specified by its density contrast $\eta$ with respect to the unperturbed wind, its opening angle $\psi$, and the location of its photospheric driver on the stellar surface.
\end{enumerate}

Here we provide a brief description of the basis of this CIR modelling approach and the results that we obtained in the case of $\zeta$~Pup, but a more detailed description of the code will be available in a dedicated forthcoming paper (St-Louis et al., submitted).

%%%%%%%%%%%%%%%%%%%%%%%%%%%%%%%%%%%%%%%%%%%%%%%%%%%%%%%%%
\paragraph{The emergent line profile for a spherically symmetric wind\\\\}
\label{paragraph:Naos_Results_Spectro_CIRs_models_lineform}

\begin{figure}
\includegraphics[width=8.4cm]{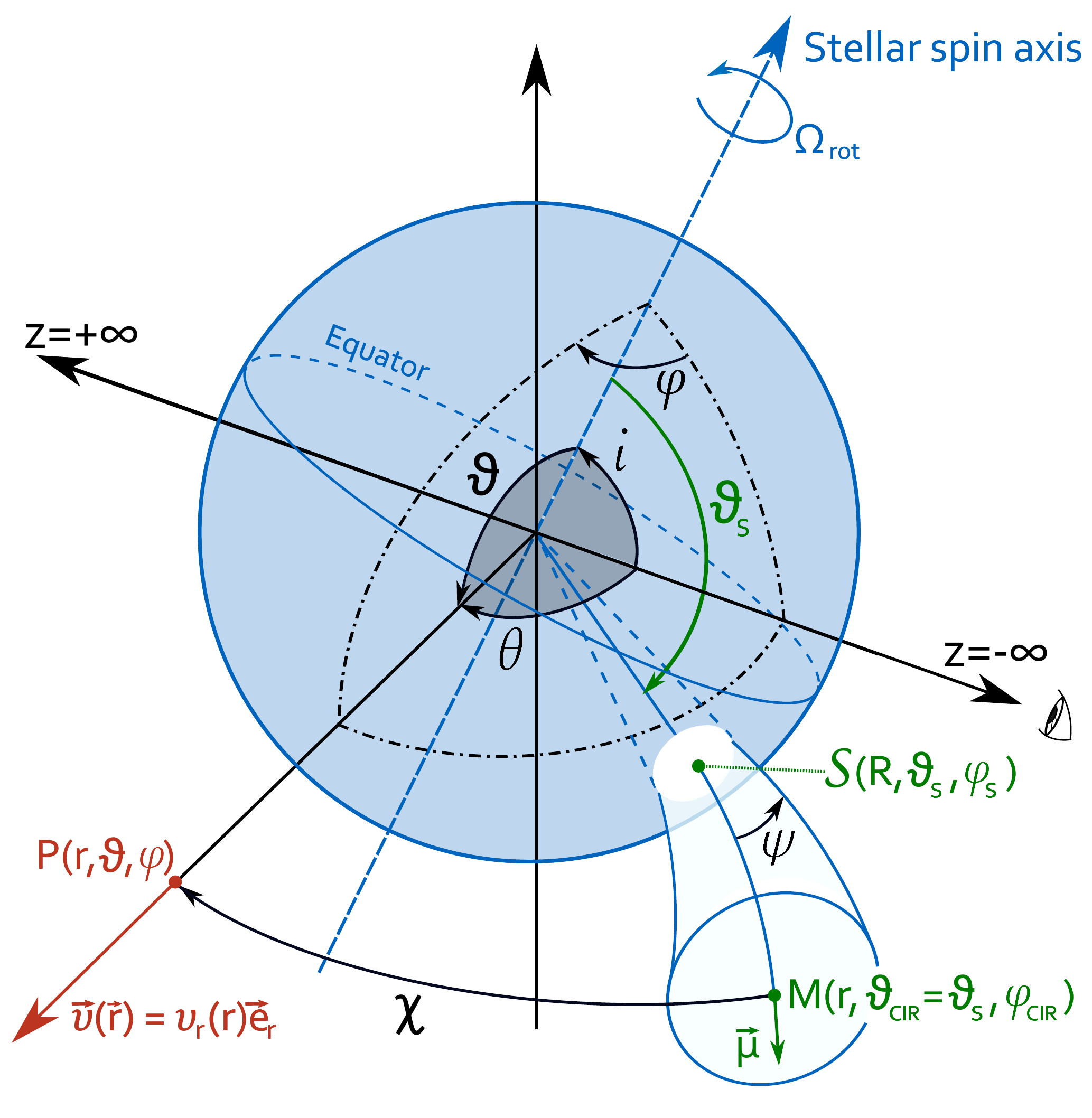}
 %\vspace{-0.5cm}
 \caption{Geometry adopted for the simulation of hot-star wind emission line profiles from a spherically symmetric smooth stationary stellar wind and calculation of LPVs induced by the presence of a spiral-shaped CIR compression.}
  \label{fig:Naos_CIRs_geometry}
\end{figure}

In the absence of CIR arms, our calculations of hot-star wind emission line profiles is performed under the assumption of spherical symmetry, stationarity, and smoothness of the outflow, the latter meaning a stellar wind free of small-scale inhomogeneities. Also in terms of the velocity field of the wind material, the polar and azimuthal components are assumed to be negligible compared to the radial expansion. Thus, with the geometry depicted in Figure~\ref{fig:Naos_CIRs_geometry}, a portion of wind material located at point $P(r,\vartheta,\varphi)$ in terms of its spherical coordinates in the stellar reference frame moves outward with a velocity $\vec{\varv}(\vec{r})=\varv_r(r)\vec{e}_r$, assumed to follow the usual $\beta$-law:   
\begin{equation}
\varv_r(r) = \varv_{\infty}\left( 1 - b\frac{r_{\textnormal{\scriptsize \textsc{i}}}}{r} \right)^\beta,~~b=1 - \left( \frac{\varv_{\textnormal{\scriptsize \textsc{i}}}}{\varv_{\infty}} \right)^{\frac{1}{\beta}}
\label{eq:Naos_CIRs_VelocityLaw}
\end{equation}
with $\varv_{\textnormal{\scriptsize \textsc{i}}}$ the radial speed at which the stellar wind initiates. As mentioned in Section~\ref{subsec:Naos_BRITE_probesphotosphere}, $\varv_{\textnormal{\scriptsize \textsc{i}}} = \varv_{\rm sonic}$ and is reached at $r_{\textnormal{\scriptsize \textsc{i}}}\simeq1.027R$ (Figure~\ref{tab:Naos_PoWR_tauscale}). Thus, the wind domain being in the supersonic regime, calculations of line profiles originating from the wind can be performed under the Sobolev approximation: at a given frequency displacement $\nu$ with respect to the line-center frequency $\nu_0$, the emission is dominated by the contribution from locations in the stellar wind where the local component of the velocity field is in resonance with the frequency $\nu$ \citep[see e.g. Equations~38~and~39~in][]{1996A&A...312..195P}. At a given frequency $\nu$, the monochromatic line flux as received by the observer at distance D from the star is:
\begin{equation}
F_\nu = \frac{1}{D^2} \int\limits_{0}^{2\pi} \int\limits_{0}^{\pi} \int\limits_{0}^{\infty} j_\nu(r) r^2dr d\cos\vartheta d\varphi,
\label{eq:Naos_CIRs_LineFlux}
\end{equation}
involving the emissivity $j_{\nu}$, which, under the Sobolev approximation, is expressed as \citep{1983ApJ...274..380R}:
\begin{equation}
j_\nu(r) = \mathcal{F}(r) \mathcal{H}\left( \frac{\varv_r(r)}{c}\nu_0  - |\nu - \nu_0| \right),
\label{eq:Naos_CIRs_emissivity}
\end{equation}
where the presence of the Heaviside function $\mathcal{H}$ accounts for the resonance zones mentioned previously. This also reflects the fact that at a given radial distance $r$, the spherical shell of radius $r$ expanding at velocity $\varv_r(r)$ contributes to the emission as a flat-top profile spanning $\nu_0[1 \pm \varv_r(r)/c]$, weighted by the quantity:
\begin{equation}
\mathcal{F}(r) = \frac{1}{2\nu_0} \frac{c}{\varv_r(r)} \kappa(r) S(r) g(r) \mathcal{P}(r),
\label{eq:Naos_CIRs_FrFunction}
\end{equation}
the factor $c/2\nu_0\varv_r(r)$ arising from wind broadening, $\kappa(r)$ being the integrated line opacity, $S(r)$ the source function, $g(r)$ a function that incorporates all dependences on temperature variations and changes in ionization throughout the wind, while $\mathcal{P}(r)$ denotes the Sobolev escape probability.

For recombination lines such as the He~{\sc ii}~$\lambda4686$ line in $\zeta$~Pup, the product $\kappa(r) S(r)$ scales as the square of the density $\rho(r)$ of the wind. By means of the equation of mass continuity, $\dot{M} = 4 \pi r^2 \varv_r(r) \rho(r)$, the product $\kappa(r) S(r)$ therefore ultimately scales as $1/r^2\varv_r^2(r)$. As for the function $g(r)$, we assume a dependence of the form $[1-\varv_r(r)]^{\alpha_0}$, where the exponent $\alpha_0$ would then counter the $1/\varv_r^2(r)$ dependence to create an asymmetric bell curve in velocity space that localizes the LER. From the investigations of \citet{2012MNRAS.426.1043H}, we know that one has to go at least up to $\sim2.5-3.5R$ to account for the actual shape of the He~{\sc ii}~$\lambda4686$ line profile in $\zeta$~Pup, and the emission falls off slowly beyond that bulk of the wind (D. J. Hillier, private communication). Here we adjust the exponent $\alpha_0$ to roughly localize the LER, although we note that in reality the departure coefficients from conditions of local thermodynamic equilibrium (LTE) may change rapidly and may not be well represented by the simple form of the $g(r)$ function that we adopted. However, it is a reasonable approximation to adopt in the absence of any detailed constraints on the LER. 

In addition, the Sobolev escape probability $\mathcal{P}(r)$ is given by:
\begin{equation}
\mathcal{P}(r) = \frac{1 - e^{-\tau(\vec{r},\vec{n})}}{\tau(\vec{r},\vec{n})},
\label{eq:Naos_CIRs_SobolevEscProb}
\end{equation}
$\tau(\vec{r},\vec{n})$ being the optical depth evaluated at the point of consideration along the direction specified by the unit vector $\vec{n}$, expressed in the Sobolev approximation as:
\begin{equation}
\tau(\vec{r},\vec{n}) = G(r) \times \frac{\rho^2(r)}{| \nabla_{\vec{n}} (\vec{\varv} \cdot \vec{n}) |},
\label{eq:Naos_CIRs_SobolevOptDepth1}
\end{equation}
where $| \nabla_{\vec{n}} (\vec{\varv} \cdot \vec{n}) |$ denotes the velocity gradient along direction $\vec{n}$, while the function $G(r)$ accounts for ionization gradients through the wind. Following \citet{1987ApJ...314..726L}, we adopt the generalized form $G(r) = \varv_r^{\alpha_1}(r)\left[1-\varv_r(r)\right]_{}^{\alpha_2}$, in which the exponents $\alpha_1$ and $\alpha_2$ are in practice adjusted to match the observed shape of the line profile. Thus, taking also into account the equation of mass continuity, the optical depth takes the form:  
\begin{equation}
\tau(\vec{r},\vec{n}) = \mathcal{M} \times \frac{\varv_r^{\alpha_1-2}(r)\left[1-\varv_r(r)\right]_{}^{\alpha_2}}{r^4 | \nabla_{\vec{n}} (\vec{\varv} \cdot \vec{n}) |},
\label{eq:Naos_CIRs_SobolevOptDepth2}
\end{equation}
with $\mathcal{M}=(\dot{M}/4\pi)^2$. Then, in order to evaluate the Sobolev optical depth, the velocity gradient is measured along the line-of-sight ($\vec{n}=\vec{e}_z$ here) and takes the form \citep[e.g.][]{1983ApJ...274..380R,1987ApJ...314..726L,1990A&A...238..191M,1996A&A...312..195P,2000ApJ...537L.131I,2003MNRAS.341..179I}:
\begin{equation}
\nabla_{\vec{n}} (\vec{v} \cdot \vec{n}) = \frac{\partial \varv_z}{\partial z} = \frac{\partial \varv_r(r)}{\partial r} \cos^2 \theta + \frac{\varv_r(r)}{r} \sin^2 \theta,
\label{eq:Naos_CIRs_directionalDerivativeVelocity}
\end{equation}
with $\theta$ denoting the observer polar angle (i.e. the angle between the line of sight, which is the $z$ axis here, and the position vector $\vec{r}$; Figure~\ref{fig:Naos_CIRs_geometry}).

Finally, we note that the calculation of the line flux from Equations~\ref{eq:Naos_CIRs_LineFlux}~and~\ref{eq:Naos_CIRs_emissivity} has to be performed excluding part of the stellar wind material located behind the star which remains unseen by the observer. Occultation happens if the angle $\theta$ is such that:
\begin{equation}
\cos \theta < - \sqrt{1 - \left( \frac{R}{r} \right)^2 },
\label{eq:Naos_CIRs_occultation}
\end{equation}
in which $\cos\theta$ can be expressed in terms of the azimuth angle $\varphi$, the colatitude $\vartheta$ and the stellar inclination angle $i$ through the cosine rule:
\begin{equation}
\cos \theta = \cos\vartheta \cos i + \sin\vartheta \sin i \cos\varphi.
\label{eq:Naos_CIRs_occultation_calpha}
\end{equation}

%%%%%%%%%%%%%%%%%%%%%%%%%%%%%%%%%%%%%%%%%%%%%%%%%%%%%%%%%
\paragraph{Presence of CIR arms\\\\}
\label{paragraph:Naos_Results_Spectro_CIRs_models_CIRarms}

In our simulations, the stellar surface is assumed to rotate rigidly at a constant angular frequency $\Omega=2\pi/P$, and the photospheric source of a CIR arm is assumed to be a bright surface perturbation having a circular shape located at $\mathcal{S}(R,\vartheta_{\rm s},\varphi_{\rm s}+\Omega t)$. The half-opening angle $\psi$ of the CIR is defined to be that of the cone formed by the circular spot and the center of the star (Figure~\ref{fig:Naos_CIRs_geometry}). As already pointed out by \citet{2002A&A...395..209D}, the feature in the CIR that is at the origin of the variability observed in the wind emission lines is the spiral-shaped region of density compression associated with the CIR \citep[region III in Figure~5 of][]{1996ApJ...462..469C}, which in turn can be perceived as a region where the emissivity is increased by a factor $(\eta+1)^2$, with $\eta = (n_{\textnormal{\scriptsize \textsc{cir}}} - n_{\rm sph})/n_{\rm sph}$ being the density contrast between the region of CIR compression of density $n_{\textnormal{\scriptsize \textsc{cir}}}$ and the spherically symmetric smooth wind of density $n_{\rm sph}$. Thus, within the region of CIR compression, the integrand in the calculation of the line flux (Equation~\ref{eq:Naos_CIRs_LineFlux}) becomes $r^2(\eta+1)^2j_{\nu}$. Also, in terms of its shape, the region of CIR compression is taken to follow a spiral with increasing radial distance from the center of the star, while at a given radial distance $r$ the intersection of the spiral-shaped CIR compression and a spherical shell of radius $r$ is assumed to form a circle. The axis of the region of CIR compression is then defined to be the locus composed by the centers of these circlular cross-sections with increasing radial distance $r$. From all these considerations and from the geometry described in Figure~\ref{fig:Naos_CIRs_geometry}, the condition for falling within the region of CIR compression is straightforward:
\begin{equation}
\arccos \chi < \psi,
\label{eq:Naos_CIRs_ConditionInCIRcompression}
\end{equation}
where $\chi$ is the angle between the position vector $\vec{r}$ and the outward normal at the center M of the circular cross-section of the CIR compression at radial distance $r$ (unit vector $\vec{\mu}$ at point M in Figure~\ref{fig:Naos_CIRs_geometry}). Again, by means of the cosine rule, this angle can be expressed in terms of the colatidude $\vartheta$ and the azimuth angle $\varphi$ of the position vector, as well as the colatitude $\vartheta_{\textnormal{\scriptsize \textsc{cir}}}$ and the azimuth angle $\varphi_{\textnormal{\scriptsize \textsc{cir}}}$ of the center of the cross-section of the CIR arm at distance $r$ from the center of the star:  
\begin{equation}
\cos\chi = \cos\vartheta \cos\vartheta_{\textnormal{\scriptsize \textsc{cir}}} + \sin\vartheta \sin\vartheta_{\textnormal{\scriptsize \textsc{cir}}} \cos\left(\varphi_{\textnormal{\scriptsize \textsc{cir}}} - \varphi \right), 
\label{eq:Naos_CIRs_cosChi}
\end{equation}
in which $\vartheta_{\textnormal{\scriptsize \textsc{cir}}}=\vartheta_{\rm s}$ at all times and radial distances, whereas the azimuth angle $\varphi_{\textnormal{\scriptsize \textsc{cir}}}$ carries all the information on the time dependence (owing to the stellar rotation) and the shape of the CIR arm: 
\begin{eqnarray}
&&\varphi_{\textnormal{\scriptsize \textsc{cir}}} = \varphi_{\rm s} + \Omega t - \frac{\sin\vartheta_{\rm s}}{x_{\rm p}} \bigg( \frac{r}{R} - 1 + b \ln \left[ \frac{r}{R} \right] \bigg. \nonumber  \\  
&&~~~~~~~\bigg. + \frac{1-b^2}{b} \ln \left[ (1-b) \frac{\varv_{\infty}}{\varv_r(r)} \right] \bigg),
\label{eq:Naos_CIRs_phiCIR}
\end{eqnarray}
a closed-form expression that follows \citet{2009AJ....137.3339I,2015A&A...575A.129I} in which it is assumed that $\varv_\varphi(r)=0$ for the purpose of solving the spiral shape, due to the complexity of non-radial force considerations and because we do not perform hydrodynamical calculations in our simulations and do not have any real guidance on $\varv_\varphi(r)$. Also, it is worth noting that in Equation~\ref{eq:Naos_CIRs_phiCIR}, the dimensionless quantity $x_{\rm p}=\varv_{\infty}/(\Omega R)$ measures the ratio of the distance travelled by the outer boundary of the stellar wind in one rotation period with respect to the stellar equatorial diameter, thus indicative of the asymptotic pitch angle of the spiral-shaped CIR compression.   
 
%%%%%%%%%%%%%%%%%%%%%%%%%%%%%%%%%%%%%%%%%%%%%%%%%%%%%%%%%
\paragraph{Application to the case of $\zeta$~Pup\\\\}
\label{paragraph:Naos_Results_Spectro_CIRs_models_Results}

First and foremost, we emphasize again that we generate all our CIR models to check the consistency of the observed CIR patterns (left panel of Figure~\ref{fig:Naos_HeII4686_CIRs}) with the bright photospheric spots observed by \emph{BRITE} and mapped by LI. No hydrodynamical calculations and no formal fit of the modelled LPVs to the observed ones are performed here. Moreover, the models have infinite $S/N$, as no noise of instrumental origin or intrinsic to the star (e.g. due to clumps) was added to the models.

Regarding the observed shape of the He~{\sc ii}~$\lambda4686$ emission line profile in $\zeta$~Pup, the presence of a notch close to the line center is noteworthy. Since the line is mainly formed in the stellar wind, the radial expansion of the wind itself can contribute to a blue-shifted P Cygni-type absorption in the line. In that respect, the shape of the line profile can be considered as of type III in \citeauthor{1953PDAO....9....1B}'~(\citeyear{1953PDAO....9....1B}) classification of P Cygni-type line profiles. Additionally, $\zeta$~Pup being a fast rotator, it has been shown that a proper consideration of the azimuthal velocity field can account for the actual shape of some of its photospheric emission line profiles as well as its wind emission lines such as H$\alpha$ and He~{\sc ii}~$\lambda4686$ \citep{2012MNRAS.426.1043H}, the main mechanism involved being the so-called ``resonance zone effect'' as described in detail by \citet{1996A&A...312..195P}, who found that the introduction of the azimuthal velocity field induces a decrease of the optical depth for Doppler velocities near the line center as the corresponding resonance zones are twisted away from the star, whereas the optical depth is increased for higher Doppler velocities as the corresponding resonance zones are twisted towards the star \citep[Figure~6~in][]{1996A&A...312..195P}. As a consequence, the line emission flux is redistributed towards higher Doppler velocities, resulting in the actual shape of the He~{\sc ii}~$\lambda4686$ line having two maxima near $\pm v \sin i$ and a (slightly blue-shifted) notch close to the line center. Therefore, that notch is a feature associated with the unperturbed wind, and not related to the presence of variable small-scale or large-scale wind structures. Thus, in our simulations we do not take into account the resonance zone effect and consider a velocity field dominated by radial expansion as already stated in Section~\ref{paragraph:Naos_Results_Spectro_CIRs_models_lineform}.

The right panel of Figure~\ref{fig:Naos_HeII4686_CIRs} depicts the modelled patterns of LPVs corresponding to CIR arms that would be induced by the bright spots detected by the light curve inversion algorithm for Part~I, Part~III and Part~IV of the observing campaign (Section~\ref{paragraph:Naos_LI_BRITE20142015}, Figure~\ref{fig:Naos_BRITE1415_LI_maps_inc24}). These models correspond to a stellar inclination angle of $24\degr$ and the stellar wind parameters of $\zeta$~Pup listed in Table~\ref{tab:Naos_StellarParams}, the values of the density contrast $\eta$ ranging from $1.5-4$ to match with the typical amplitudes of the observed LPVs, and the exponents $\left(\alpha_0,\alpha_1,\alpha_2\right)=(1,-14,5)$. 

The modelled LPVs are in good agreement with the observed ones, especially in Part~IV where the best phase coverage was achieved in the observations, allowing for the clear detection of the double ``S'' pattern due to the two CIR arms. In the modelled CIR patterns corresponding to Part~I, we notice that we do not reproduce the apparent bump happening around the line center around phase $\sim0.8$. This could be due to the fact that the light curve inversion algorithm might have missed detecting a faint/small spot that would have caused a CIR arm responsible for that bump during Part~I. However, it has to be noted and emphasized that this part of the observing campaign has the worst phase coverage, which plays an important role in the clear detection and characterization of the LPVs.

\begin{figure}
\includegraphics[width=8.4cm]{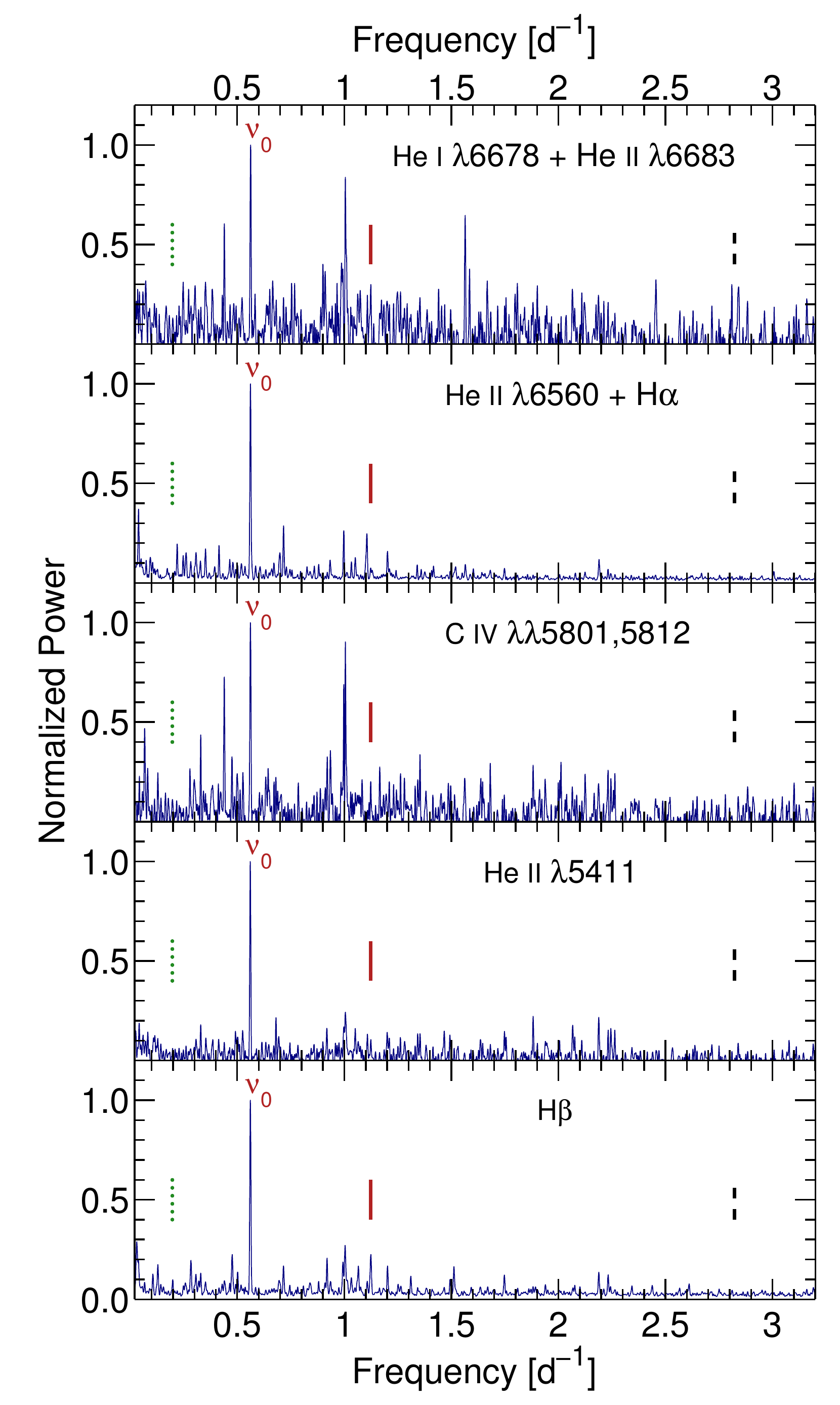}
 \caption{Normalized CLEAN power spectra of the time series of difference profiles of some optical lines in $\zeta$~Pup, integrated over $\pm500$~km~s$^{-1}$ except for He~{\sc ii}~$\lambda6560$+H$\alpha$ for which integration was performed over $\pm1000$~km~s$^{-1}$. The $1.78$~d period due to the stellar rotation is detected in all these lines. The vertical red line points to the location of the first harmonic of the rotation period (here detected in the Balmer lines), while the vertical dotted green line indicates the location of the $5.075 \pm 0.003$~d period (no detection here), and the vertical black dashed line indicates the location of the $8.54 \pm 0.054$~h period (only a possible marginal detection in the noisy and blended He~{\sc i}~$\lambda6678$+He~{\sc ii}~$\lambda6683$ lines).}
   \label{fig:Naos_CLEAN_DFT}
\end{figure}

\begin{figure}
\includegraphics[width=8.4cm]{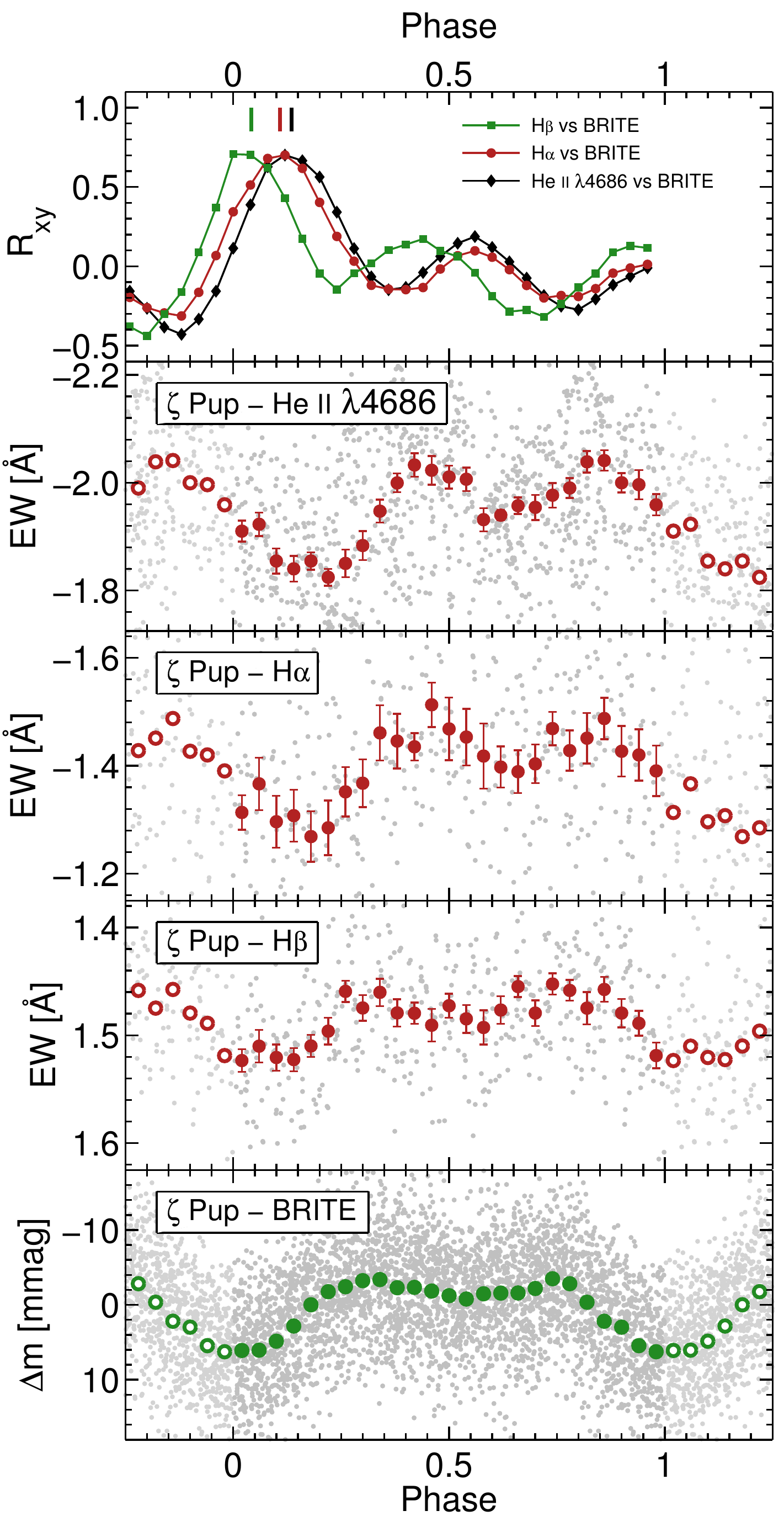}
 %\vspace{-0.5cm}
 \caption{First four panels from bottom to top: light variations of $\zeta$~Pup as observed by \emph{BRITE} phased with the stellar rotation period, along with the phased time series of equivalent widths of the H$\beta$, H$\alpha$, and He~{\sc ii}~$\lambda4686$ line profiles of the star. The vertical scale on the panel for H$\beta$ is exactly twice larger than the vertical scales on He~{\sc ii}~$\lambda4686$ and H$\alpha$. \emph{Upper panel:} Cross-correlation sequences of the equivalent width variations with respect to the light variations observed by \emph{BRITE}. The measured phase delays extracted through Gaussian fits to the first main cross-correlation peaks are indicated by the vertical markers (the Gaussian fits themselves are not overplotted for the sake of visibility).}
   \label{fig:Naos_BRITEvsEW}
\end{figure}

%%%%%%%%%%%%%%%%%%%%%%%%%%%%%%%%%%%%%%%%%%%%%%%%%%%%%%%%%
\subsubsection{Variability in other lines}
\label{subsubsec:Naos_Results_Spectro_CIRs_otherlines}

In view of the behaviour in the phased diagrams for He~{\sc ii}~$\lambda4686$, we explored the variability in other line profiles, some photospheric (e.g. He~{\sc ii}~$\lambda5411$) and some forming in the wind (H$\alpha$). Our CLEAN Fourier analyses across the lines detected the rotation period in most of them (Figure~\ref{fig:Naos_CLEAN_DFT}), and even the first harmonic of the rotation period is detected in the Balmer lines. However, the phased dynamic difference diagrams for these lines, either by considering the entire time series or by splitting it into different parts, does not show any obvious pattern, which might be an indication on the weakness of the signal in these lines, requiring more $S/N$ for their characterization in phased dynamic difference plots. Particularly, in the case of photospheric lines we do not detect any sign of the $\sim8.5$~h periodicity reported by \citet{1986ASIC..169..465B} and \citet{1996A&A...311..616R}, except a possible marginal detection in the blended He~{\sc i}~$\lambda6678$+He~{\sc ii}~$\lambda6683$ line for which the noise level in its CLEAN periodogram is much higher than those of the other lines. 

To further inspect the correlation between the variations observed by \emph{BRITE} and the line profile variations, we extracted the time series of the equivalent widths (EW) of He~{\sc ii}~$\lambda4686$, H$\alpha$ and H$\beta$ (Figure~\ref{fig:Naos_BRITEvsEW}). Not only is the behaviour of the phased EW variations and the phased light variations strongly correlated, but more importantly a noticeable phase delay is observed in the EW variations in the lines most sensitive to the wind (He~{\sc ii}~$\lambda4686$ and H$\alpha$) with respect to the photospheric light variations probed by  \emph{BRITE}. Table~\ref{tab:Naos_BRITEvsEW_XCORR_CIR} lists the outcome of a cross-correlation analysis of the phased EW measurements with the phased \emph{BRITE} light curve averaged in $0.04$ phase bins. From these cross-correlation analyses, we conclude that, besides the noticeable phase delays in He~{\sc ii}~$\lambda4686$ and H$\alpha$ with respect to the light variations probed by \emph{BRITE}, there appears to be a real phase delay even between He~{\sc ii}~$\lambda4686$ and H$\alpha$, and between H$\beta$ and the \emph{BRITE} observations. More importantly, the phase delays observed in Figure~\ref{fig:Naos_BRITEvsEW} and quantified by the cross-correlation analyses are indicative of the azimuthal difference $\Delta\varphi = \varphi_{\textnormal{\scriptsize \textsc{cir}}} - \varphi_{\rm s}$ involved in Equation~\ref{eq:Naos_CIRs_phiCIR}, such that we can solve for $r$ in that relation and thus extract the typical radial distance in the wind at which we probe the CIR arms through H$\beta$, H$\alpha$ and He~{\sc ii}~$\lambda4686$. The resulting values of $r$ are also listed in Table~\ref{tab:Naos_BRITEvsEW_XCORR_CIR}, in which the error bars are the consequences of accounting for the uncertainties on $R$, $\varv_{\infty}$, $\Omega$ and $\Delta\varphi$. It is worth noting that the values that we find for $r$ are in good agreement with the extent of the regions of line formation (up to $\sim2.5-3.5R$ for He~{\sc ii}~$\lambda4686$ and H$\alpha$ as previously mentioned, and closer to the stellar surface for H$\beta$).

\begin{table}
\caption{Phase delays measured from the cross-correlations of the phased equivalent width variations of the H$\beta$, H$\alpha$, and He~{\sc ii}~$\lambda4686$ lines of $\zeta$~Pup with respect to its phased light variations observed by \emph{BRITE}. The values of the phase delays $\Delta \phi$ were assessed through Gaussian fits to the first main cross-correlation peaks, then translated via Equation~\ref{eq:Naos_CIRs_phiCIR} into the radial extent of the CIR arm (measured from the center of the star) probed by each line profile.}
\centering
{\normalsize
\begin{center}
\begin{tabular}{l c c}
\hline
\hline
Line & $\Delta \phi$ & $r$ [R] \\
\hline

H$\beta$ 					& $0.04\pm0.03$ & $1.7\pm0.7$\\
H$\alpha$ 				& $0.11\pm0.01$ & $2.9\pm0.9$\\
He~{\sc ii}~$\lambda4686$	& $0.14\pm0.02$ & $3.5\pm1.1$
\\
\hline
\end{tabular}
\end{center}}
\label{tab:Naos_BRITEvsEW_XCORR_CIR}
\end{table}

Thus, from all these considerations we conclude that the $1.78$~d cyclic intrinsic light variations observed by \emph{BRITE} in $\zeta$~Pup, which we found to occur at the photosphere (Section~\ref{subsec:Naos_BRITE_probesphotosphere}) and are due to rotational modulation related to the presence of bright photospheric inhomogeneities (Section~\ref{subsec:Naos_1.78dperiod}), are the photospheric drivers of the large-scale CIR arms in the stellar wind as traced by the cyclic variations that we detect in He~{\sc ii}~$\lambda4686$ (and H$\alpha$). This implies that our findings constitute the first observational evidence for a direct  link between CIRs in the wind of an O-type star and their photospheric drivers.

%%%%%%%%%%%%%%%%%%%%%%%%%%%%%%%%%%%%%%%%%%%%%%%%%%%%%%%%%
\subsubsection{CIR/DAC recurrence timescales}
\label{subsubsec:Naos_Results_Spectro_CIRs_timescales}

\begin{figure}
\includegraphics[width=8.4cm]{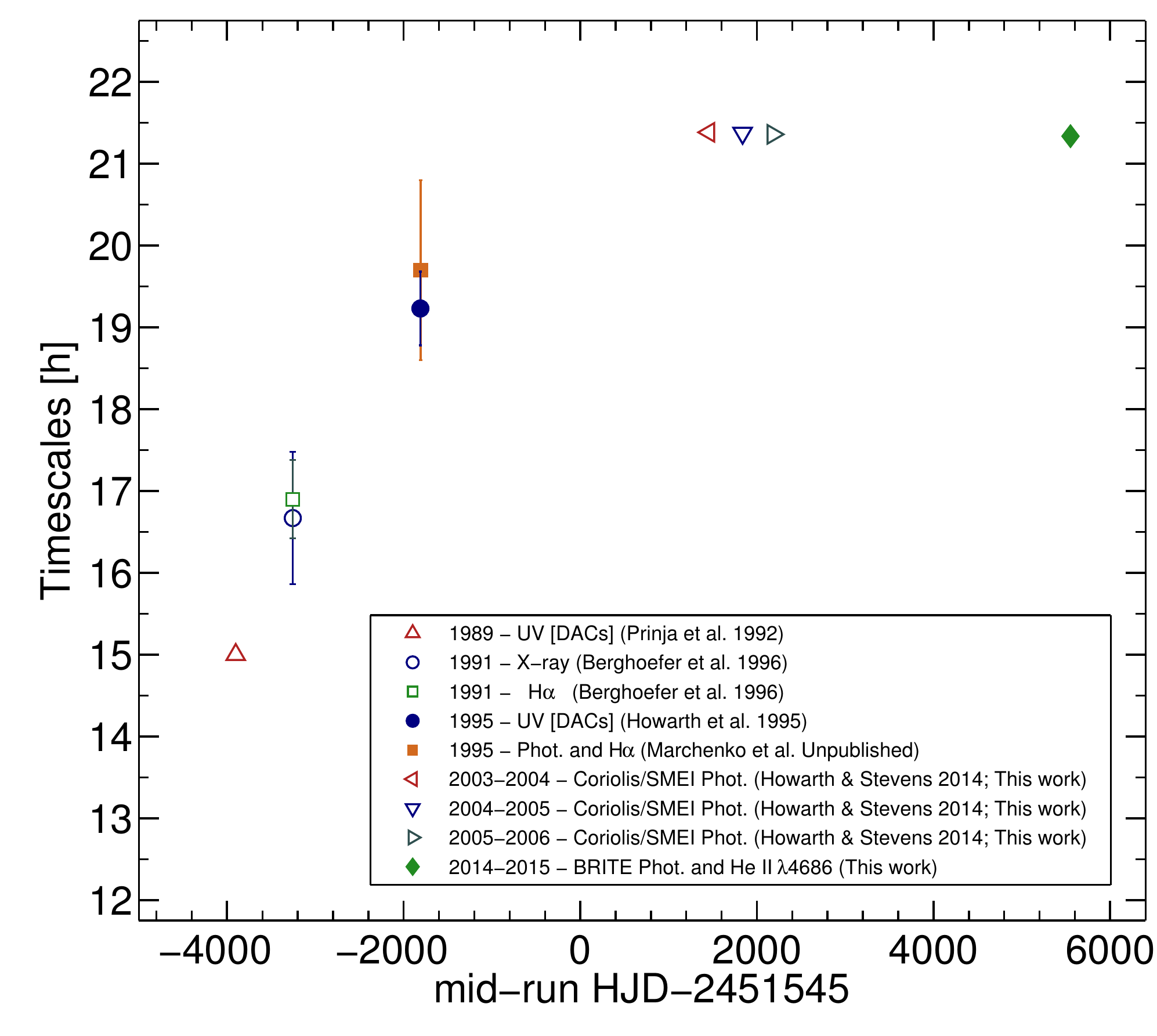}
 \caption{Known ``periodicities'' that might be related to CIRs/DACs in $\zeta$~Pup. Note: The $\sim15$~h timescale reported by \citet{1992ApJ...390..266P} does not have an error bar. The results of the simultaneous ground-based photometry and H$\alpha$ spectroscopy in support of the \emph{IUE} MEGA campaign observations remain unpublished, although they reveal a $19.7(1.1)$~h periodicity that is consistent with the results of the \emph{IUE} observations.}
   \label{fig:Naos_PDACs}
\end{figure}

As mentioned in Section~\ref{subsec:Naos_Intro_spacephot_Ostars}, the presence and blueward recurrent propagation of DACs observed in the absorption troughs of unsaturated resonance lines of most O-type stars is best interpreted as the spectroscopic manifestation of the presence of CIRs in their winds \citep{1984ApJ...283..303M,1996ApJ...462..469C}. Also, previous \emph{IUE} monitoring of $\zeta$~Pup unveiled that the mean DAC recurrence period of the star was $19.23(45)$~h, so that when it was believed that the stellar rotation period is $5.1$~d, the star would exhibit on average $5-6$ DACs per rotation cycle \citep[Figure~1 of ][]{1995ApJ...452L..65H}, a behaviour that deviates from that found in most O-type stars for which an average of two DACs per rotation cycle are generally observed \citep{1999A&A...344..231K}. Now our results that the true rotation period of $\zeta$~Pup is much shorter, $P = 1.78063(25)$~d, changes that picture. It is worth noting that the consecutive bumps observed in the phased \emph{BRITE} light curves and in the EW variations of the wind emission lines have a separation that varies between $0.4-0.5$ in phase. For instance, a separation of $\Delta \phi = 0.45$ in rotational phase would correspond to two ``events'' separated by a time interval of $\Delta t = 19.2308(27)$~h, which is compatible with the DAC recurrence timescale of $19.23(45)$~h reported by \citet{1995ApJ...452L..65H}, but then in that configuration the next ``event'' would have to happen after $\Delta \phi^{\prime} = 0.55$, i.e $\Delta t^{\prime} = 23.5043(33)$~h later, such that the Fourier analysis of the time series over several cycles would still give rise to a peak at the first harmonic ($21.3676(57)$~h), not at the $\sim19.23$~h timescale. Additionally, it has to be considered that \citet{1992ApJ...390..266P} identified a DAC recurrence timescale of $\sim15$~h (no error bar available) from the examination of $\sim2.2$~d of intensive observations within a $5.5$-day contiguous \emph{IUE} monitoring of $\zeta$~Pup in $1989$.  Thus, all the information available to date appears to point out that the average inter-DAC ``period'' of $\zeta$~Pup has not always been strictly the same over the past three decades. In Figure~\ref{fig:Naos_PDACs} we plot the known ``periodicities'' that \emph{might be associated} with CIRs/DACs in $\zeta$~Pup. The change in ``periodicity'' seems gradual over about two decades, increasing from $\sim15$ to $21$~h, although rapidly at first, then asymptotic later. In this plot we have also marked the location of the $16.67\pm0.81$~h and $16.90 \pm 0.48$~h periodicities reported by \citet{1996A&A...306..899B} from a simultaneous X-ray and H$\alpha$ monitoring of $\zeta$~Pup spanning $11$~d in $1991$, as they appear to follow the global trend in the plot, although at this point the physical origin of these periodicities is not well-established and more importantly has never been formally proven to be related to CIRs/DACs. Also, for the epoch of the \emph{Coriolis}/SMEI observations, we have explicitly marked the location of the first harmonic of the $1.78$~d signal, but one has to keep in mind that only the fundamental frequency has a high statistical significance in these datasets.

The global trend in Figure~\ref{fig:Naos_PDACs} is unlikely related to a slowing down (especially if non-uniform in time) of the stellar rotation period: the mass-loss rate is simply too low to allow this and there is no plausible or conceivable internal mechanism to cause such a large change in rotation over such a short time.  On the other hand, looking at the behaviour of the Sun in terms of rotation and spot activity, we see clear surface differential rotation from a period of $\sim25$~d at the equator to $\sim35$~d at the poles, with spot migration from mid to low latitude over an $11$-year cycle.  Among late-type stars in general, one does see both solar-type surface differential rotation \citep[faster rotation at the equator: e.g.][]{1997MNRAS.291....1D,2004A&A...417.1047K,2004MNRAS.351..826P,2011AJ....141..138R} and anti-solar type of differential rotation \citep[slower rotation at the equator: e.g.][]{2001A&A...374..171H,2003A&A...408.1103S,2016A&A...592A.117H}. Inspired by this, we \emph{speculate} that $\zeta$~Pup with its very rapid overall rotation could also exhibit differential rotation with the \emph{dominant} spots wandering quickly first from higher latitudes with faster rotation to spots wandering slowly to lower latitudes with slower rotation at the bulging equator. In that scenario, the \emph{dominant} spots observed at a given epoch of observation would be located around the same latitude, forming an ``active latitude'', analogous to the Sun's active latitudes that change over time to form the so-called ``Butterfly diagram'' of the Sun's activity cycle. That being said, the surface maps that we obtained through the light curve inversion of the \emph{Coriolis}/SMEI and \emph{BRITE} light curves of $\zeta$~Pup also reveal the presence of spots at mid-latitudes. However, one important point that we emphasize here is that the light curve inversion algorithm as it stands currently allows for rigid rotation only, although it still can reveal differential rotation from the relative drifts in spot longitudes between different data sets \citep{2011AJ....141..138R}. Nevertheless, on this note we leave further speculation in anticipation of future long-term monitoring of this ever-more interesting massive star.

%%%%%%%%%%%%%%%%%%%%%%%%%%%%%%%%%%%%%%%%%%%%%%%%%%%%%%%%%
\subsection{Small-scale wind structures}
\label{subsec:Naos_Results_Spectro_Clumps}

\begin{figure*}
\includegraphics[width=18cm]{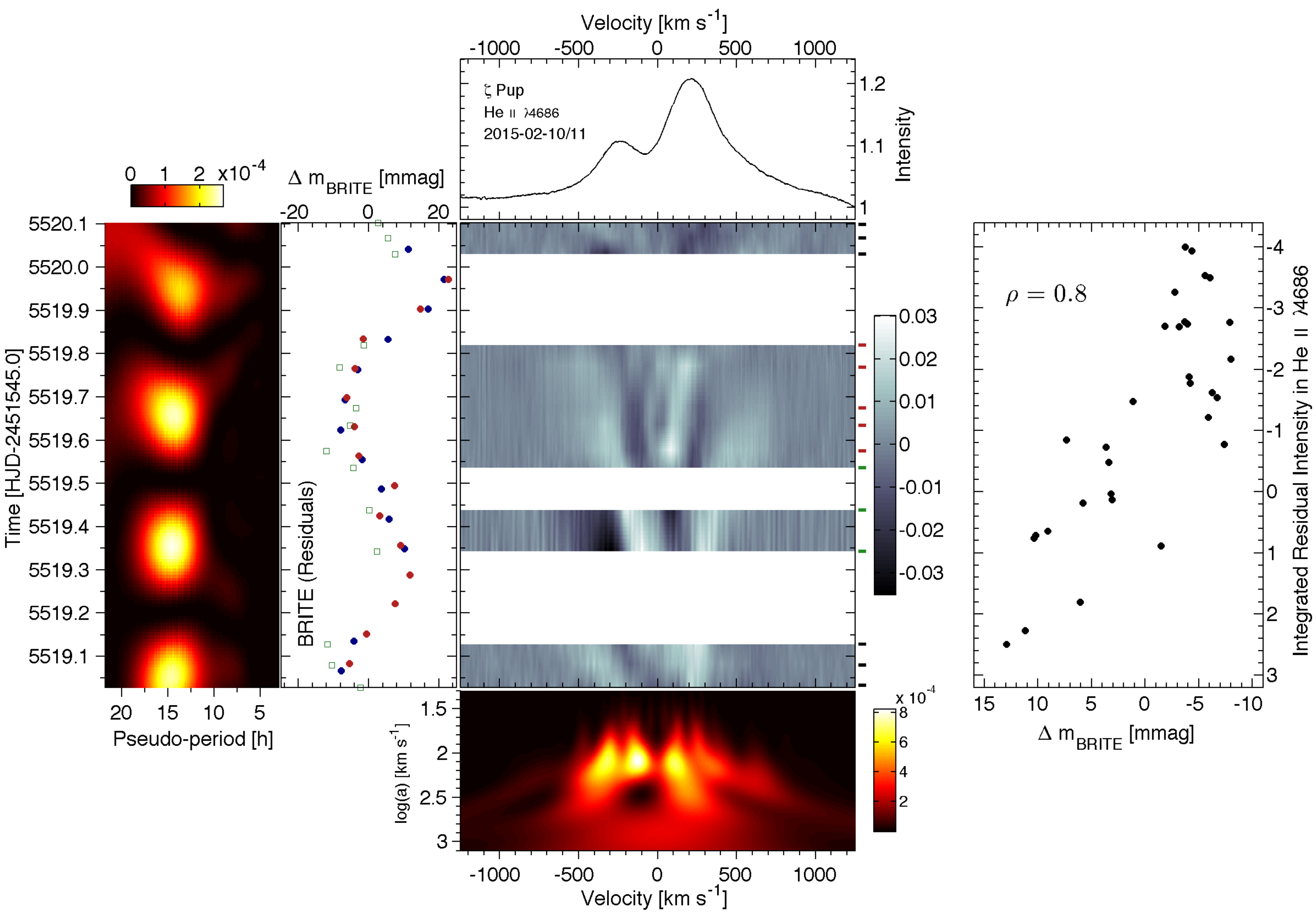}
 %\vspace{-0.5cm}
 \caption{Kinematics of outmoving wind clumps in $\zeta$~Pup as traced in He~{\sc ii}~$\lambda4686$, along with their photospheric drivers as observed by \emph{BRITE} during the night of February 10/11, 2015. \emph{Top:} Average profile of the He~{\sc ii}~$\lambda4686$ line for the night. \emph{Middle:} Dynamic difference spectra from the average spectrum after removal of the effects of the $1.78$~d cyclic variations due to large-scale wind structures (Section~\ref{subsubsec:Naos_Results_Spectro_Clumps_dynamicspectra}). Small horizontal ticks on the right side of the diagram indicate the time sampling, with observations performed from Shenton Park Observatory (Black), SAAO~1.9m/GIRAFFE (Green) and CTIO-SMARTS~1.5m/CHIRON (Red). \emph{Bottom:} Average scalogram from the Ricker wavelet-based CWTs of the spectra (in the spatial/wavelength domain). \emph{Left:} Light variations of $\zeta$~Pup recorded by \emph{BRITE} in the blue (Blue points) and red (Red points) filters during the night of February 10/11, 2015. The light curves illustrated here are those free of the effects of the $1.78$~d signal associated with rotational modulation (Section~\ref{subsec:Naos_stochastic_variability}, Figures~\ref{fig:Naos_BRITE_residuals}~and~\ref{fig:Naos_BRITE_residuals_CWT}). Green squares represent the integrated intensity variations in the difference spectra within the velocity range $\pm1000$~km~s$^{-1}$, with the scaling factor $C=-3$ (Equation~\ref{eq:Naos_IntegratedResidualIntensity}, Section~\ref{subsubsec:Naos_Results_Spectro_Clumps_rhoBRITELPVs}). \emph{Far Left:} Portion of the scalogram of the residual \emph{BRITE} light curve restricted to that specific night (taken from Figure~\ref{fig:Naos_BRITE_residuals_CWT}), showing that the dominant timescale during that night was $\sim14-15$~h. There seems to be a slight pseudo-period drift during the night, as already noted in Section~\ref{subsubsec:Naos_stochastic_variability_timescales}. \emph{Right:} Strong correlation between the integrated intensity variations in He~{\sc ii}~$\lambda4686$ and the light variations observed by \emph{BRITE}.}
  \label{fig:Naos_HeII4686_Clumps_0210}
\end{figure*}

The densest temporal coverage for the multi-site ground-based spectroscopic monitoring of $\zeta$~Pup that we conducted in parallel with the \emph{BRITE} observing run was achieved during $\sim5$~d in February 2015 (HJD $2,457,059.94 - 2,457,065.10$) when we had optimal longitude coverage from facilities at CTIO-SMARTS~1.5m/CHIRON, SAAO~1.9m/GIRAFFE and SASER. We took this opportunity to look for signatures of small-scale wind structures in He~{\sc ii}~$\lambda4686$, as previously found by \citet{1998ApJ...494..799E}.

%%%%%%%%%%%%%%%%%%%%%%%%%%%%%%%%%%%%%%%%%%%%%%%%%%%%%%%%%
\subsubsection{Dynamic difference spectra}
\label{subsubsec:Naos_Results_Spectro_Clumps_dynamicspectra}

\begin{figure*}
\includegraphics[width=18cm]{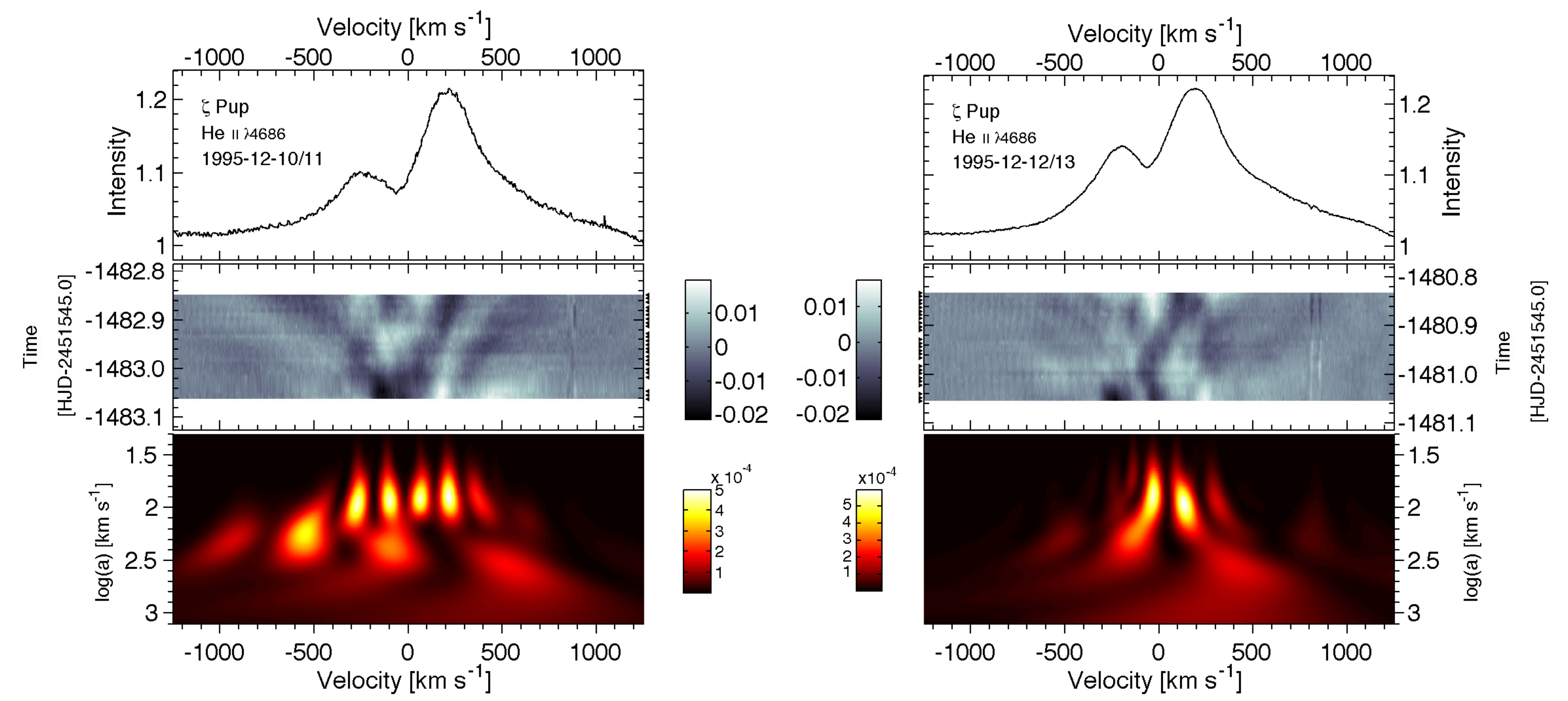}
 %\vspace{0.35cm}
 \caption{Outmoving wind clumps in $\zeta$~Pup as first reported by \citet{1998ApJ...494..799E} from two nights of CFHT~3.6m/Reticon observations (archival data) in December 10/11, 1995 (\emph{Left}) and December 12/13, 1995 (\emph{Right}). For a given night, the \emph{top panel} illustrates the mean spectrum for the night, the \emph{middle panel} shows the dynamic difference spectra from the average spectrum, and the \emph{bottom panel} depicts the average scalogram. The scalogram signatures of the clumps are the same as observed during the \emph{BRITE} $2014-2015$ observing run, with a dominant scale at $a\sim10^{2.0}-10^{2.1}$~km~s$^{-1}$.}
  \label{fig:Naos_HeII4686_Clumps_CFHT95}
\end{figure*}

Our spectra have a typical $S/N\sim500$ in the continuum close to He~{\sc ii}~$\lambda4686$, which is about a factor two lower than the $S/N$ in the CFHT~3.6m/Reticon spectra of $\zeta$~Pup in which \citet{1998ApJ...494..799E} found manifestations of wind clumps. Still, we detect signatures of wind clumps in our observations, as clearly revealed by the dynamic difference spectra in the main panel of Figure~\ref{fig:Naos_HeII4686_Clumps_0210} for the night of February 10/11, 2015 (HJD $ 2,457,064.03 - 065.10$) during which the signatures are the most discernible: features in emission excess that appear at random both in Doppler velocity and in time, and propagate away from the line center towards the red/blue edges while smearing out and dropping in intensity with time until they reach Doppler velocities $\sim\pm500$~km~s$^{-1}$ from where they seem to fade away. That the features seem to disappear beyond velocities $\sim\pm500$~km~s$^{-1}$ does not necessarily mean that the actual clumps causing these signatures physically vanish, as the LER in which we see their manifestations only spans a limited Doppler velocity range that clearly does not reach the terminal wind speed in the case of He~{\sc ii}~$\lambda4686$ for $\zeta$~Pup which has $\varv_{\infty} = 2300\pm100$~km~s$^{-1}$. Also, another striking property that  we notice in the dynamic difference spectra on the main panel of Figure~\ref{fig:Naos_HeII4686_Clumps_0210} is that the velocity widths are narrower for features located close to the line center (e.g. those that appear at time $\sim 5519.50$) compared to those features that appear at the wings of the emission line, a behaviour also seen in other hot stars \citep{1994RvMA....7...51M,1994Ap&SS.216...55M,1999ApJ...514..909L} and best interpreted as the manifestation of anisotropic velocity dispersion seen in Doppler projection.   

We note that, at first, from a preliminary look at the dynamic difference spectra we wondered how the cyclic $1.78$~d signal associated with large-scale wind structures influences the shape of the features related to clumps. The major difficulty here is the evolutive nature of the CIRs in time (ultimately as a consequence of the evolution of their photospheric drivers as discussed in Sections~\ref{subsec:Naos_1.78dperiod}~and~\ref{subsec:Naos_Results_Spectro_CIRs}), but more importantly the $\sim5$~d intensive observations including the night of February 10/11, 2015, turn out to be right in the middle of the transition phase in Part~III where we only see mild signature of CIRs in the phased diagrams, unlike in Part~IV where the CIR signatures are more clearly defined. Nevertheless, we proceeded with the removal of the observed mild CIR patterns in Part~III by adopting the same approach as we used for the removal of the $1.78$~d signal in the light curves (Section~\ref{subsec:Naos_stochastic_variability}), but this time performed for each Doppler velocity in the range $\pm1275$~km~s$^{-1}$: we phase-folded the time series spectra in Part~III with the rotation period, rebinned them in phase, interpolated over the original phase sampling, subtracted the resulting smoothed pattern from the original phased time series, and then went back in the time domain to compare the resulting corrected time series with the original time series in order to check for possible overcorrections in which case we restarted again the whole procedure with another phase bin size. In general, during the comparison step we noted that the effect of the $1.78$~d variability on the $\sim1.08$~d time span of the February 10/11, 2015 night was a slight increase of intensity between times $\sim 5519.70 - 5520.10$. The phase bin size is the key parameter for controlling overcorrections and undercorrections, since too small phase bin sizes result in templates with relatively large noisy fluctuations (leading to overcorrection), while too large phase bin sizes yield too smooth templates (yielding undercorrection). We found that a phase bin size of $0.05$ was a good compromise, yielding the dynamic difference spectra illustrated on the main panel of Figure~\ref{fig:Naos_HeII4686_Clumps_0210}.

%%%%%%%%%%%%%%%%%%%%%%%%%%%%%%%%%%%%%%%%%%%%%%%%%%%%%%%%%
\subsubsection{Wavelet analysis}
\label{subsubsec:Naos_Results_Spectro_Clumps_CWT}

Given the stochastic nature of the appearance of the spectroscopic features associated with the wind clumps, and also their morphology as small features in emission excess superimposed on the wind emission line, the wavelet formalism (here in the spatial/wavelength domain) is very appropriate for the extraction of the properties of the observed clumps, as was already performed in other investigations on spectroscopic observations of WR stars \citep{1994Ap&SS.221..371L,1994RvMA....7...51M,1994Ap&SS.216...55M,1996ApJ...466..392L,1999ApJ...514..909L}. Also, unlike the case of the residual light curve on which we performed a wavelet analysis to characterize the timescales of the features composing the observed stochastic signal intrinsic to $\zeta$~Pup (Section~\ref{subsubsec:Naos_stochastic_variability_timescales}), our spectra have a constant dispersion in wavelength\footnote{In general, raw echelle spectra  do not have a constant dispersion through the different orders, but are resampled with a constant dispersion during the process of merging of the orders and extraction of the one-dimensional wavelength-calibrated spectra.} and thus are intrinsically equally sampled, such that the continuous formalism can be applied and the CWTs readily calculated. Still, the choice of an appropriate wavelet family to be used needs to be considered. Since the subpeaks that we want to characterize in the difference spectra are excesses in emission often accompanied by less strong side depressions (unlike the features in the light curve that have roughly equal bumps and dips between $\pm10$~mmag), the simplest and most appropriate family of wavelets for this situation is the family of Ricker wavelets, also often called Mexican hat wavelets. The Ricker wavelet function is constructed from a normalization of the second derivative of a Gaussian function. It was also the type of family of wavelets adopted by the studies on observations of clumping in hot stars mentioned previously. Under these considerations we calculated the CWTs of the $14$ spectra during the night of February 10/11, 2015 and extracted the mean scalogram which is depicted in the lower panel of Figure~\ref{fig:Naos_HeII4686_Clumps_0210}, confirming that the substructures are mostly located between $\sim\pm500$~km~s$^{-1}$, but more importantly that the typical widths of the detected features as indicated in the scalogram are of the order of $10^{2.0} - 10^{2.1}$~km~s$^{-1}$ which is slightly higher but remains consistent with the value of $\sim10^{2.0}$~km~s$^{-1}$ found by \citet{1999ApJ...514..909L} for the clumps in WR stars.

Also in order to fully compare the characteristics of the clumps of $\zeta$~Pup observed in our campaign with the clumps detected in the star at the epoch of the observations performed by \citet{1998ApJ...494..799E}, we re-extracted the archival CFHT~3.6m/Reticon optical spectra of the star taken during the nights of December 10/11, 1995 (HJD $2,450,061.94 - 062.15$) and December 12/13, 1995 (HJD $ 2,450,063.95 - 064.17$), and calculated the corresponding average Ricker wavelet-based scalograms, which are displayed in Figure~\ref{fig:Naos_HeII4686_Clumps_CFHT95} below the corresponding dynamic difference spectra. The average scalograms for the CFHT~3.6m/Reticon spectra indicate that the typical widths of the features are also $\sim10^{2.0}$~km~s$^{-1}$, with some isolated minor power detected at larger scales $\sim10^{2.3}$~km~s$^{-1}$. The presence of such broader features could be due to either nesting effects or contamination from large-scale wind structures, the latter being impossible to remove in the CFHT~3.6m/Reticon spectra since the gap between the two nights of observations is of the order of one rotational cycle and each night only spanned $\sim5$~h.

%%%%%%%%%%%%%%%%%%%%%%%%%%%%%%%%%%%%%%%%%%%%%%%%%%%%%%%%%
\subsubsection{Correlated light variations and LPVs}
\label{subsubsec:Naos_Results_Spectro_Clumps_rhoBRITELPVs}  

An inspection of the behaviour of the stochastic light variations of $\zeta$~Pup as recorded by \emph{BRITE} (Section~\ref{subsec:Naos_stochastic_variability}) during the night of February 10/11, 2015 shows a relatively wide bump that reaches a maximum at time $\sim 5519.70$ (Left panel of Figure~\ref{fig:Naos_HeII4686_Clumps_0210}), and the Morlet wavelet-based scalogram of the \emph{BRITE} light curve indicates that the dominant pseudo-period during that night is $\sim14-15$~h with a slight pseudo-period drift towards shorter timescales during the night (Far Left panel of Figure~\ref{fig:Naos_HeII4686_Clumps_0210}). As already noted in Section~\ref{subsubsec:Naos_stochastic_variability_timescales}, that pseudo-period drift, of unexplained origin, is part of a drift that starts from time $\sim 5517.5$ at timescales $\sim16$~h and finishes at time $\sim 5521.0$ at timescales $\sim11$~h. Nevertheless, in order to quantify the correlation between the stochastic light variations observed by \emph{BRITE} in $\zeta$~Pup and the observed clumps in the wind, we assessed the integrated residual intensity variations in each difference spectrum over Doppler velocities $\pm1000$~km~s$^{-1}$:
\begin{equation}
\mathcal{I}_{k}= \frac{C}{\varv_2-\varv_1}\int_{\varv_1}^{\varv_2} I_r(\varv,t_{k}) d\varv
\label{eq:Naos_IntegratedResidualIntensity}
\end{equation}
in which $I_r\left(\varv,t_{k}\right)$ is the residual intensity at Doppler velocity $\varv$ in the difference spectrum observed at time $t_{k}$, and the integration is performed over velocities ranging from $\varv_1=-1000$~km~s$^{-1}$ to $\varv_2=+1000$~km~s$^{-1}$. The constant $C$ is a negative normalization factor accounting for the comparison with the \emph{BRITE} light variations which are expressed in magnitudes. The individual values of $\mathcal{I}_{k}$ for the $14$ spectra in the night are displayed as green squares on the left panel of Figure~\ref{fig:Naos_HeII4686_Clumps_0210} along with the residual \emph{BRITE} light curves, taking $C=-3$. In view of that panel it appears that correlation exists between the two, an impression that is confirmed by the resulting Pearson correlation coefficient value $\rho=0.8$ (Right panel of Figure~\ref{fig:Naos_HeII4686_Clumps_0210}). 

Since \emph{BRITE} probes the light variations of $\zeta$~Pup at the level of its photosphere (Section~\ref{subsec:Naos_BRITE_probesphotosphere}), we conclude that the stochastic light variations detected by \emph{BRITE} in the star, which we characterized in Section~\ref{subsec:Naos_stochastic_variability}, are the manifestations of the photospheric drivers of the clumps in the stellar wind as traced in He~{\sc ii}~$\lambda4686$.

%%%%%%%%%%%%%%%%%%%%%%%%%%%%%%%%%%%%%%%%%%%%%%%%%%%%%%%%%
\section{Discussion and conclusion}
\label{sec:Naos_Discussion}

\begin{figure}
\begin{center}
\includegraphics[width=7.1cm]{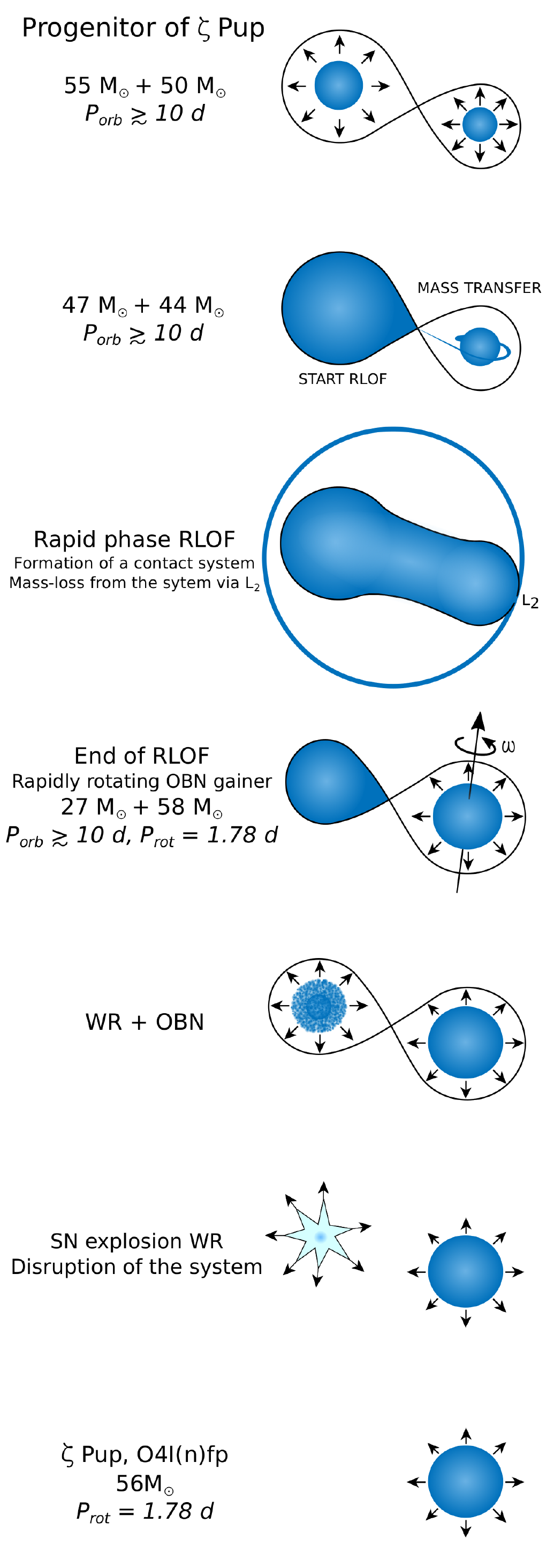}
\end{center}
 %\vspace{-0.5cm}
 \caption{Evolutionary history of $\zeta$~Pup in the scenario in which the star could have been the secondary component of a massive binary and would have been spun-up during a Roche lobe overflow phase of the primary component (fourth row from the top). The latter later underwent a supernova explosion, leading to the current fast rotator runaway status of the $\zeta$~Pup. Figure adapted from \citet{1998ASSL..232.....V}, updated with a rotation period of $1.78$~d, and satisfying the observed runaway velocity $\| \vec{v}_{\rm rw} \| = 60.0\pm16.9$~km~s$^{-1}$ and the stellar mass $M=56.1^{+14.5}_{-11.6} M_{\sun}$ listed in Table~\ref{tab:Naos_StellarParams}.}
   \label{fig:Naos_EvolutionaryScenario}
\end{figure}

The relatively short rotation period of $\zeta$~Pup, $P = 1.78063(25)$~d, that we inferred from the photometric observations has a strong impact on our current understanding of the properties of the star. Most importantly the now better constrained relatively low stellar inclination angle $i\simeq24\degr$ and high equatorial velocity (Section~\ref{subsubsec:Naos_pulsations_vs_rotational_modulation}) completely change the previous beliefs that the star has a rotation period of $\sim5.1$~d, is seen almost equator on and spins at $\varv_{\rm e}\simeq \varv_{\rm e} \sin i = 219\pm18$~km~s$^{-1}$. Moreover, as mentioned earlier (Section~\ref{subsubsec:Naos_pulsations_vs_rotational_modulation}), the short rotation period of $\zeta$~Pup is in good agreement with the suspected rotational evolution of the star, with two proposed scenarios so far: either spin-up during a Roche lobe overflow phase in a WR+O close massive binary followed by ejection from the supernova explosion of the primary component leading to the current status of $\zeta$~Pup as a single fast rotator runaway \citep{1998ASSL..232.....V}, or a spin-up resulting from the merger of at least two massive stars during dynamical interactions within the Vela R2 R-association \citep{2009Ap&SS.324..271V,2012ASPC..465..342V,2012A&A...538A..75P}. In the former case, \citet{1998ASSL..232.....V} estimated that, with initial binary masses of $40M_{\sun}$ and $38M_{\sun}$ and an initial orbital period of $4$~d, the resulting orbital period of the system at the end of the Roche-lobe overflow phase is $2.6$~d, in which case tidal effects would be very large and synchronize the components of the system such that the rotation period of the secondary would also be tuned to $2.6$~d \citep[Figure~14.15 in][page 232]{1998ASSL..232.....V}. This is of the same order of magnitude as the actual value of the rotation period that we inferred from the photometric observations in this investigations, $P = 1.78063(25)$~d. The exact values of the initial binary masses implied by this post-RLOF rotation period depend on the exact values of the physical parameters involved during the RLOF, e.g. the exact amount of mass and angular momentum loss from the binary. However, in that scenario, the runaway velocity acquired by the secondary after the supernova explosion (asymmetrical) of the primary has to satisfy \citep[][Equations~11.4~and~11.17]{1998ASSL..232.....V}:
\begin{equation}
\| \vec{\varv}_{\rm rw} \| = \frac{M_{\rm WR}}{\sqrt{M_{\rm WR} + M_{\rm O}}} \left( \frac{\mathcal{G}}{A} \right)^\frac{1}{2}
\label{eq:Naos_RunawayV}
\end{equation}
in which the masses $M_{\rm WR}$ and $M_{\rm O}$, as well as the semi-major axis $A$ are values just before the supernova explosion (here $\mathcal{G}$ denotes the gravitational constant). Assuming pre-supernova masses $M_{\rm WR}=10M_{\sun}$ and $M_{\rm O}=56M_{\sun}$, the runaway velocity corresponding to an orbital period of $\sim1.78$~d would be $\sim 107$~km~s$^{-1}$, which is too high compared to the observed runaway velocity of the star ($\| \vec{\varv}_{\rm rw} \| = 60.0\pm16.9$~km~s$^{-1}$), while only an orbital period $\gtrsim10$~d would yield a runaway velocity that is more compatible with the observed one. The latter case remains a plausible scenario since in that situation tidal effects are very small and the rotation period of the gainer (which was spun-up to $\sim1.78$~d during the RLOF phase) remains unchanged. Additionally, in order to satisfy the observed current mass of $\zeta$~Pup, $M=56.1^{+14.5}_{-11.6} M_{\sun}$ (despite the huge uncertainty on that value), the initial binary masses have to be much larger than those assumed in Figure~14.15~of~\citet{1998ASSL..232.....V}, e.g. $55M_{\sun}$ and $50M_{\sun}$. In that case, it has to be assumed that the $55M_{\sun}$ primary ends its life with a supernova explosion that disrupted the binary, which is not entirely excluded as there is some evidence that black hole formation (at least for the lowest mass black holes) can be accompanied by a supernova explosion \citep[e.g.][]{1999Natur.401..142I}. Under these circumstances, the updated scenario is illustrated in Figure~\ref{fig:Naos_EvolutionaryScenario}. We emphasize that we performed these calculations only to update this scenario to satisfy the most pertinent values of the stellar parameters  of $\zeta$~Pup, but at this point no strong constraints is available to help decide which of the two proposed evolutionary scenarios is the most plausible one.   

With respect to the physical mechanism at the origin of the photospheric bright spots that give rise to the rotationally modulated light variations observed by \emph{BRITE} and \emph{Coriolis}/SMEI, one possible candidate is the surface emergence of small-scale magnetic fields generated through dynamo action within a subsurface convection zone associated with the iron-opacity bump at $T\sim150000$~K \citep[FeCZ:][]{2009A&A...499..279C,2011A&A...534A.140C}. The detection of such small-scale magnetic fields requires high-resolution, high-S/N circular spectropolarimetric observations imposed by the fast rotation of $\zeta$~Pup. All previous spectropolarimetric observations of the star ended up with no detection, with an upper limit of $121$~G for a dipolar field strength, and longitudinal field error bar of $21$~G \citep{2014MNRAS.444..429D}. Also since most of the spots that we detect during the \emph{BRITE} and \emph{Coriolis}/SMEI observing runs lie near the equator and are separated by $\sim 144-180\degr$ in longitude, their magnetic signature (if they are related to small-scale magnetic fields) might cancel out in low-resolution mode as was the case of the spectropolarimetric monitoring of $\zeta$~Pup performed by \citet{2016ApJ...822..104H} who, despite no magnetic detection and no period detection, found marginal signs of modulation of the longitudinal magnetic field measurements when they phase-folded their data with the $1.78$~d period (rather than the $5.1$~d period). More importantly, as mentioned in Section~\ref{subsec:Naos_Intro_spacephot_Ostars} previous photometric monitoring of O stars revealed light variations best interpreted as signs of rotational modulation possibly due to surface inhomogeneities, with only the case of $\xi$~Per showing potential link to its CIR/DAC behaviour \citep{2014MNRAS.441..910R}, but our present findings that the bright localized photospheric spots observed by \emph{BRITE} and \emph{Coriolis}/SMEI in $\zeta$~Pup are the photospheric drivers of its CIRs/DACs constitutes the first observational evidence for a link between large-scale wind structures in an O-type star and their photospheric origin. This reinforces the need to observe more O-type stars with known CIR/DAC recurrence timescale \citep[e.g. those studied by][]{1999A&A...344..231K} on a long-term through high-precision photometric monitoring and contemporaneous high-S/N mid/high-resolution spectroscopic/spectropolarimetric monitoring, because the near-ubiquity of CIRs/DACs in O-type stars implies that their photospheric drivers must also be a common phenomenon.

Lastly, the detection of the stochastically excited (but coherent for several hours) light variations intrinsic to $\zeta$~Pup during the \emph{BRITE} and \emph{Coriolis}/SMEI observations is as important as the cyclic variations related to rotational modulation itself, since we also find for the first time that those randomly excited (but coherent for several hours) light variations are the photospheric drivers of stellar wind clumps. Although the physical nature of these stochastic intrinsic variations at the photosphere remains unknown, there is a possibility that such variations arise from randomly excited stellar oscillations with finite lifetimes such as IGWs, generated within a FeCZ \citep{2009A&A...499..279C} or even from the convective core in the case of IGWs as found by \citet{2015ApJ...806L..33A}. Nevertheless, the fact that these stochastic photospheric variations are linked to the variability associated with wind clumps suggests that the latter are formed even at the very base of the wind. Previous X-ray observations of $\zeta$~Pup suggested that only a large number of clumps in the wind ($N>10^5$) are able to reproduce the observed X-ray variability of the star \citep{2013ApJ...763..143N}. Since the variability scales with the inverse of the square root of the number of clumps involved, we can infer by the optical line flux variability that we observe ($\sim6\%$) that it would arise from the contribution of $N\simeq300$ clumps. The question therefore remains  as whether those clumps are of equal size or follow a power law distribution in which there is a relatively small number of strongly emitting clumps which would dominate the variability. The latter case fits naturally into the notion of the scaling laws of turbulence (compressible for hot winds) with a power law of negative slope, making few large clumps along with a rapidly increasing number of smaller ones each of which emits less and less as the size is diminished \citep[e.g.][]{1994RvMA....7...51M}. This is further supported by the fact that turbulence is a very common phenomenon in astrophysics, e.g. large-scale structures in the Universe, the intergalactic medium (Ly-$\alpha$ clouds towards quasars), galaxies, cloudlets in giant molecular clouds, the discrete nature of stars themselves (as implied by the IMF) or surface granulation in the Sun itself. Also, the fairly clear and distinct observed spectral emission-line excess streaks over several hours must be spatially confined and not due to e.g. a sum of smaller random perturbations anywhere in the wind. However it has also been shown that the first scenario, that is variability that arises from a large number of clumps of equal size, cannot be ruled out \citep[e.g.][]{2007A&A...469.1045D}. To further test the hypothesis of turbulence, high-S/N high-resolution spectroscopic observations of a WR star with strong emission line and high terminal wind speed (e.g. up to $\sim5000$~km~s$^{-1}$) needs to be performed, the best candidates being one of the WO2 stars in the Milky Way. But given the faintness of these objects, observations with the upcoming generation of extremely large telescopes such as the Thirty Meter Telescope (TMT) would be the most ideal.

\section*{Acknowledgements}

We gratefully acknowledge fruitful discussions with D. John Hillier, Ya\"{e}l Naz\'e and Rados\l{}aw Smolec in relation to this investigation. We also thank the reviewer, Richard Townsend, for his insights and for providing us with useful constructive remarks. T. Ramiaramanantsoa acknowledges support from the Canadian Space Agency (CSA) grant FAST. A. F. J. Moffat is grateful to the Natural Sciences and Engineering Research Council of Canada (NSERC) and the Fonds de Recherche du Qu\'{e}bec - Nature et Technologies (FRQNT) for general financial aid, and to CSA for funding the training of highly qualified personnel associated with the \emph{BRITE} project. N. D. Richardson acknowledges postdoctoral support by the University of Toledo and by the Helen Luedtke Brooks Endowed Professorship. A. Pigulski acknowledges support from the National Science Centre grant No. 2016/21/B/ST9/01126. A. Popowicz acknowledges NCN grant  2016/21/D/ST9/00656. G. Handler acknowledges support from NCN grant 2015/18/A/ST9/00578. The Polish contribution to the \emph{BRITE} mission is supported by the Polish NCN grant 2011/01/M/ST9/05914. G. A. Wade and S. M. Rucinski acknowledge Discovery Grant support from NSERC. K. Zwintz acknowledges support from the Austrian Fonds zur F\"{o}rderung derwissenschaftlichen Forschung (FWF; project V431-NBL). This investigation is based on data collected by the \emph{BRITE-Constellation} satellite mission, designed, built, launched, operated and supported by the Austrian Research Promotion Agency (FFG), the University of Vienna, the Technical University of Graz, the Canadian Space Agency (CSA), the University of Toronto Institute for Aerospace Studies (UTIAS), the Foundation for Polish Science \& Technology (FNiTP MNiSW), and National Science Centre (NCN). This project also uses observations made from the South African Astronomical Observatory (SAAO). The professional authors of this paper are grateful to the amateur astronomers of the SASER team (also co-authors of this paper), who invested personal time and enthusiasm in this project. Part of the research leading to these results has received funding from the European Research Council (ERC) under the European Union's Horizon 2020 research and innovation programme (grant agreement N$^\circ$670519: MAMSIE).

%%%%%%%%%%%%%%%%%%%%%%%%%%%%%%%%%%%%%%%%%%%%%%%%%%



%\bibliographystyle{mnras}


\noindent \textbf{AUTHORS' AFFILIATIONS:}

\noindent $^{1}$ D\'epartement de physique, Universit\'e de Montr\'eal, CP 6128, Succursale Centre-Ville, Montr\'eal, Qu\'ebec, H3C 3J7\\
$^{2}$ Centre de Recherche en Astrophysique du Qu\'ebec (CRAQ), Canada\\
$^{3}$ Department of Physics and Astronomy, Ohio Wesleyan University, Delaware, OH 43015, USA\\
$^{4}$ East Tennessee State University, Department of Physics \& Astronomy, Johnson City, TN, 37614, USA\\
$^{5}$ Astrophysical Institute, Vrije Universiteit Brussel, Pleinlaan 2, B-1050 Brussels, Belgium\\
$^{6}$ Institut f\"{u}r Physik und Astronomie, Universit\"{a}t Potsdam, Karl-Liebknecht-Str. 24/25, D-14476 Potsdam, Germany\\
$^{7}$ Ritter Observatory, Department of Physics and Astronomy, The University of Toledo, Toledo, OH 43606-3390, USA\\
$^{8}$ Department of Physics \& Astronomy, University College London, Gower St, London WC1E 6BT, United Kingdom\\
$^{9}$ School of Physics and Astronomy, University of Birmingham, Edgbaston, Birmingham B15 2TT, United Kingdom\\
$^{10}$ Schn\"{o}rringen Telescope Science Institute, Waldbr\"{o}l, Germany\\
$^{11}$ Instytut Astronomiczny, Uniwersytet Wroc\l{}awski, Kopernika 11, 51-622 Wroc\l{}aw, Poland\\
$^{12}$ Instytut Automatyki, Politechnika \'Sl\c{a}ska, Akademicka 16, 44-100 Gliwice, Poland\\
$^{13}$ Institut f\"{u}r Astrophysik, Universit\"{a}t Wien, T\"{u}rkenschanzstrasse 17, 1180 Wien, Austria \\
$^{14}$ Nicolaus Copernicus Astronomical Center, Bartycka 18, 00-716 Warsaw, Poland\\
$^{15}$ LESIA, Observatoire de Paris, PSL Research University, CNRS, Sorbonne Universit\'es, UPMC Univ. Paris 06, Univ. Paris Diderot, Sorbonne Paris Cit\'e, 5 place Jules Janssen, F-92195 Meudon, France\\
$^{16}$ Instituut voor Sterrenkunde, KU Leuven, Celestijnenlaan 200D, 3001 Leuven, Belgium\\
$^{17}$ Department of Physics, Royal Military College of Canada, PO Box 17000, Station Forces, Kingston, Ontario, K7K 7B4, Canada\\
$^{18}$ Department of Astronomy \& Astrophysics, University of Toronto, 50 St. George Street, Toronto, Ontario, M5S 3H4, Canada\\
$^{19}$ Institut f\"{u}r Astro- und Teilchenphysik, Universit\"{a}t Innsbruck, Technikerstrasse 25/8, 6020 Innsbruck, Austria\\
$^{20}$ International Centre for Radio Astronomy Research, The University of Western Australia, 35 Stirling Hwy, Crawley, Western Australia 6009\\
$^{21}$ SASER Team, Domain Observatory, 269 Domain Road, South Yarra, Vic 3141, Australia\\
$^{22}$ SASER Team, Dogsheaven Observatory, SMPW Q25 CJ1 LT10B, Brasilia, Brazil\\
$^{23}$ SASER Team, Latham Observatory, Australia\\
$^{24}$ SASER Team, R. F. Joyce Observatory, Canterbury Astronomical Society, West Melton, New Zealand\\
$^{25}$ SASER Team, Mirranook Observatory, Booroolong Road, Armidale, NSW, Australia\\
$^{26}$ Gemini Observatory, Northern Operations Center, 670 N. A'ohoku Place, Hilo, Hawaii, 96720, USA\\
$^{27}$ South African Astronomical Observatory, PO Box 9, Observatory, 7935 Cape, South Africa\\
$^{28}$ Southern African Large Telescope, PO Box 9, Observatory, 7935 Cape, South Africa\\
$^{29}$ Eureka Scientific, Inc., 2452 Delmer Street, Oakland, CA 94602, USA\\
$^{30}$ Universit\'e de Toulouse; UPS-OMP; IRAP; Toulouse, France\\
$^{31}$ CNRS; IRAP; 14, avenue Edouard Belin, 31400 Toulouse, France\\

%%%%%%%%%%%%%%%%%%%%%%%%%%%%%%%%%%%%%%%%%%%%%%%%%%

%%%%%%%%%%%%%%%%% APPENDICES %%%%%%%%%%%%%%%%%%%%%

%%%%%%%%%%%%%%%%%%%%%%%%%%%%%%%%%%%%%%%%%%%%%%%%%%%%%%%%%
%%%%%%%%%%%%%%%%%%%%%%%%%%%%%%%%%%%%%%%%%%%%%%%%%%%%%%%%%
\appendix
%%%%%%%%%%%%%%%%%%%%%%%%%%%%%%%%%%%%%%%%%%%%%%%%%%%%%%%%%
%%%%%%%%%%%%%%%%%%%%%%%%%%%%%%%%%%%%%%%%%%%%%%%%%%%%%%%%%

%%%%%%%%%%%%%%%%%%%%%%%%%%%%%%%%%%%%%%%%%%%%%%%%%%%%%%%%%
\section{Evolution of the $1.78$-day signal during the \emph{Coriolis}/SMEI observing run}
\label{sec:Naos_Evolution178d_SMEI}

\begin{figure*}
\includegraphics[width=18cm]{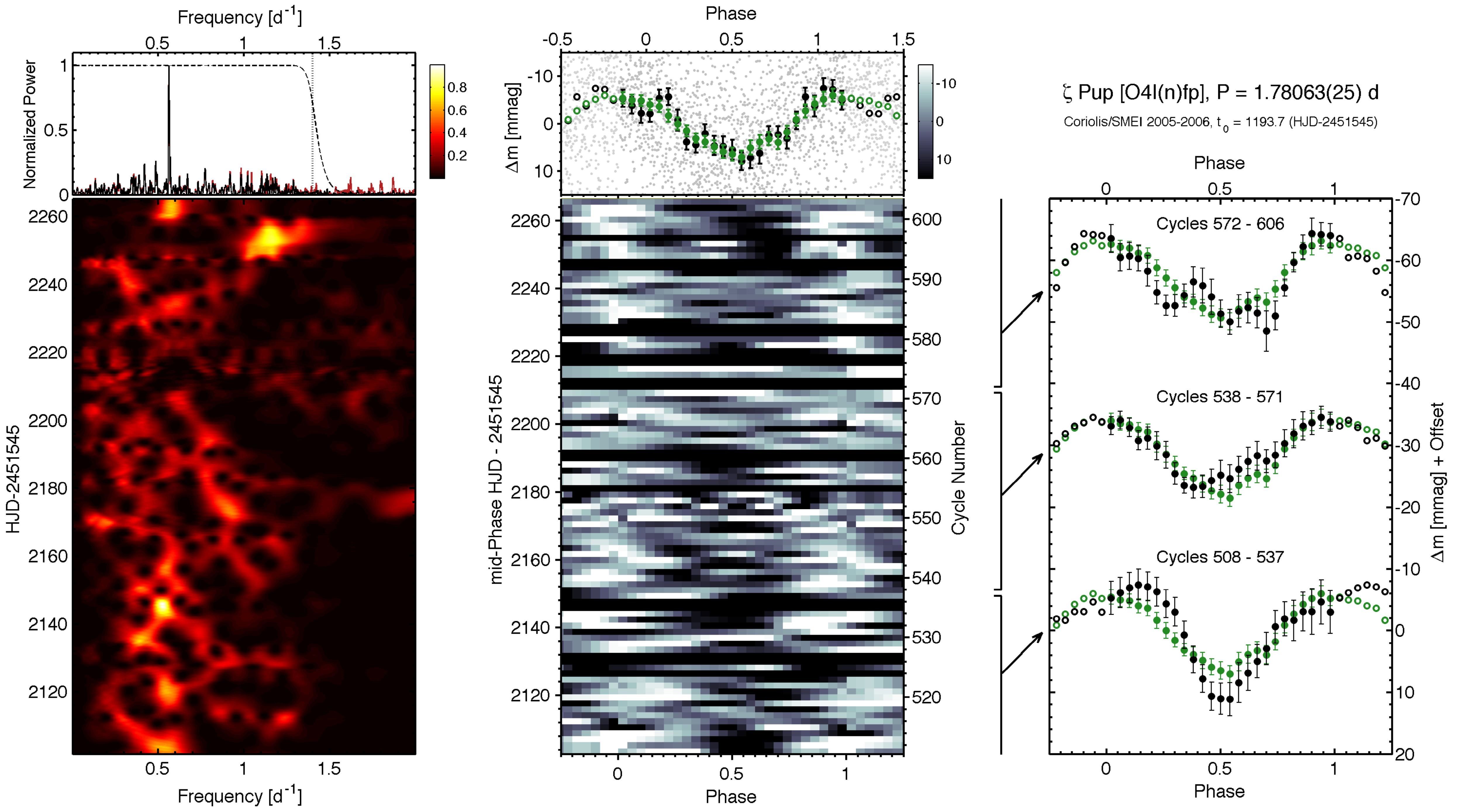}
 %\vspace{-0.5cm}
 \caption{Same as Figure~\ref{fig:Naos_BRITE_lcs_dynamic} but for the \emph{Coriolis}/SMEI observing run in 2005-2006.}
  \label{fig:Naos_SMEI20052006_lcs_dynamic}
\end{figure*}

\begin{figure*}
\includegraphics[width=18cm]{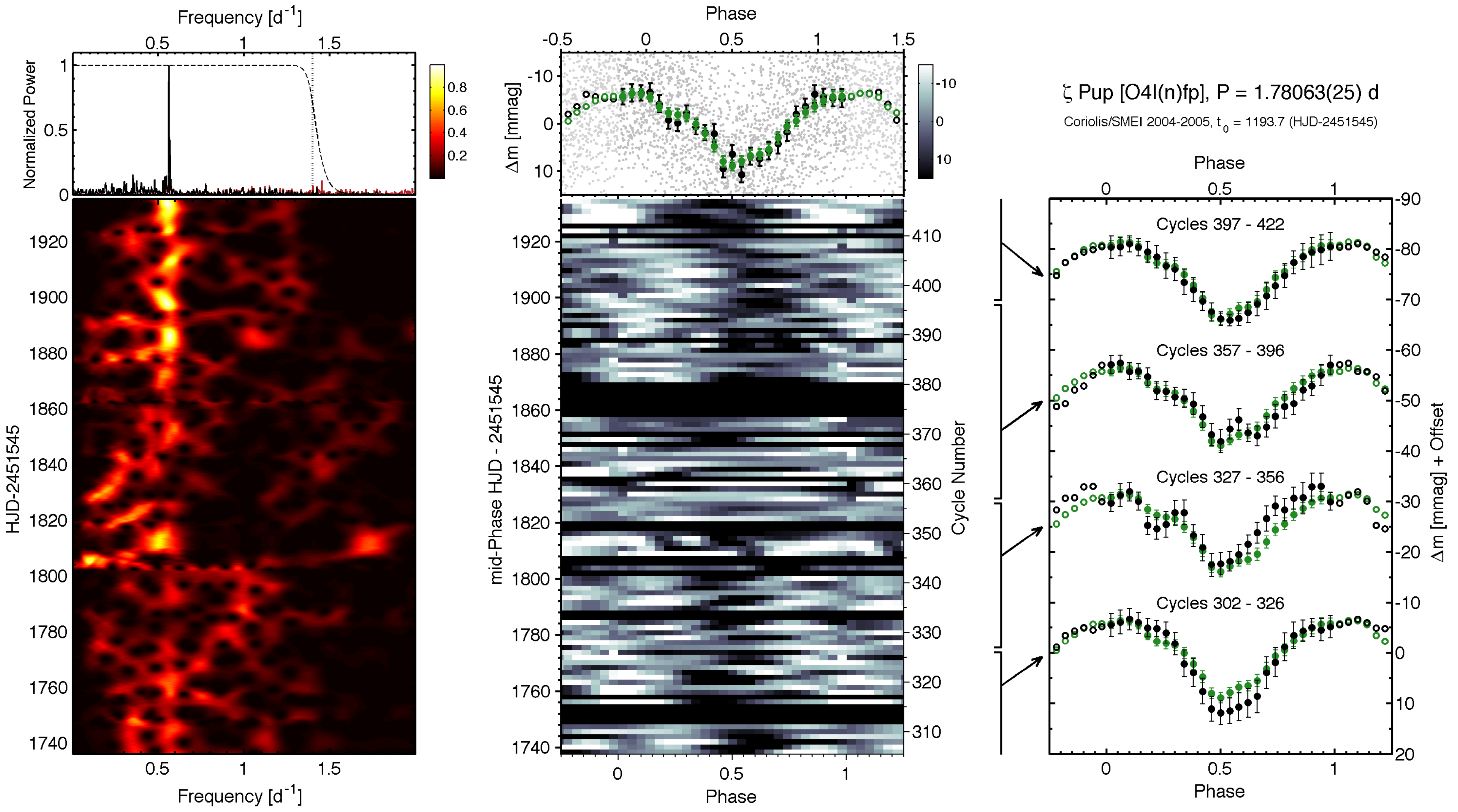}
% \vspace{-0.5cm}
 \caption{Same as Figure~\ref{fig:Naos_BRITE_lcs_dynamic} but for the \emph{Coriolis}/SMEI observing run in 2004-2005.}
  \label{fig:Naos_SMEI20042005_lcs_dynamic}
\end{figure*}

\begin{figure*}
\includegraphics[width=18cm]{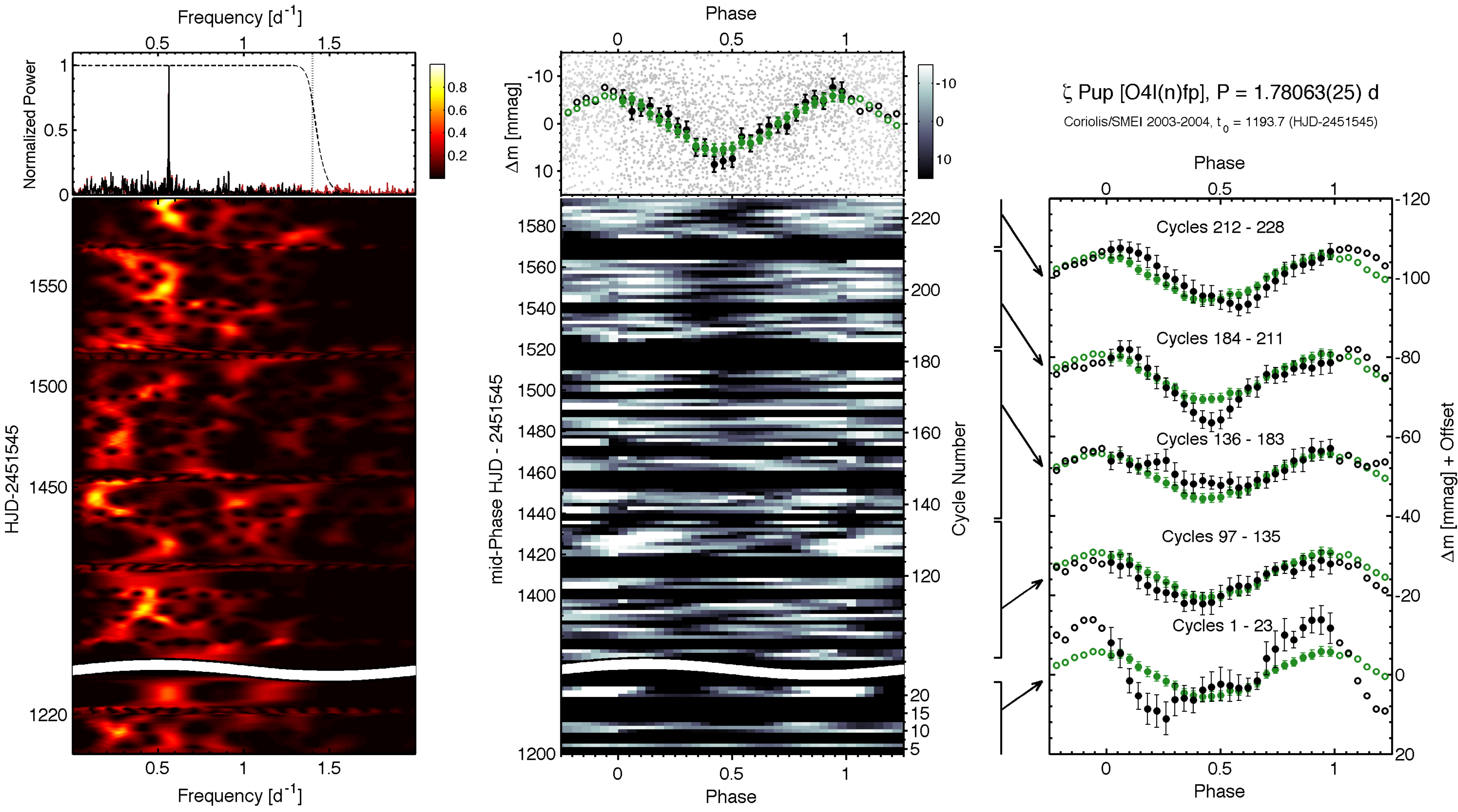}
 %\vspace{-0.5cm}
 \caption{Same as Figure~\ref{fig:Naos_BRITE_lcs_dynamic} but for the \emph{Coriolis}/SMEI observing run in 2003-2004.}
  \label{fig:Naos_SMEI20032004_lcs_dynamic}
\end{figure*}

\clearpage
%%%%%%%%%%%%%%%%%%%%%%%%%%%%%%%%%%%%%%%%%%%%%%%%%%%%%%%%%
\section{Light curve inversion: Surface maps of $\zeta$~Pup during the \emph{Coriolis}/SMEI seasonal observing runs}
\label{sec:Naos_LI_SMEI}

\begin{figure*}
\includegraphics[width=18cm]{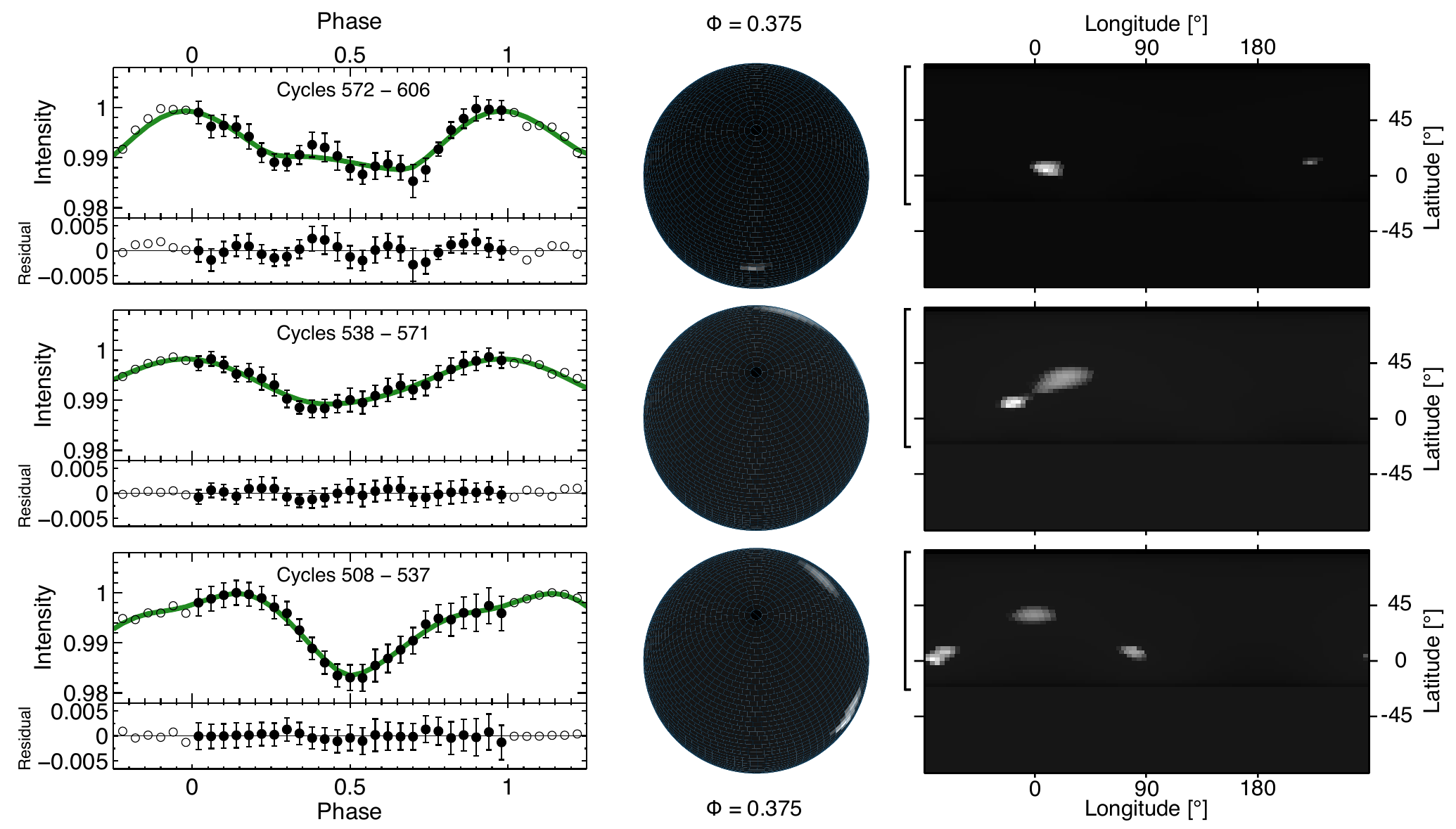}
% \vspace{-0.5cm}
 \caption{Light curve inversion: mapping the photosphere of $\zeta$~Pup as observed by \emph{Coriolis}/SMEI in $2005-2006$, for $T^{\rm (s)}=42.5$~kK and a stellar inclination angle $i=24\degr$. Time increases upwards.}
  \label{fig:Naos_SMEI0506_LI_maps}
\end{figure*}

\begin{figure*}
\includegraphics[width=18cm]{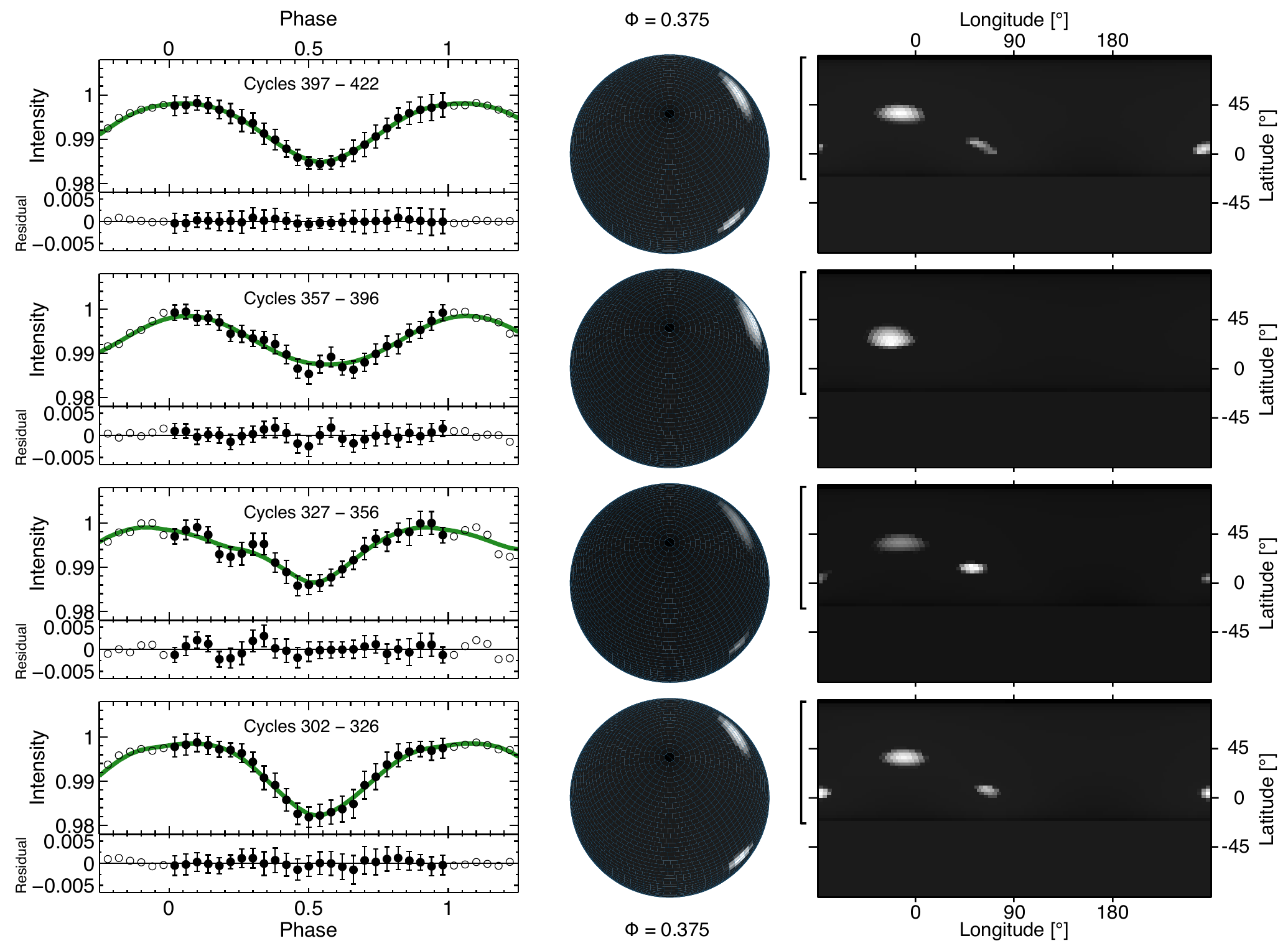}
 %\vspace{-0.5cm}
 \caption{Light curve inversion: mapping the photosphere of $\zeta$~Pup as observed by \emph{Coriolis}/SMEI in $2004-2005$, for $T^{\rm (s)}=42.5$~kK and a stellar inclination angle $i=24\degr$. Time increases upwards.}
  \label{fig:Naos_SMEI0405_LI_maps}
\end{figure*}

\begin{figure*}
\includegraphics[width=18cm]{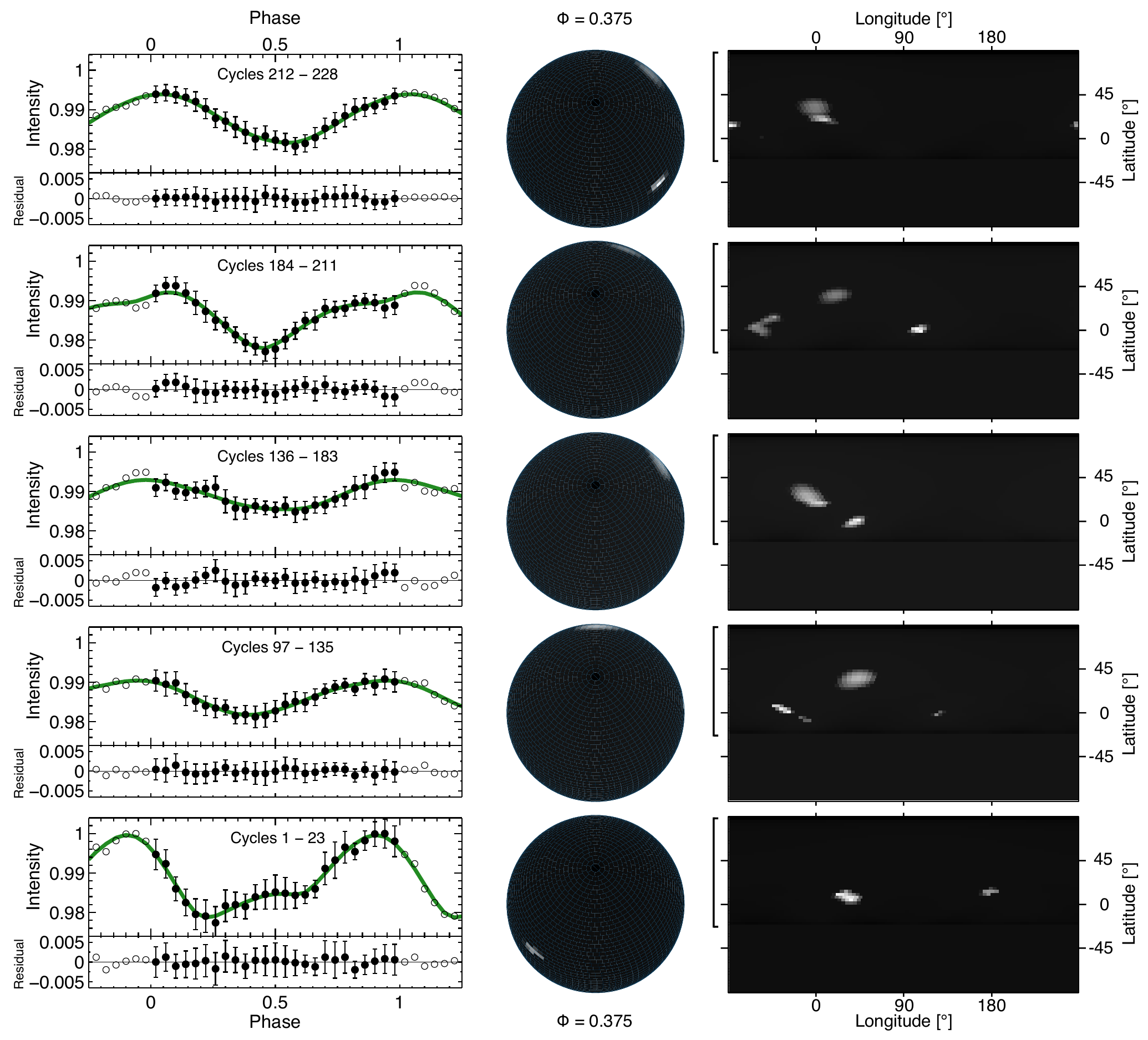}
 %\vspace{-0.5cm}
 \caption{Light curve inversion: mapping the photosphere of $\zeta$~Pup as observed by \emph{Coriolis}/SMEI in $2003-2004$, for $T^{\rm (s)}=42.5$~kK and a stellar inclination angle $i=24\degr$. Time increases upwards.}
  \label{fig:Naos_SMEI0304_LI_maps}
\end{figure*}

%%%%%%%%%%%%%%%%%%%%%%%%%%%%%%%%%%%%%%%%%%%%%%%%%%

% Don't change these lines
\bsp	% typesetting comment
\label{lastpage}
\end{document}